\documentclass{jfm}
\usepackage{natbib}
\usepackage{upmath}
\usepackage[british]{babel}
\usepackage{amsmath,bm}
\usepackage{amssymb}
\usepackage{amsbsy}
\usepackage{amscd}
\usepackage{amstext}
\usepackage{tabularx}
\usepackage{float}
\usepackage{makeidx}
\usepackage{amsmath}
\usepackage{subfigure}
\usepackage{afterpage}
\usepackage[T1]{fontenc}
\usepackage[latin1]{inputenc}
\usepackage{multirow}
\usepackage{color}

\usepackage{graphicx} 
\usepackage{graphics,epsfig}
\usepackage{amsfonts}
\usepackage{psfrag}

\ifCUPmtlplainloaded \else
  \checkfont{eurm10}
  \iffontfound
    \IfFileExists{upmath.sty}
      {\typeout{^^JFound AMS Euler Roman fonts on the system,
                   using the 'upmath' package.^^J}%
       \usepackage{upmath}}
      {\typeout{^^JFound AMS Euler Roman fonts on the system, but you
                   dont seem to have the}%
       \typeout{'upmath' package installed. JFM.cls can take advantage
                 of these fonts,^^Jif you use 'upmath' package.^^J}%
      }
  \else
  \fi
\fi

\ifCUPmtlplainloaded \else
  \checkfont{msam10}
  \iffontfound
    \IfFileExists{amssymb.sty}
      {\typeout{^^JFound AMS Symbol fonts on the system, using the
                'amssymb' package.^^J}%
       \usepackage{amssymb}%

      }{}
  \fi
\fi

\ifCUPmtlplainloaded \else
  \IfFileExists{amsbsy.sty}
    {\typeout{^^JFound the 'amsbsy' package on the system, using it.^^J}%
     \usepackage{amsbsy}}
    {\providecommand\boldsymbol[1]{\mbox{\boldmath $##1$}}}
\fi

\newsavebox{\astrutbox}
\sbox{\astrutbox}{\rule[-5pt]{0pt}{20pt}}

\title[Modal and nonmodal stability analysis of electrohydrodynamic flow]
{Modal and nonmodal stability analysis of electrohydrodynamic flow with and without cross-flow}

\author[M. Zhang, F. Martinelli, J. Wu, P.J. Schmid and M. Quadrio]%
{
\ns M\ls E\ls N\ls G\ls Q\ls I\ns Z\ls H\ls A\ls N\ls G$^1$\thanks{Email address for correspondence: mengqi.zhang@univ-poitiers.fr},
\ns F\ls U\ls L\ls V\ls I\ls O\ns M\ls A\ls R\ls T\ls I\ls N\ls E\ls L\ls L\ls I\ls $^2$,
\break
\ns J\ls I\ls A\ls N\ns W\ls U$^1$,\ns
\ns P\ls E\ls T\ls E\ls R\ns J.\ns S\ls  C\ls H\ls M\ls I\ls D\ls$^3$\ns
\break
\and M\ls A\ls U\ls R\ls I\ls Z\ls  I\ls O\ns Q\ls U\ls A\ls D\ls R\ls  I\ls O$^2$
}

\affiliation{
  $^1$ D{\'e}partement Fluides, Thermique, Combustion, Institut PPrime,
  CNRS-Universit{\'e} de Poitiers-ENSMA, Poitiers, France\\
  $^2$ Dipartimento di Scienze e Tecnologie Aerospaziali del Politecnico di Milano, 
  via La Masa 34, 20156 Milano, Italy\\
  $^3$ Department of Mathematics, Imperial College London, London, SW7 2AZ, United Kingdom
  \\[\affilskip]
}

\date{\today}

\graphicspath{{figures/}}
\newlength\savewidth

\begin{document}
\maketitle

\begin{abstract}
  We report the results of a complete modal and nonmodal linear
  stability analysis of the electrohydrodynamic flow (EHD) for the
  problem of electroconvection in the strong injection
  region. Convective cells are formed by Coulomb force in an insulating liquid residing between two plane
  electrodes subject to unipolar injection. Besides pure electroconvection, 
  we also consider the case where a cross-flow is present, generated by a streamwise
  pressure gradient, in the form of a laminar Poiseuille flow. The
  effect of charge diffusion, often neglected in previous linear
  stability analyses, is included in the present study and a transient
  growth analysis, rarely considered in EHD, is carried out. In the
  case without cross-flow, a non-zero charge diffusion leads to a
  lower linear stability threshold and thus to a more unstable
  flow. The transient growth, though enhanced by increasing charge
  diffusion, remains small and hence cannot fully account for the
  discrepancy of the linear stability threshold between theoretical
  and experimental results. When a cross-flow is present, increasing
  the strength of the electric field in the high-$Re$ Poiseuille flow
  yields a more unstable flow in both modal and nonmodal stability
  analyses. Even though the energy analysis and the input-output
  analysis both indicate that the energy growth directly related to
  the electric field is small, the electric effect enhances the
  lift-up mechanism. The symmetry of channel flow with respect to the
  centerline is broken due to the additional electric field acting in
  the wall-normal direction. As a result, the centers of the
  streamwise rolls are shifted towards the injector electrode, and the
  optimal spanwise wavenumber achieving maximum transient energy
  growth increases with the strength of the electric field.
\end{abstract}

\section{Introduction}

\subsection{General description of EHD flow}

Electrohydrodynamics (EHD) is concerned with the interaction between
an electric field and a flow field. Such configurations have broad
applications in a range of industrial and biological devices. EHD
effects can be used to enhance the heat transfer
efficiency~\citep{Jones1978,Allen1995}, to design microscale
electrohydrodynamic pumps~\citep{Bart1990,Darabi2002}, to fabricate
diagnostic devices and drug delivery systems~\citep{Chakraborty2009}
and DNA microarrays~\citep{Lee2006}, and to design new strategies for
active flow control~\citep{Bushnell1989}. Physically, EHD flow is
characterized by a strong nonlinear interaction between the velocity
field, the electric field and space charges: the electric force
results in flow motion, which in turn affects the charge
transport. The intricate nature of this nonlinearity defies a
fundamental understanding of EHD flow. Moreover, as we will see, there
still remains a mismatch or discrepancy between experimental
observations and a theoretical analysis.

One classic problem in EHD, named electroconvection, deals with the
convective motions induced by unipolar charge injection into a
dielectric liquid (of very low conductivity) which fills the gap
between two parallel rigid plane electrodes. The Coulomb force acting
on the free charge carriers tends to destabilize the
system. Electroconvection is often compared to Rayleigh-B{\'e}nard
convection (RBC) because of their similar geometry and convection
patterns. Moreover, RBC is known to be analogous to the Taylor-Couette
(TC) flow in the gap between two concentric rotating cylinders, where
thermal energy transport in RBC corresponds to the transport of
angular momentum in TC flow \citep{Bradshaw1969,GROSSMANN2000}. In the
linear regime of RBC, the flow is destabilized by the buoyancy force
caused by the continued heating of the lower wall (an analogous role
is played by centrifugal force in TC flow). As the thermal gradient
exceeds a critical value, chaotic motion sets in. In EHD flow, the
destabilizing factor is the electric force, acting in the wall-normal
direction. However, the analogy between the two flows ends, as soon as
nonlinearities arise, especially as diffusive effects are concerned:
in RBC, molecular diffusion constitutes the principal dissipative
mechanism whereas in EHD flow, it is the ion drift velocity $K E$
(with $K$ being the ionic mobility) which diffuses perturbations in
the fluid. It is well-known that RBC is of a supercritical nature,
i.e., transition from the hydrostatic state to the finite-amplitude
state occurs continuously as the controlling parameter, i.e., the
Rayleigh number, is increased. For EHD flow, on the other hand, the
bifurcation is subcritical, characterized (i) by an abrupt jump in
motion amplitude from zero to a finite value, as a critical parameter
is crossed, and (ii) by the existence of a hysteresis loop. It is
interesting to mention an analogy between EHD flow and polymeric flow:
polymeric flow shows a hysteresis loop as well, as the first
bifurcation is considered. In fact, the counterpart of EHD flow, i.e.,
magneto-hydrodynamics (MHD) flow, has been compared to polymeric flow
in \cite{Ogilvie2003}.

Most studies in the EHD literature address electroconvection in the
hydrostatic condition, i.e., without cross-flow. In this work we also
investigate the EHD stability properties in the presence of
cross-flow. Our interest is two-fold. First, the potential of this
flow configuration resides in the possibility of using the electric
field to create large-scale rollers for flow manipulation; turbulent
drag reduction designed in the spirit of \cite{Schoppa1998} and
investigated by \cite{Soldati1998} in the nonlinear regime is an
example of this type. Secondly, EHD with cross-flow has been applied
to wire-plate electrostatic precipitators, but due to the complex
nature of the chaotic interaction between wall turbulence and the
electric field, our current understanding of such flows is rather
limited. Nonlinear EHD simulations with a cross-flow component have
been reported in \cite{Soldati1998}. More relevant to our linear
problem is the unipolar-injection-induced instabilities in plane
parallel flows studied by \cite{Atten1982} and
\cite{Castellanos1992}. The former work focused on so-called
electroviscous effects, defined by an increase of viscosity due to the
applied electric field compared to the canonical channel flow. The
latter work found that, at high Reynolds numbers, the destabilizing
mechanism is linked to inertia, while, at sufficiently low Reynolds
numbers, EHD instability are dominant. In this article, we will not
only address the modal stability problem of EHD channel flow, as those
two previous studies did, we will also take into account transient
effects, discussed shortly below, of the high-$Re$ number channel flow
in the presence of an electric field. Our results would be interesting 
to the researchers in the flow instability and transition to turbulence, 
especially for high-$Re$ flow. The results will also 
shed light on the study of flow control strategy using EHD effects.

\subsection{Stability of EHD flow}

The endeavor to understand the stability and transition to turbulence
in EHD flow dates back to the 1970's, when \cite{Schneider1970} and
\cite{Atten1972}, among the first, performed a linear stability
analysis on the flow of dielectric liquids confined between two
parallel electrodes with unipolar injection of charges. The mechanism
for linear instability could be explained via the formation of an
electric torque engendered by the convective motion when the driving
electric force is sufficiently strong to overcome viscous
diffusion. It was established in \cite{Atten1972} that, in the weak
injection limit, $C \ll 1$, with $C$ as the charge injection level,
the flow is characterized by the criterion $T_c C^2 \approx 220.7$,
where $T_c$ is the linear stability criterion for the stability
parameter $T$, defined in the mathematical modeling section
\ref{governingequation}, and, in the case of space-charge-limited
(SCL) injection, $C \rightarrow \infty$, they found $T_c \approx
160.75$. However, according to \cite{Lacroix1975} and
\cite{Atten1979}, the experimentally determined stability criterion
was notably different from the theoretical calculations. In the
experiments performed by \cite{Atten1979}, the linear criterion was
found to be $T_c \approx 100$ in the case of SCL, which is far lower
than the theoretically predicted value. It was argued then that this
disagreement might be due to neglecting charge diffusion
\citep{Atten1976}. We will address this discrepancy in the SCL 
in this paper, and confirm that charge diffusion is indeed an
important factor influencing the linear stability criterion in this
case.

The first nonlinear stability analysis was performed by
\cite{Felici1971}, who assumed a two-dimensional, \textit{a priori}
hydraulic model for the velocity field in the case of weak injection
between two parallel plates. It was found that within the interval
$[T_{nl}, T_c]$, where $T_{nl}$ is the nonlinear stability criterion
for $T$, two solutions exist, namely, a stable state and an unstable
finite-amplitude state. This finding corroborated the fact that the
bifurcation in the unipolar injection problem is of a subcritical
nature and that the flow has a hysteresis loop, as experimentally
verified by \cite{Atten1979}. Physically, this subcritical bifurcation
is related to the formation of a region of zero charge
\citep{Perez1989}. Later, this simple hydraulic model was extended to
three-dimensional, hexagonal convective cells for the case of SCL by
\cite{Atten1979}, and it was shown that the most unstable hydrodynamic
mode consists of hexagonal cells with the interior liquids flowing
towards the injector. The nonlinear stability criterion for
three-dimensional, hexagonal cells, according to \cite{Atten1979}, was
$T_{nl} \approx 90$ in the experiments, but theoretical studies
produced $T_{nl} \approx 110$.

Most of the previous linear stability analyses of EHD flow focus on the most
unstable mode of the linear system, which is insufficient for a
comprehensive flow analysis. In fact, theoretically, the linear
stability analysis is linked to the characteristics of the linearized
Navier-Stokes (N-S) operator $\bm{\mathcal{L}}$ which, in the case of
shear flows (in this paper, the cross-flow case), may be highly
nonnormal, i.e., $\bm{\mathcal{L}}^+ \bm{\mathcal{L}} \neq
\bm{\mathcal{L}} \bm{\mathcal{L}}^+$ (with $\bm{\mathcal{L}}^+$
denoting the adjoint of $\bm{\mathcal{L}}$) or, expressed differently,
the eigenvectors of the linear operator are mutually
nonorthogonal~\cite[see][]{Trefethen1993,Schmid2001}. For a normal
operator ($\bm{\mathcal{L}}^+ \bm{\mathcal{L}} = \bm{\mathcal{L}}
\bm{\mathcal{L}}^+$), the dynamics of the perturbations is governed by
the most unstable mode over the entire time horizon. In contrast, a
nonnormal operator has the potential for large transient amplification
of the disturbance energy in the early linear phase, even though the
most unstable mode is stable. The theory of nonmodal stability
analysis \citep{Farrell1996, Schmid2007}, the main tool to be used in this work,
has been applied successfully to explain processes active during
transition to turbulence in several shear flows. The fact that the
bifurcation of EHD flow is subcritical, a trait often observed in
shear flows governed by nonnormal linearized operators, tempts one to
think that the discrepancy between the experimental value $T_c \approx
100$ and the theoretical value $T_c \approx 161$ in the SCL regime of
EHD flow might be examined in the light of nonmodal stability
theory. In fact, it seems surprising that this type of stability
analysis has so far only rarely been applied to EHD flows, except for
the work of \cite{Atten1974} in the case of hydrostatic flow. The
method we employ here is different from Atten's quasistationary
approach: nonmodal stability theory, based on solving the
initial-value problem, seeks the maximum disturbance energy growth over entire time horizon
when considering all admissible initial conditions and identifies the
optimal initial condition for achieving this maximum energy growth. In
\cite{Atten1974}, a quasistationary approach was taken that proposed
that disturbances grow rapidly, compared to the time variation of the
thickness of the unipolar layer; however, transient energy growth due
to the nonnormality of the linearized operator in hydrostatic EHD has
been found to be rather limited in this work. This is in contrast with EHD
Poiseuille flow, where nonnormality is prevalent and should be
considered from the outset.

The present paper extends the work by \cite{Martinelli2011} and is
organized as follows. In \textsection~\ref{problemformulation}, we
present the mathematical model, the governing equations and the
framework of the linear stability analysis. In
\textsection~\ref{numerics}, numerical details are given and a code
validation is provided in the appendix. We then present in \textsection~\ref{results} the
results of the modal and nonmodal stability analysis and in 
\textsection~\ref{energyanalysis} the energy analysis. Finally, in
\textsection~\ref{discussion}, we summarize our findings and conclude
with a discussion.


\section{Problem formulation}\label{problemformulation}

\subsection{Mathematical modeling}

\begin{figure}
  \centering
  \subfigure{\hspace{5em}\includegraphics[width=.8\textwidth]
    {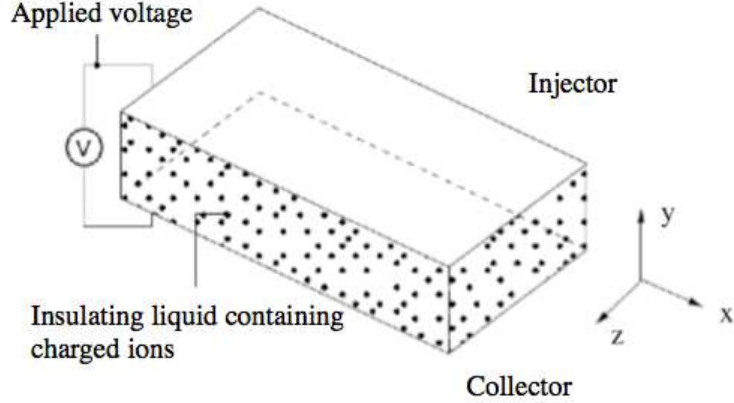}}
  \caption{Sketch of the electroconvection problem with coordinate
    system ($x$, $y$, $z$). In the non-hydrostatic case, a flow rate
    is induced along the streamwise ($x$) direction.}
  \label{ehdflow}
\end{figure}

We consider the planar geometry sketched in figure~\ref{ehdflow},
where the Cartesian coordinate system used in this work is ($x$, $y$,
$z$) or ($\bm{1}_x$, $\bm{1}_y$, $\bm{1}_z$) as the streamwise,
wall-normal and spanwise directions, respectively. The two flat
electrodes extend infinitely in the $x$- and $z$-directions, and the
applied voltage only varies in the $y$-direction. The distance between
the two electrodes is $2 L^*$. The dimensional variables and parameters
are denoted with a superscript $^*$.
The electric field satisfies the reduced Maxwell equations. The
charges are generated through electrochemical reactions on the
charge-injecting electrode~\citep{Alj1985}. Since the electric
conductivity is very low, conduction currents are negligible even in
the presence of large electric fields. Therefore, magnetic effects in
the Maxwell equations can be
neglected~\citep{Melcher1981,Castellanos1998}, leading to the
quasi-electrostatic limit of the Maxwell equations
\begin{subeqnarray}
  \nabla^* \times \bm{E}^* &=0, \label{maxwell1}\\
  \nabla^* \cdot \bm{D}^* &=Q^*, \label{maxwell2} \\
  \frac{\partial Q^*}{\partial t^*} + \nabla^* \cdot \bm{J^*} &=0,
\end{subeqnarray}
where $\bm{E}^*$ is the electric field, $\bm{D}^*=\epsilon^* \bm{E}^*$
denotes the electric displacement, $\epsilon^*$ stands for the fluid
permittivity which we assume constant here, $Q^*$ represents the
charge density and $\bm{J}^*$ is the current density. Considering
equation~(\ref{maxwell1}a), it is a well-known practice to define a
potential field $\phi^*$ according to $\bm{E}^*=-\bm{\nabla}^*
\phi^*$. Combining the first two equations~(\ref{maxwell1}a)
and~(\ref{maxwell2}b), we can write the governing equation for
$\phi^*$ as
\begin{equation}
  \nabla^{*2} \phi^* = -\frac{Q^*}{\epsilon^*}.
\end{equation}
The current density $\bm{J}^*$ arises from several sources. By
modeling the EHD flow with only one ionic species in a perfectly
insulating fluid (conductivity $\sigma^*=0$), one can express
$\bm{J}^*$ as~\citep{Castellanos1998}
\begin{equation}
  \bm{J}^*= K^*\bm{E}^*Q^*+ \bm{U}^*Q^* - D^*_{\nu} \bm{\nabla}^* Q^*
\end{equation}
where the first term accounts for the drift of ions (with respect to
the fluid) under the effect of the electric field, moving at the
relative velocity $K^*\bm{E}^*$, with $K^*$ as the ionic mobility, the
second term represents the convection of ions due to the
fluid velocity $\bm{U}^*$, and the last term takes into account the
charge diffusion, with $D^*_{\nu}$ as the diffusion coefficient. Since
the work of~\cite{Perez1989}, the vast body of literature, with the
exception of~\cite{Kourmatzis2012} for turbulent EHD flow, neglects
the charge-diffusion term because of its small value when compared to
the drift terms. However, we will show that, even though the numerical
value of $D^*_{\nu}$ is very small, its impact on the flow dynamics is
undeniable.

The flow field is incompressible, viscous and Newtonian and governed
by the Navier-Stokes equations, which, in vector notation, read
\begin{subeqnarray}
  \nabla^* \cdot \bm{U}^*&=&0, \\
  \rho^* \frac{\partial  \bm{U}^*}{\partial t^*} + \rho^*
  ( \bm{U}^* \cdot \nabla^*)  \bm{U}^*&=&-\bm{\nabla}^* P^* +
  \mu^* \nabla^{*2}  \bm{U}^*+ \bm{F_q}^* ,
\end{subeqnarray}
where $\bm{U}^*$ is the velocity field, $P^*$ the pressure, $\rho^*$
the density, $\mu^*=\rho^* \nu^*$ the dynamic viscosity ($\nu^*$ the
kinematic viscosity) and $\bm{F_q}^*$ the volumic density of electric force,
which expresses the coupling between the fluid and the electric
field. In general, $\bm{F_q}^*$ can be written as
\begin{equation}
  \bm{F_q}^*=Q^*\bm{E}^* - \frac{1}{2} |\bm{E}^*|^2\bm{\nabla}^*
  \epsilon^* + \bm{\nabla}^* \Big{[} \frac{|\bm{E}^*|^2}{2}\rho^*
    \frac{\partial \epsilon^*}{\partial \rho^*} \Big{]},
\end{equation}
where the three terms on the right-hand side represent,
respectively, the Coulomb force, the dielectric force and the
electrostrictive force. The Coulomb force is
commonly the strongest force when a DC voltage is applied. As we
assume an isothermal and homogeneous fluid, the permittivity
$\epsilon$ is constant in space. As a result, the dielectric force is
zero (however, it would be dominant in the case of an AC voltage). The
electrostrictive force can be incorporated into the pressure term of
the Navier-Stokes equation as it can be expressed as the gradient of a
scalar field. Therefore, the only remaining term of interest in
our formulation is the Coulomb force.

The system is supplemented by suitable boundary conditions. In our
problem, we assume periodic boundary conditions in the wall-parallel
directions. The no-slip and no-penetration conditions for the
velocities are assumed at the channel walls. For the potential field,
we require Dirichlet conditions on both walls, on the injector
$\phi^*(L^*)=\phi^*_0$ and the collector $\phi^*(-L^*)=0$ in order to
fix the potential drop $\Delta \phi_0^*$ between the electrodes. The injection 
mechanism is autonomous and homogeneous, meaning that
 the charge density is constant on the injector, not influenced by the nearby electric
field and has a zero wall-normal flux of charge on the collector,
i.e., $Q^*(L^*)=-Q^*_0$ and $\frac{\partial Q^*}{\partial
  y^*}(-L^*)=0$. Owing to the homogeneity in the wall-parallel
directions, there is no requirement for boundary conditions in the
$x$- and $z$-direction.


\subsection{Nondimensionalized governing equations}\label{governingequation}

In the no-crossflow case, as we are interested in the effect of the
electric field on the flow dynamics, we nondimensionalize the full
system with the characteristics of the electric field, i.e., $L^*$
(half distance between the electrodes), $\Delta\phi_0^*$ (voltage
difference applied to the electrodes) and $Q^*_0$ (injected charge
density). Accordingly, the time $t^*$ is nondimensionalized by
$L^{*2}/(K^*\Delta \phi_0^*)$, the velocity $\bm{U}^*$ by $K^*\Delta
\phi_0^*/L^*$, the pressure $P^*$ by $\rho^*_0 K^{*2} \Delta
\phi_0^{*2}/L^{*2}$, the electric field $\bm{E}^*$ by $\Delta \phi_0^*
/ L^*$ and the electric density $Q^*$ by $Q^*_0$. Therefore, the
nondimensional equations read
\begin{subeqnarray}
  \nabla \cdot \bm{U}&=& 0, \label{EHD1}\\
  \frac{\partial  \bm{U}}{\partial t} + ( \bm{U} \cdot \nabla)
  \bm{U} &=& -\bm{\nabla} P + \frac{M^2}{T} \nabla^2  \bm{U}+
  CM^2 Q\bm{E}, \label{EHD2}\\
  \frac{\partial Q}{\partial t} + \bm{\nabla} \cdot
       [(\bm{E}+  \bm{U})Q] &=& \frac{1}{Fe}\nabla^2 Q, \label{EHD3}\\
       \nabla^2 \phi &=& -CQ, \\
       \bm{E} &=& -\bm{\nabla} \phi \label{EHD5}
\end{subeqnarray}
where
\begin{equation}
  M=\frac{(\epsilon^*/\rho^*_0)^{\frac{1}{2}}}{K^*}, \qquad
  T=\frac{\epsilon^* \Delta \phi_0^*}{K^* \mu^*}, \qquad
  C=\frac{Q_0^* L^{*2}}{\Delta\phi^*_0 \epsilon^*}, \qquad
  Fe=\frac{K^*\Delta\phi^*_0}{D_{\nu}^*}.
\end{equation}
Additionally, the nondimensional boundary conditions are
$\bm{U}(\pm1)=0$, $\phi(1)=1$, $\phi(-1)=0$, $Q(1)=-1$ and
$\frac{\partial Q}{\partial y}(-1)=0$.

Various dimensionless groups appear in the equations as written
above. $M$ is the ratio between the hydrodynamic mobility
$(\epsilon/\rho_0)^{\frac{1}{2}}$ and the true ion mobility $K$. Gases
usually take on a value of $M$ less than $0.1$ and liquids have values
of $M$ greater than $1$~\citep{Castellanos1992}. $T$ (Taylor's
parameter) represents the ratio of the Coulomb force to the viscous
force. It is the principal stability parameter, assuming a similar
role as the Rayleigh number in Rayleigh-B{\'e}nard convection. $C$ measures
the injection level. When $C \gg 1$, the system is in a
strong-injection regime, and when $C \ll 1,$ it is in a weak-injection
regime. $Fe$ is the reciprocal of the charge diffusivity
coefficient. The factor ${M^2}/{T}$ appearing in
equation~(\ref{EHD2}b) can be interpreted as the ratio between the
charge relaxation time $L^{*2}/(K^*\Delta \phi^*_0)$ by drift and the
momentum relaxation time $L^{*2}/\nu^*$. This mathematical model for
EHD flow has been assumed and studied in many previous investigations
of the linear stability and turbulence analyses for a dielectric
liquid subject to unipolar injection of
ions~\citep{Lacroix1975,Traore2012,Wu2013}, except that the diffusion
term in equation~(\ref{EHD3}c) is usually neglected (excluding the
study of~\cite{Kourmatzis2012}).


\subsection{Linear stability problem}\label{linearstabilityproblem}

The linear problem is obtained by decomposing the flow variable as a
sum of base state and perturbation, i.e., $\bm{U} = \bar{\bm{U}} +
\bm{u}$, $P = \bar{P} + p$, $\bm{E} = \bar{\bm{E}} + \bm{e}$, $\bm{D}
= \bar{\bm{D}} + \bm{d}$, $Q = \bar{Q} + q$ and $\phi = \bar{\phi} +
\varphi$. For the vector fields, we have $\bm{u} = (u, v, w)$ and
$\bm{e} = (e_1, e_2, e_3)$ along the three Cartesian coordinate
directions. After substituting the decompositions into the governing
equations~(\ref{EHD1}a-e), subtracting from them the governing
equations for the base states and retaining the terms of first order,
the linear system reads
\begin{subeqnarray}
  \bm{\nabla} \cdot \bm{u} &=& 0, \label{nd_ehd1}\\
  \frac{\partial \bm{u}}{\partial t} + (\bm{u} \cdot \bm{\nabla})
  \bar{\bm{U}} + (\bar{\bm{U}} \cdot \bm{\nabla}) \bm{u}
  &=& -\bm{\nabla} p + \frac{M^2}{T} \nabla^2 \bm{u}+
  CM^2 (q\bar{\bm{E}}+\bar{Q}\bm{e}), \label{nd_ehd2} \\
  \frac{\partial q}{\partial t} + \bm{\nabla} \cdot
       [(\bar{\bm{E}}+ \bar{\bm{U}})q +(\bm{e}+ \bm{u})\bar{Q} ]
       &=& \frac{1}{Fe}\nabla^2 q, \label{nd_ehd3}\\
       \nabla^2 \varphi &=& -Cq, \label{nd_ehd4}\\
       \bm{e} &=& -\bm{\nabla} \varphi \label{nd_ehd5},
\end{subeqnarray}
with the boundary conditions for the fluctuations $\bm{u}(\pm 1)=0$,
$\varphi(\pm1)=0$ and $q(1)=0, \frac{\partial q}{\partial y}(-1)=0$.

\subsubsection{Base states}

The base states are the solutions to equations~(\ref{EHD1}a-e) in the
case of no time dependence. Owing to the periodicity 
in the wall-parallel directions, we can reduce the shape of the base states as functions of $y$
only, that is, $\bar{\bm{U}} = \bar{U}(y)\bm{1}_y$ and $\bar{\bm{E}} =
\bar{E}(y)\bm{1}_y$. For the base flow $\bar{U}(y)$, we are interested
in the hydrostatic and pressure-driven Poiseuille flows which, after
nondimensionalization, are given by
\begin{equation}
  \bar{U}(y)=0, \qquad \qquad \bar{U}(y)=Re\frac{M^2}{T}(1-y^2)=(1-y^2),
  \label{baseflowequation}
\end{equation}
respectively, in which the (electric) Reynolds number is defined as
$Re=\frac{T}{M^2}=K^* \Delta \phi^*_0/\nu^*$ (in order to enforce
the same constant flow rate). It is a passive parameter in the
hydrostatic case, but becomes a free parameter in the presence of
high-$Re$ cross-flow, in which, consequently, $M$ would be the passive
parameter. Therefore, in the Poiseuille flow case, we modify the
governing equation~(\ref{nd_ehd2}b) by substituting the relation
$Re=T/M^2$ to obtain
\begin{equation}
  \frac{\partial \bm{u}}{\partial t} + (\bm{u} \cdot \bm{\nabla})
  \bar{\bm{U}} + (\bar{\bm{U}} \cdot \bm{\nabla}) \bm{u} = -\bm{\nabla} p +
  \frac{1}{Re} \nabla^2 \bm{u}+ \frac{CT}{Re} (q\bar{\bm{E}}+\bar{Q}\bm{e}).
  \label{nd_ehd2_highRe}
\end{equation}
By doing so, it is more obvious to identify the effects of $T$ and $C$
on the electric force term. The parameter $Re=K^*\Delta
  \phi^*_0/\nu^*$ here coincides with the canonical hydrodynamic
equivalent $Re_h=U^*L^*/\nu^*$ because of the electric scaling
we chose. However, in a general sense, the two may not necessarily be
identical. The nondimensional quantity
\begin{equation}
  \frac{Re}{Re_h}=\frac{K^* \Delta \phi^*_0}{U^*L^*}=\frac{L^*/U^*}
       {L^{*2}/(K^*\Delta \phi^*_0)}
\end{equation}
relates the eddy turn-over time and the charge relaxation time by the
drift. According to the equality $T=Re\cdot M^2$, when $Re$ is near
linear criticality at $5772$ and $T$ is around $10^2$, $M \approx
0.1$. It implies that the working liquid is gas. Moreover, in contrast
to the nonlinear constitutive modeling for polymers in viscoelastic
flow, the base flow is not modified under the influence of the base
electric field, even though the coupling between $\bm{U}$ and $Q$ is nonlinear in equation~(\ref{EHD3}c). This is because the
directions of the base flow and the base electric field are
perpendicular. Nevertheless, the base pressure gradient in the wall-normal
direction is no longer zero.

\begin{figure}
  \centering
  \subfigure {\includegraphics[width=.5\textwidth]{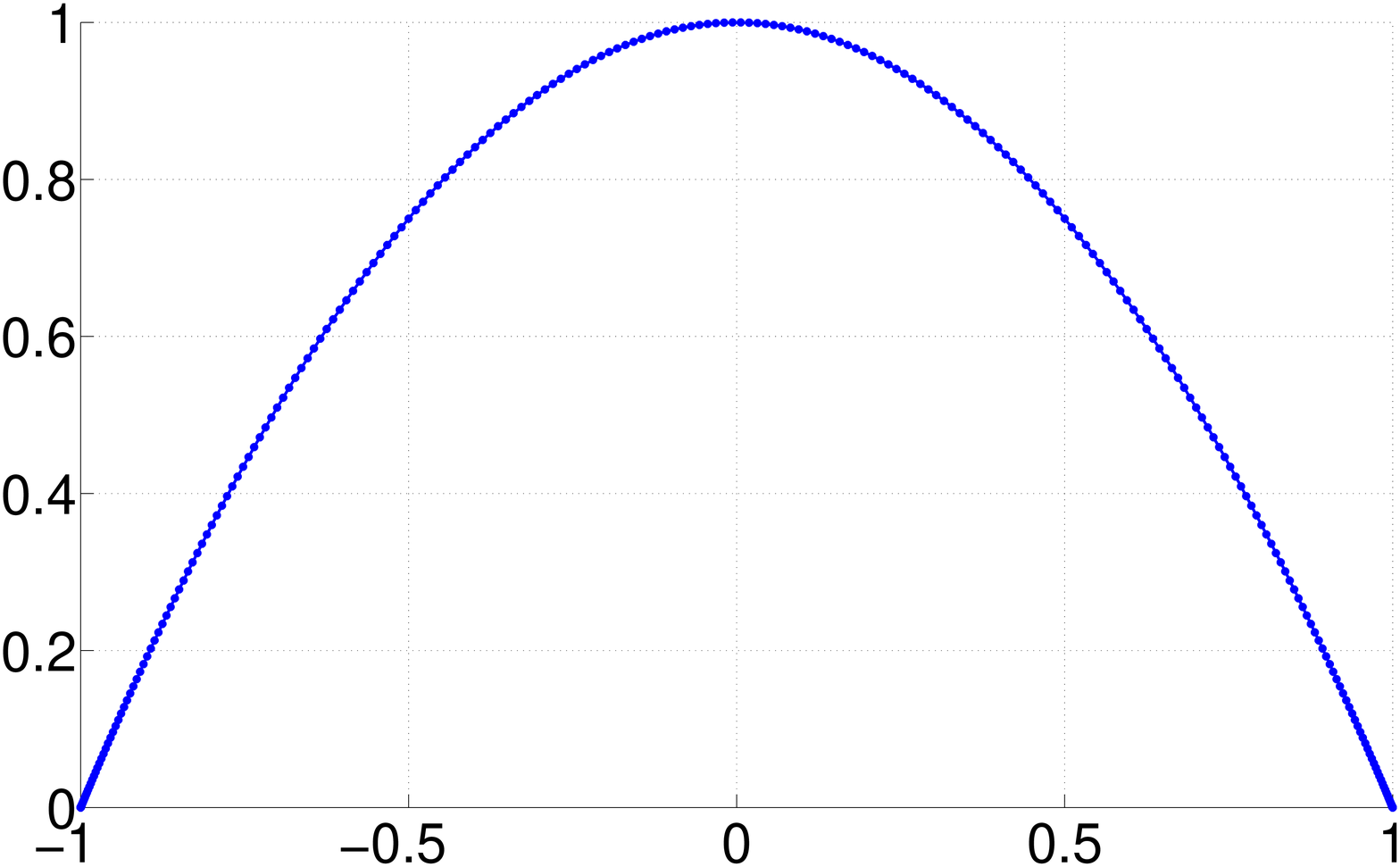}
    \put(-200,100){{\large $(a)$}}
    \put(-190,50){{\large $\bar{U}$}}
    \put(-95,-5){{\large $y$}}
    \includegraphics[width=.5\textwidth]{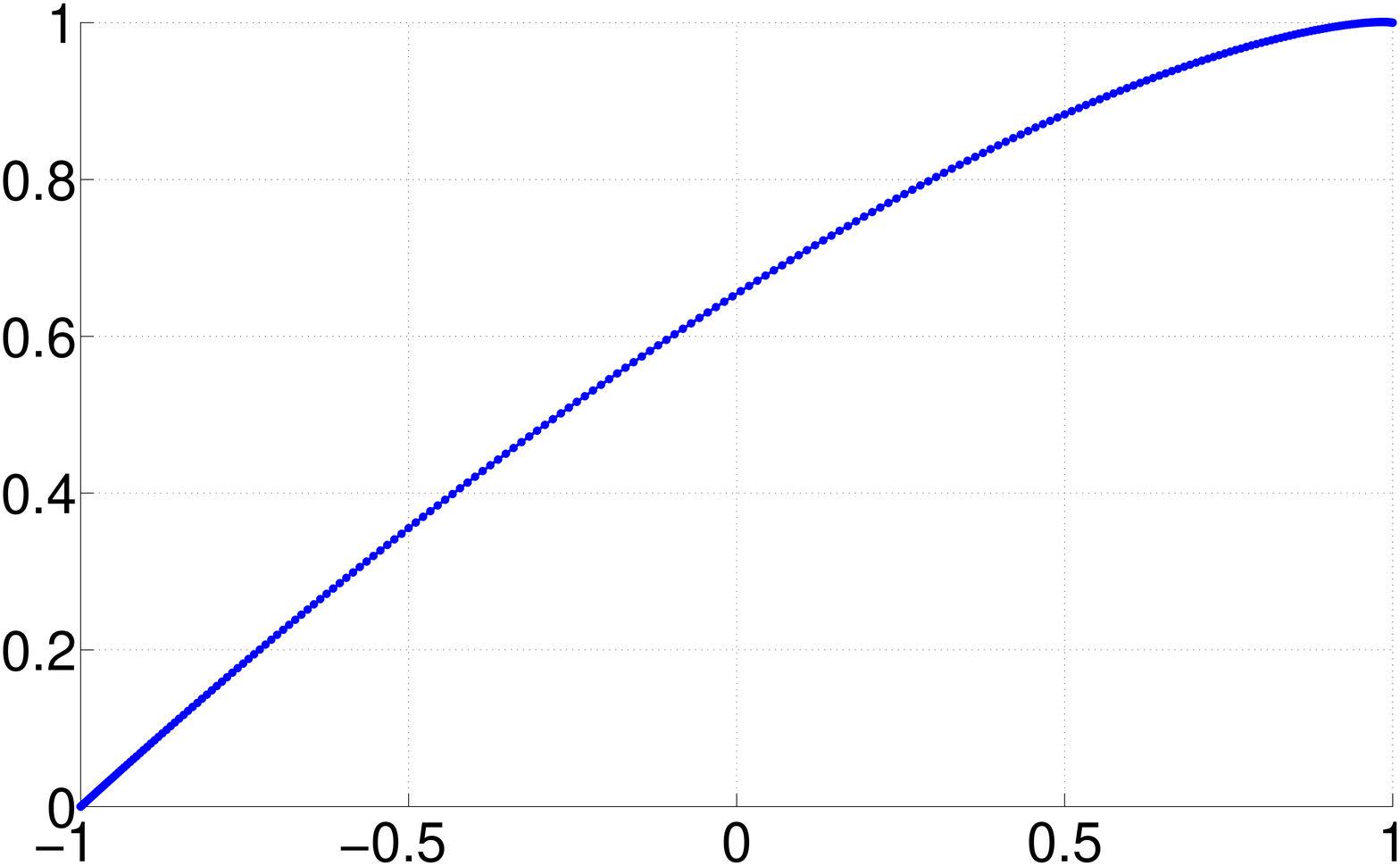}
    \put(-200,100){{\large $(b)$}}
    \put(-190,50){{\large $\bar{\phi}$}}
    \put(-95,-5){{\large $y$}}}\\
  \subfigure {\includegraphics[width=.5\textwidth]{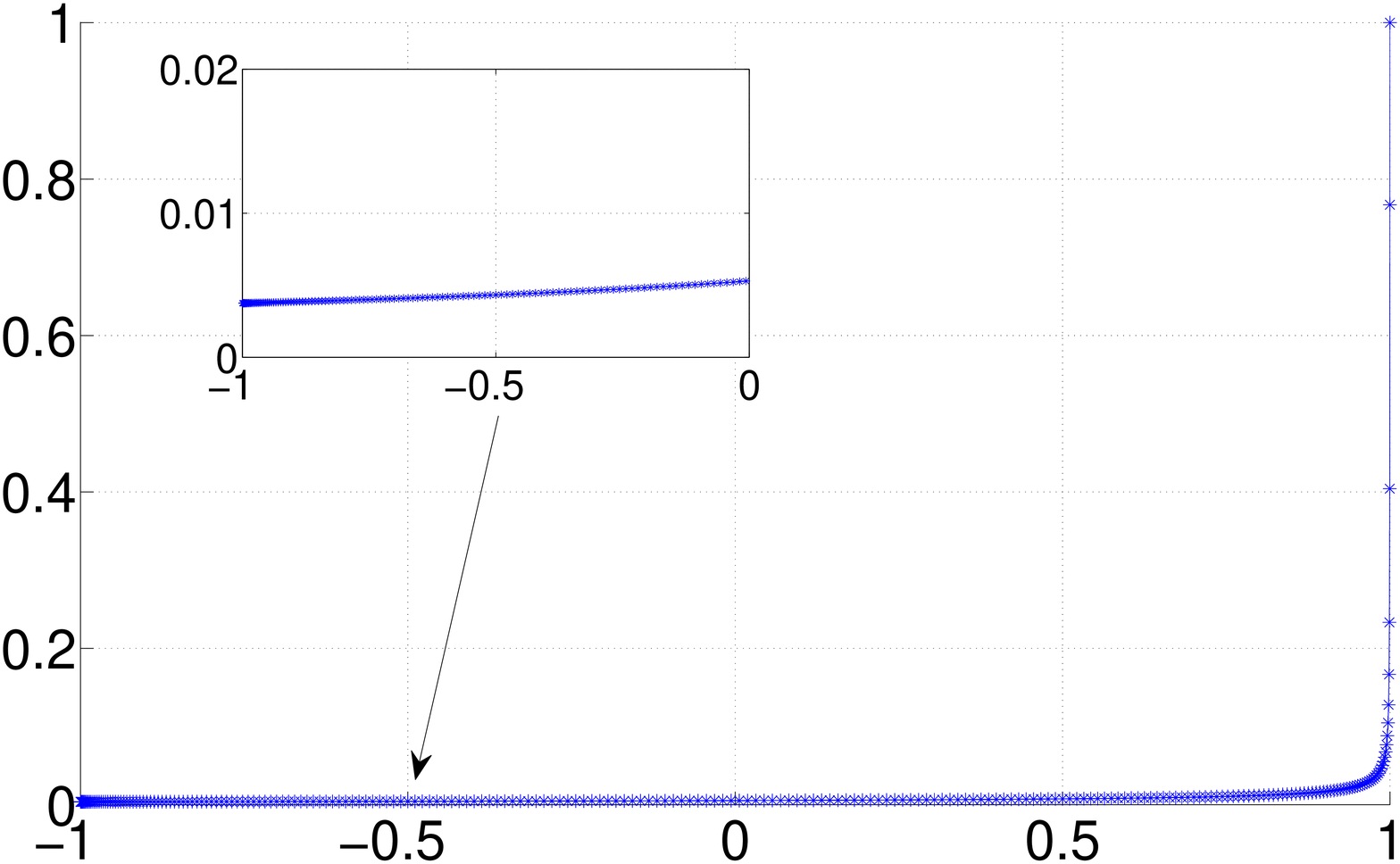}
    \put(-200,100){{\large $(c)$}}
    \put(-190,50){{\large $\bar{Q}$}}
    \put(-95,-5){{\large $y$}}
  }
  \caption{The base states: (a) $\bar{U}$, (b) $\bar{\phi}$, (c) $\bar{Q}$.}
  \label{baseflow}
\end{figure}

The base electric field $\bar{\bm{E}}(y)$ can be solved from
equations~(\ref{EHD3}c-e), recast into an equation only for
$\bar{\phi}$ which reads
\begin{equation}
  \bar{\phi}' \bar{\phi}''' +  (\bar{\phi}'')^2 +
  \frac{1}{Fe}\bar{\phi}''''=0,
\end{equation}
where prime $'$ denotes the spatial derivative with respect to the
$y$-direction. The boundary conditions are $\bar{\phi}(1)=1$,
$\bar{\phi}(-1)=0$, $\bar{\phi}''(1)=-C$ and
$\bar{\phi}'''(-1)=0$. Analytical solutions to this fourth-order
ordinary differential equation can be obtained by observing that the
equation can be transformed into a Riccati equation; alternatively, as
we do here, a simple numerical integration combined with a nonlinear
gradient method provides us with the required
$\bar{\phi}(y)$-profile. The Poiseuille base flow and the base states
of the electric and charge fields are shown in figure~\ref{baseflow}.

\subsubsection{Matrix representation}

In linear stability analysis, it is a common practice to rewrite the fluid system~(\ref{nd_ehd1}a-b) in terms of the wall-normal velocity $v$ and the wall-normal vorticity $\eta=\partial_z u - \partial_x w$ by eliminating the pressure term. For the electric field, the three equations~(\ref{nd_ehd3}c-e) can be reduced to one for $\varphi$. Therefore, the governing equations~(\ref{nd_ehd1}a-e) become, in terms of a $v-\eta-\varphi$ formulation,
\begin{subeqnarray}
  \frac{\partial \nabla^2v}{\partial t}  &=& \Big{[}- \bar{U} \frac{\partial}{\partial x}\nabla^2 + \bar{U}'' \frac{\partial }{\partial x} + \frac{M^2}{T}\nabla^4 \Big{]} v \notag\\
  &&  \hspace{20mm}   + M^2\Big{[}-\bar{\phi}''' (\nabla^2-\frac{\partial^2}{\partial y^2})\varphi + \bar{\phi}' (\nabla^2-\frac{\partial^2}{\partial y^2})\nabla^2\varphi \Big{]},  \label{ehdeq1} \\
  \frac{\partial \eta}{\partial t} &=&-\bar{U}\frac{\partial }{\partial x} \eta - \bar{U}' \frac{\partial v}{\partial z}  +\frac{M^2}{T}\nabla^2\eta,  \label{ehdeq2} \\
\frac{\partial \nabla^2 \varphi}{\partial t} &=& \bar{\phi}' \frac{\partial \nabla^2 \varphi}{\partial y} + \bar{\phi}''' \frac{\partial \varphi}{\partial y} + 2\bar{\phi}''\nabla^2 \varphi - \bar{U} \frac{\partial \nabla^2 \varphi}{\partial x} - \bar{\phi}''' v + \frac{1}{Fe}\nabla^4 \varphi,  \label{ehdeq3}
\end{subeqnarray}
with boundary conditions
\begin{subeqnarray}
  &&v(\pm1)=0, v'(\pm1)=0, \\
  &&\eta(\pm1)=0, \\
  &&\varphi(\pm1)=0, \varphi''(1)=0, \varphi'''(-1)=0. \label{bcphi}
\end{subeqnarray}
For compactness, we write $\bm{\gamma} = (v,\eta,\varphi)^T,$ and the
linearized system, recast in matrix notation, becomes
\begin{equation}
  \begin{pmatrix}
    \bm{\nabla}^2 & \bm{0}  & \bm{0}   \\
    \bm{0}   &  \bm{I} & \bm{0}  \\
    \bm{0}   & \bm{0}  & \bm{\nabla}^2  \\
  \end{pmatrix}\frac{\partial}{\partial t}\begin{pmatrix}
    v \\  \eta \\ \varphi
  \end{pmatrix} = \begin{pmatrix}
    \bm{L}_{os} & \bm{0} &  \bm{L}_{v\varphi} \\
    \bm{L}_c  &  \bm{L}_{sq} & \bm{0} \\
    \bm{L}_{\varphi v} & \bm{0} &  \bm{L}_{\varphi\varphi} \\
  \end{pmatrix}\begin{pmatrix}
    v \\  \eta \\ \varphi
  \end{pmatrix}
  \label{linearmatrix}
\end{equation}
where $\bm{I}$ denotes the identity matrix and the submatrices
$\bm{L}_{os}$, $\bm{L}_{v\varphi}$, $\bm{L}_c$, $\bm{L}_{sq}$,
$\bm{L}_{\varphi v}$ and $\bm{L}_{\varphi \varphi}$ can be easily 
deduced from equations~(\ref{ehdeq1}a-c).  To represent the system
even more compactly, we can rewrite the linearized
problem~(\ref{linearmatrix}) as
\begin{equation}
  \bm{\mathcal{A}} \frac{\partial \bm{\gamma}}{\partial t} =
  \bm{\mathcal{B}}
  \bm{\gamma} \hspace{10mm}\Longrightarrow \hspace{10mm}
  \frac{\partial \bm{\gamma}}{\partial t} = \bm{\mathcal{L}}
  \bm{\gamma},
  \label{lineartimeevolution}
\end{equation}
where $\bm{\mathcal{L}} = \bm{\mathcal{A}}^{-1} \bm{\mathcal{B}}$
represents the linearized Navier-Stokes operator for EHD flow.

Since the flow is homogeneous in the wall-parallel directions, the
perturbations are assumed to take on a wave-like shape. Moreover, as
we consider a linear problem with a steady base flow, it is legitimate
to examine the frequency response of the linear system for each
frequency individually. These two simplifications lead to
\begin{equation}
  \bm{f}(x,y,z,t) = \hat{\bm{f}}(y,t)\exp(i\alpha x + i\beta z) =
  \tilde{\bm{f}}(y)\exp(-i \omega t)\exp(i\alpha x + i\beta z),
  \label{wave2}
\end{equation}
where $\bm{f}$ could represent any flow variable in $(\bm{u}, p,
\bm{e}, q, \varphi)^T$, $\hat{\bm{f}}(y,t)$ and $\tilde{\bm{f}}(y)$
are the shape functions, $\alpha$ and $\beta$ are the real-valued
streamwise and spanwise wavenumbers, and the complex-valued $\omega$
is the circular frequency of the perturbation, with its real part
$\omega_r$ representing the phase speed and its imaginary part
$\omega_i$ representing the growth rate of the linear
perturbation. Upon substitution of the above expression into the
linear problem~(\ref{lineartimeevolution}), we arrive at an eigenvalue
problem for the $v-\eta-\varphi$ formulation which reads
\begin{equation}
  -i \omega \tilde{\bm{\gamma}}= \bm{\mathcal{L}} \tilde{\bm{\gamma}},
  \label{lineareigen}
\end{equation}
where $-i\omega$ is the eigenvalue and $\tilde{\bm{\gamma}}$ is the
corresponding eigenvector. Both formulations,
(\ref{lineartimeevolution}) and~(\ref{lineareigen}), would be relevant
as discussed in a recent review by \cite{Schmid2014}. The least
unstable eigenvalues obtained from the eigenproblem
formulation~(\ref{lineareigen}) would determine the asymptotic
behavior of the linear system, while the initial-value problem
formulation~(\ref{lineartimeevolution}) could be used to examine the
dynamics of the fluid system evolving over a finite time scale.

\subsubsection{Energy norm}

In our calculation of the nonmodal transient growth, we define the
total energy density of the perturbation contained in a control volume
$\Omega$ as
\begin{eqnarray}
  \int_{\Omega} \mathcal{E}^* dV^* &=& \int_{\Omega}
  \Big{(}\mathcal{E}^* _k+\mathcal{E}^* _{\varphi} \Big{)} dV^*=\int_{\Omega}
  \frac{1}{2} \Big{(} \rho_0^* \bm{u}^*\cdot \bm{u}^* + \bm{e}^*\cdot
  \bm{d}^* \Big{)} \notag dV^* \nonumber \\
  &=&\int_{\Omega} \frac{1}{2} \Big{(} \rho_0^* (u^{*2}+v^{*2}+w^{*2}) +
  \epsilon^* |\bm{\nabla}^* \varphi^*|^2 \Big{)} dV^*.
  \label{energydefinition}
\end{eqnarray}
The perturbed electric energy $\mathcal{E}^*_{\varphi}$ follows the
definition in~\cite{Castellanos1998}. In terms of the
$v-\eta-\varphi$-formulation, the nondimensionalized energy norm in
spectral space becomes
\begin{eqnarray}
  \int_{\Omega} \mathcal{E}  dV &=& \frac{1}{2} \cdot
  \frac{1}{2}\int \hat{\bm{\gamma}}^\dag
  \begin{pmatrix}
    \bm{I}+\frac{1}{k^2}\bm{D}^\dag_1 \bm{D}_1 &  \bm{0} &  \bm{0} \\
    \bm{0} & \frac{1}{k^2}\bm{I} & \bm{0}\\
    \bm{0}&  \bm{0}& M^2(k^2 \bm{I} + \bm{D}^\dag_1 \bm{D}_1)
  \end{pmatrix} \hat{\bm{\gamma}} dy  \nonumber \\
  &=& \int_{\Omega}  \hat{\bm{\gamma}}^\dag \bm{\mathcal{M}}
  \hat{\bm{\gamma}} dy,
\end{eqnarray}
where the superscript $^\dag$ denotes the complex conjugate, $k^2 =
\alpha^2 + \beta^2$, and $\bm{D}_1$ represents the first-derivative
matrix with respect to the wall-normal direction (likewise for
$\bm{D}_2$ and $\bm{D}_3$ below). The positive definite matrix
$\bm{\mathcal{M}}$ allows us to work in the $L_2$-norm. To do so, we
apply a Cholesky decomposition to the weight matrix according to
$\bm{\mathcal{M}} = \bm{F}^\dag \bm{F}$ and define $\hat{\bm{\xi}} =
\bm{F} \hat{\bm{\gamma}}$ to arrive at
\begin{equation}
  \int_{\Omega} \mathcal{E}  dV = \int_{\Omega}  \hat{\bm{\gamma}}^\dag
  \bm{\mathcal{M}} \hat{\bm{\gamma}} dy = \int_{\Omega}
  \hat{\bm{\gamma}}^\dag \bm{F}^\dag \bm{F} \hat{\bm{\gamma}} dy =
  \int_{\Omega}  \hat{\bm{\xi}}^\dag \hat{\bm{\xi}} dy =
  || \hat{\bm{\xi}} ||_2,
\end{equation}
where $|| \cdot ||_2$ represents the $L_2$-norm and, accordingly, the
eigenvalue problem~(\ref{lineareigen}) becomes
\begin{equation}
  -i \omega (\bm{F}\hat{\bm{\gamma}}) =
  \bm{F} \bm{\mathcal{L}} \bm{F}^{-1} (\bm{F}\hat{\bm{\gamma}}).
  \label{lineareigen2}
\end{equation}
Therefore, once the linear operator is redefined as
$\bm{\mathcal{L}}_{L_2} = \bm{F}\bm{\mathcal{L}} \bm{F}^{-1}$, we can
conveniently use the $L_2$-norm and its associated inner product for all
computations. The transient growth $G$, defined as the maximum energy
growth over all possible initial conditions $\hat{\bm{\xi}}_0$, is
given below in the $L_2$ norm,
\begin{equation}
  G(t)=\underset{\hat{\bm{\xi}}_0}{max}\frac{||\hat{\bm{\xi}}(t)||_{2}}
  {||\hat{\bm{\xi}}(0)||_{2}}=\underset{\hat{\bm{\xi}}_0}{max}
  \frac{||\bm{\mathcal{T}}\hat{\bm{\xi}}(0)||_2}{||\hat{\bm{\xi}}(0)||_2} =
  || \bm{\mathcal{T}} ||_2 = ||e^{t\bm{F}\bm{\mathcal{L}} \bm{F}^{-1}}||_2,
\label{TGequation}
\end{equation}
where $\bm{\mathcal{T}}$ is the linear evolution operator, i.e., the
solution to equation~(\ref{lineartimeevolution}).

The parameters that are to be investigated include the injection level
$C$, the mobility parameter $M$, the charge diffusion coefficient
$Fe$, the Taylor parameter $T$, the Reynolds number $Re$ and the
streamwise and spanwise wavenumbers $\alpha$ and $\beta$.


\section{Numerical method and validation}\label{numerics}

\subsection{Numerical method}

To discretize the eigenvalue problem~(\ref{lineareigen}), we use a
spectral method based on collocation points chosen as the roots of
Chebyshev polynomials. The Matlab suite for partial differential
equations by~\cite{Weideman2000} is used for differentiation and
integration.

To impose the boundary condition, we employ the boundary boarding technique~\citep{Boyd2001}, in which selected rows of the linear matrices are replaced directly by the boundary conditions. 
When solving the eigenvalue problem via the Matlab routine {\tt{eig}}
with the above boundary condition enforced, we find that the
eigenvalues converge for a sufficient number $N$ of collocation points
(see figure~\ref{resolutioncheck} and table~\ref{unstablemodecheck} in
the validation section in the appendix~\ref{validation}) and approach the pure hydrodynamic results as
electric effects become negligible (see
figure~\ref{spectrumcomparison} and table~\ref{codecomparison}). The
corresponding eigenvectors, however, are incorrect since they do not
satisfy the proper boundary conditions (not shown). To overcome this
difficulty, we employ an iterative technique to obtain the eigenvector
associated with a specified eigenvalue. In the generalized eigenvalue
problem~(\ref{lineartimeevolution}), a desired eigenvalue $\omega$
(and its corresponding eigenvector) is targeted by applying the
spectral transformation
\begin{equation}
  \bm{\mathcal{S}}=(\bm{\mathcal{B}}-\omega
  \bm{\mathcal{A}}_1)\backslash \bm{\mathcal{A}}_1,
  \label{iterativemethod}
\end{equation}
where $\bm{\mathcal{A}}_1=-i \bm{\mathcal{A}}$ and $\bm{\mathcal{S}}$
will be processed by an iterative routine \citep{Saad2011}.


\subsection{Validation}

The stability problem for EHD flow is exceedingly challenging from a
numerical point of view, which warrants a careful and thorough
validation step, before results about stability characteristics, modal
and non-modal solutions and physical mechanisms are produced.
To conserve the clarity of the paper structure, we postpone the validation
steps in the appendix~\ref{validation}.


\section{Results of stability analysis} \label{results}

\subsection{EHD without cross-flow} \label{EHDwo}

As mentioned earlier, the parameter $T$ plays the main role of
determining the flow instability. The critical $T_c$ denotes the
minimum value of $T$ within the linear regime, above which
infinitesimal disturbances can grow exponentially in time; $T_c$ will
vary with the flow parameters. In the case of no cross-flow, the
effects of $Fe$, $T$, $M$ and $C$ on the flow stability are
investigated. As has already been assumed, the flow will be confined
to the SCL (space-charged-limited) regime, implying a large value for
$C.$

\begin{figure}
  \centering
  \subfigure {\includegraphics[width=.5\textwidth]
    {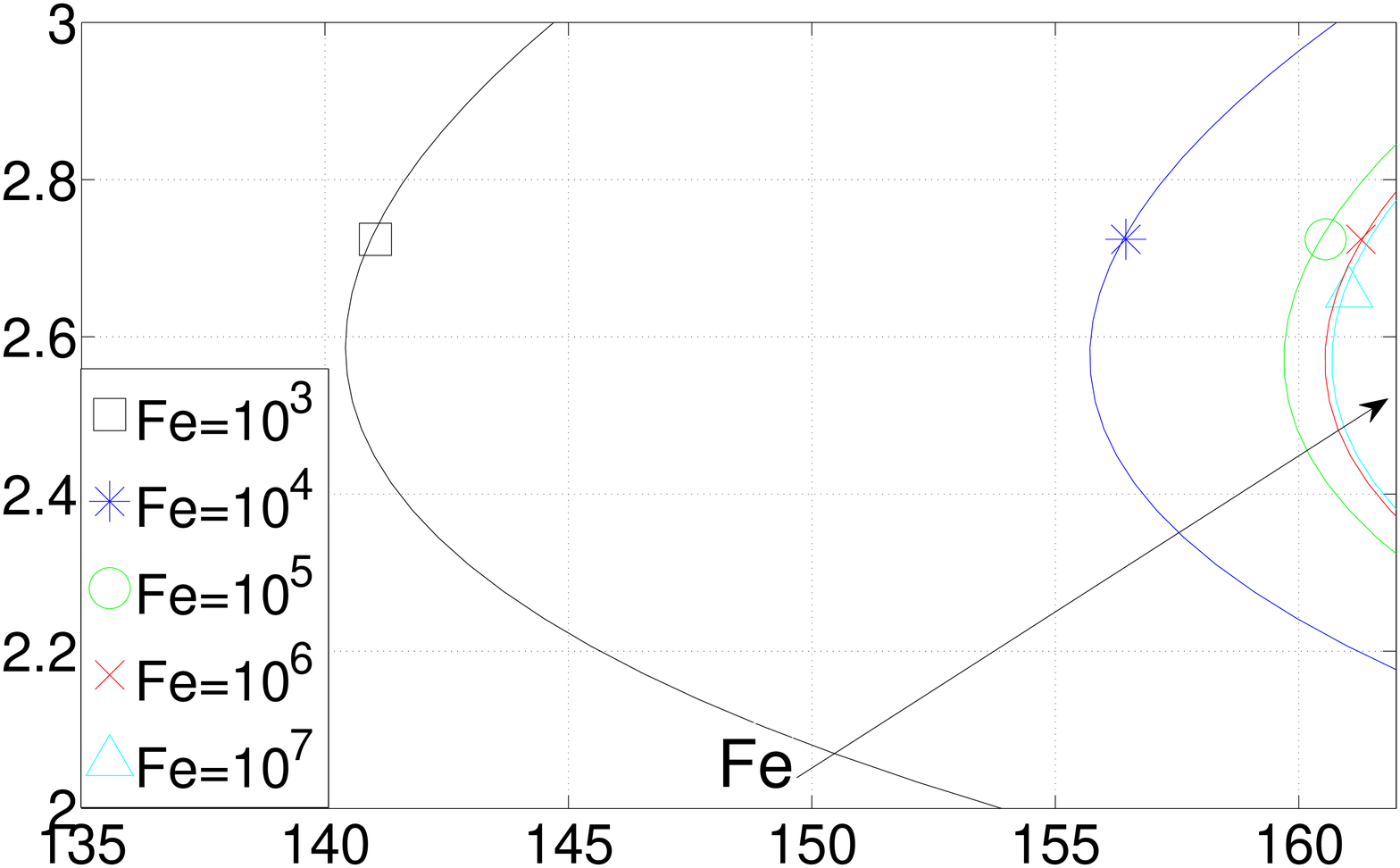}
    \put(-200,100){{\large $(a)$}}
    \put(-190,52){{\large $\alpha$}}
    \put(-95,-5){{\large $T$}}
    \includegraphics[width=.5\textwidth]
                    {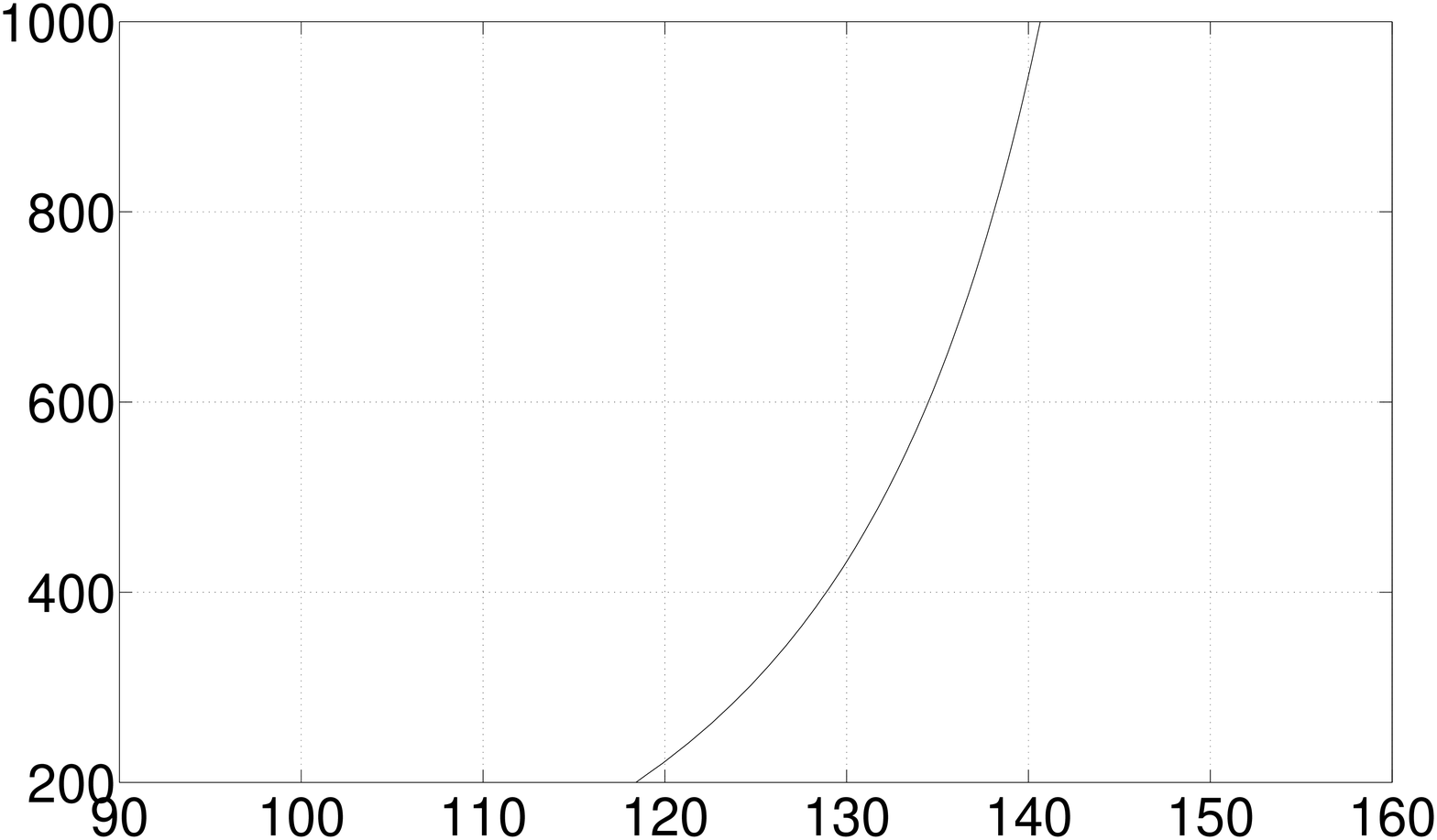}
    \put(-200,100){{\large $(b)$}}
    \put(-195,52){{\large $Fe$}}
    \put(-95,-5){{\large $T_c$}}}
  \subfigure {\includegraphics[width=.5\textwidth]
    {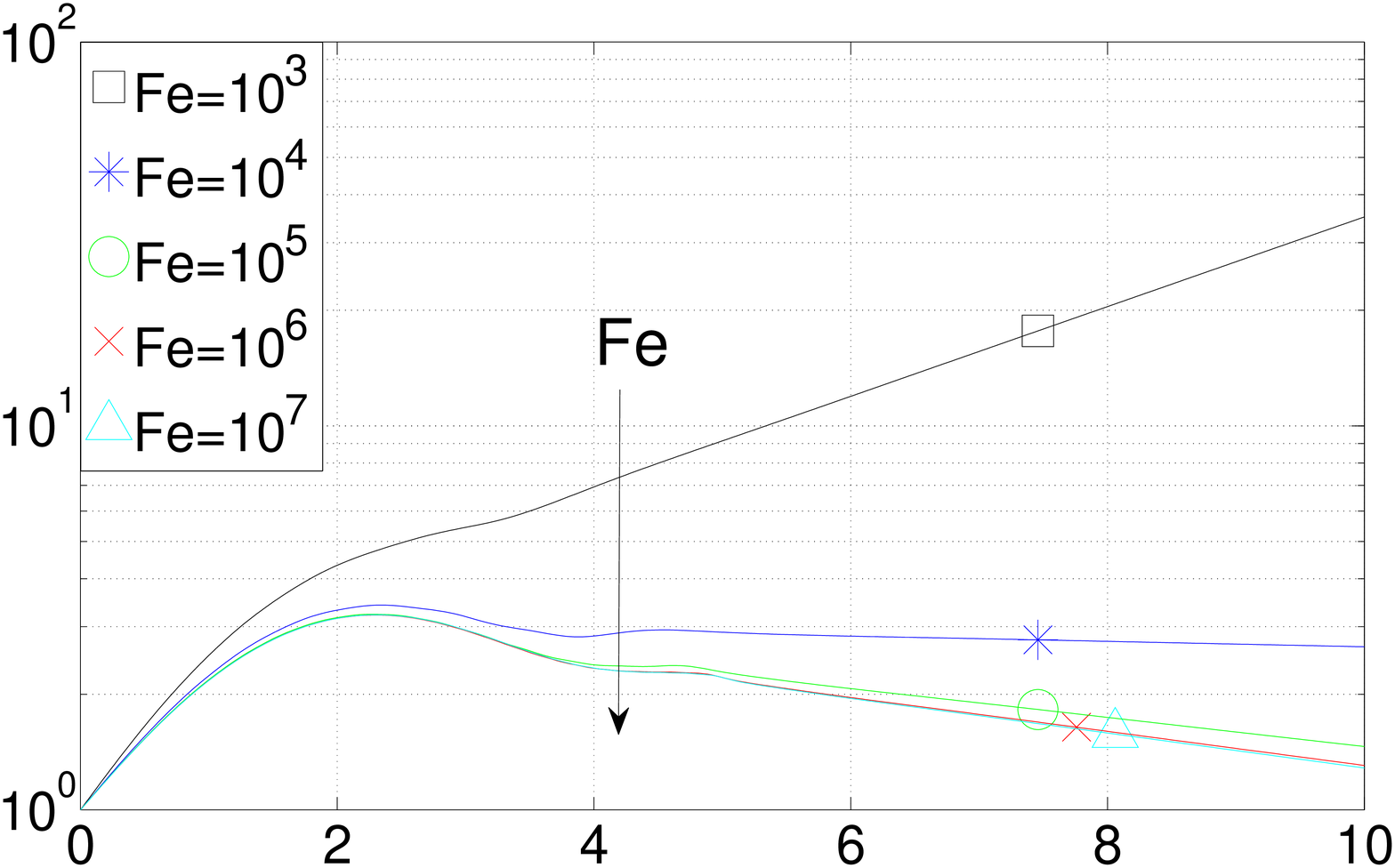}
    \put(-200,100){{\large $(c)$}}
    \put(-190,52){{\large $G$}}
    \put(-95,-5){{\large $t$}}}
  \caption{Effect of $Fe$. The parameters are $C=50$, $M=100$,
    $T=155$, $\alpha=2.5$, $\beta=0$ and $N=250$. The direction of the
    arrow indicates increasing $Fe$. (a) Neutral stability curves for
    various $Fe$. (b) $T_c$ as a function of $Fe$. (c) Transient
    energy growth versus time.}
  \label{effectofFe}
\end{figure}

We display the neutral stability curve in figure~\ref{effectofFe} for
different $Fe$ at $C=50$, $M=100$, $T=155$, $\alpha=2.5$, and
$\beta=0$. In the case without cross-flow, one does not need to
distinguish between the $x$- and the $z$-axis, since neither is
preferred by the base flow $\bar{\bm{U}}=0$; thus, we simply set
$\beta=0$. As mentioned in the validation section, results for
$Fe=10^{7}$ are very close to previous investigations. Even though the
diffusion coefficient is small, it plays an important role in
determining the critical $T_c,$ as shown in figure~\ref{effectofFe}(a)
and (b). For example, for $Fe=10^3$ the critical $T_c$ declines to
$140$. In fact, the value of $Fe$ could fall within the range $10^3
\sim 10^4$ for real liquids \citep{Perez1989}, when $Fe$ is
nondimensionalized in the same way as presented here. 
Physically, the effect of diffusion will smooth out sharp
gradients in the flow. Unlike the unidirectional electric field
pointing in the wall-normal direction, the diffusion effect act
equally in all directions. With charge diffusion considered in the
model, the discontinuous separatrix is blurred in the nonlinear phase \citep{Perez1989}.
The physical mechanism of how charge diffusion influences the critical stability parameter
$T_c$ will be discussed by using an energy analysis (see
section~\ref{energyNoC}). In addition, the transient growth of disturbance
energy has been discussed in~\cite{Atten1974} using the
quasi-stationary method; transient energy growth has been confirmed as
a minor factor in this work. This is also confirmed in our
computations, as presented in figure~\ref{effectofFe}(c):
specifically, the figure shows that disturbance energy growth $G$
reaches a value of about $3$ at $T=155$ for stable flows ($Fe>10^3$).

\begin{figure}
  \centering
  \subfigure{\includegraphics[width=.5\textwidth]
    {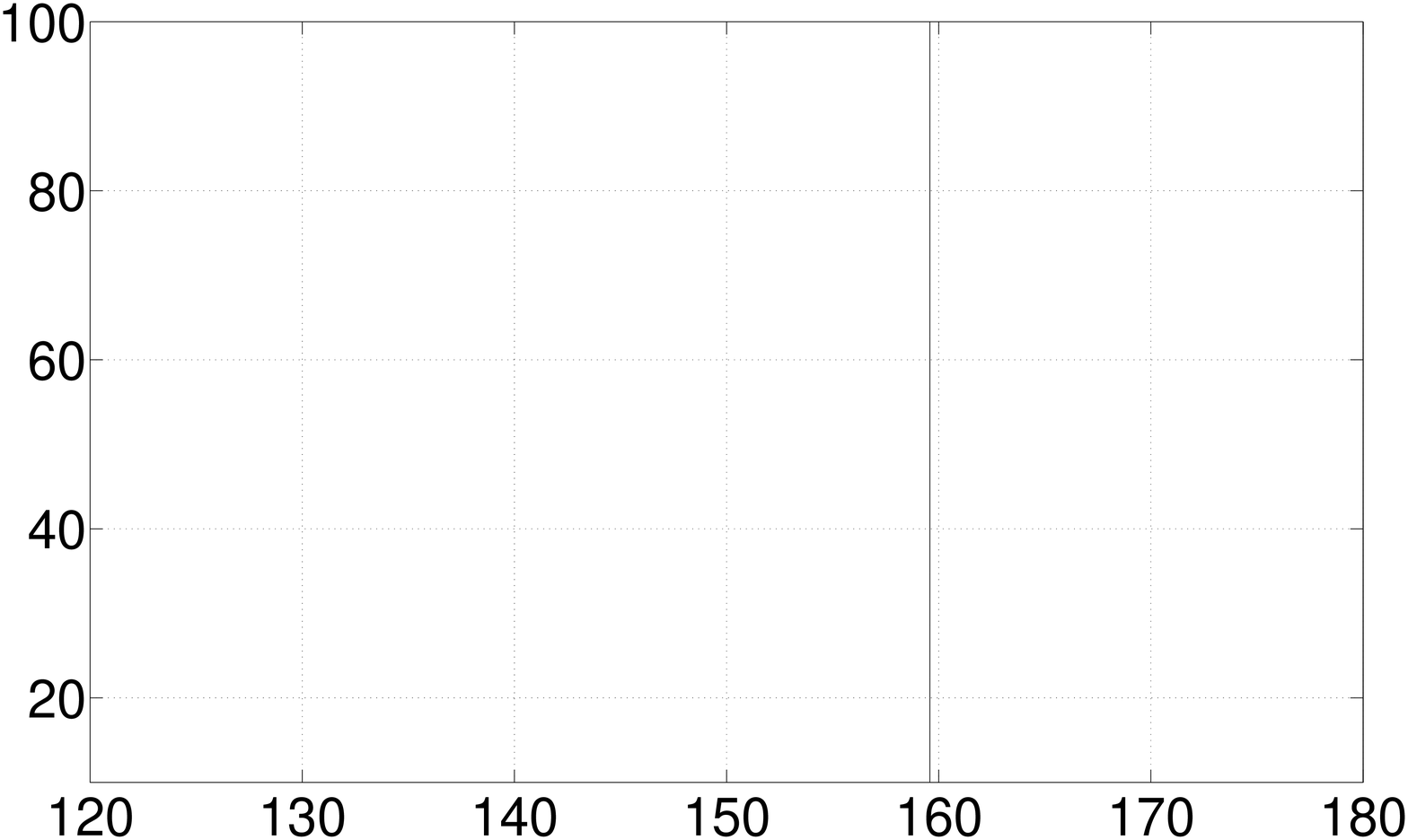}
    \put(-200,100){{\large $(a)$}}
    \put(-190,50){{\large $M$}}
    \put(-95,-5){{\large $T$}}
    \includegraphics[width=.5\textwidth]
                    {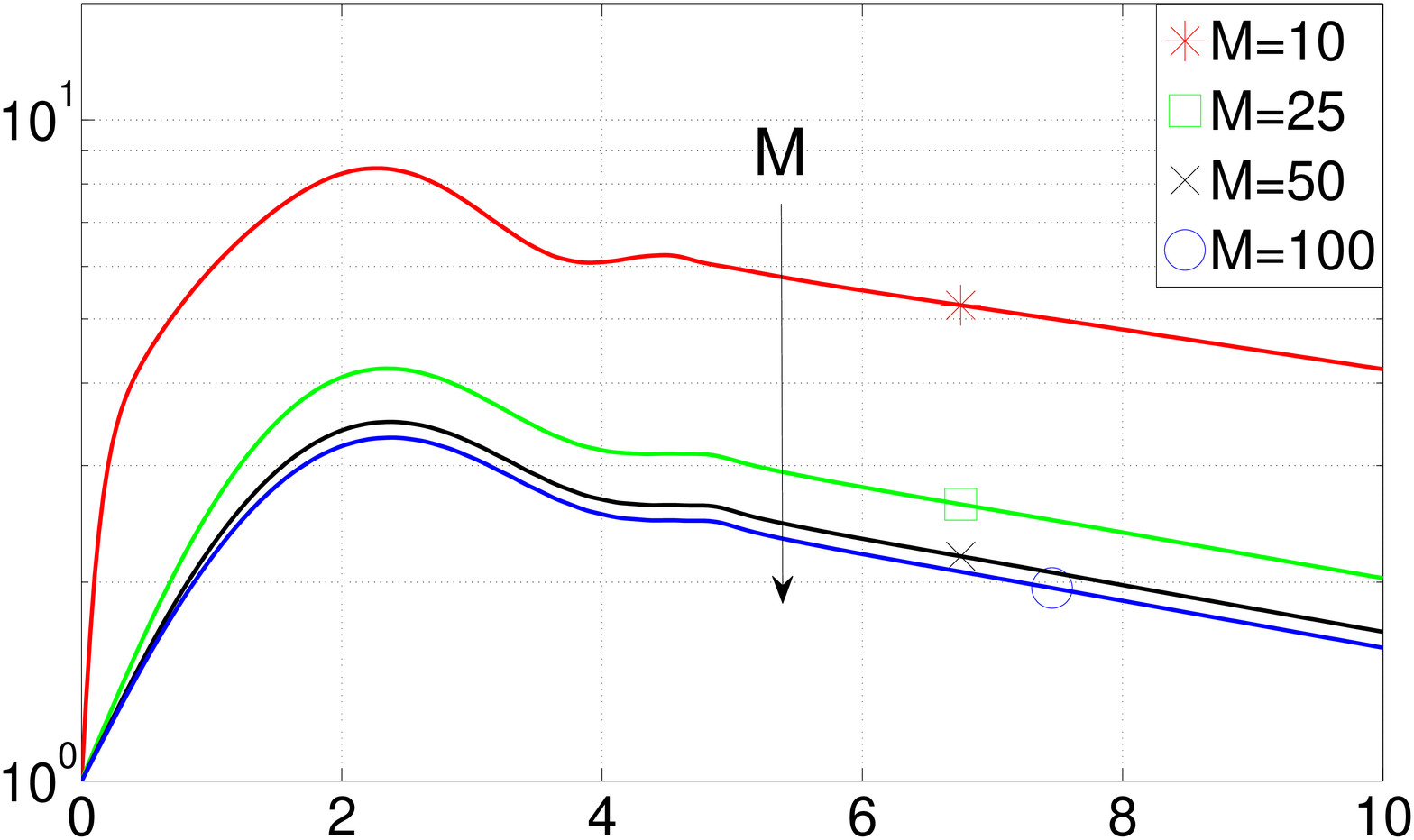}
    \put(-200,100){{\large $(b)$}}
    \put(-190,50){{\large $G$}}
    \put(-95,-5){{\large $t$}} }
  \caption{Effect of $M$. The parameters are $C=100$, $Fe=10^5$,
    $\alpha=2.57$, $\beta=0$ and $N=250$ without cross-flow. (a) The
    neutral stability curve. (b) Transient energy growth versus time
    for different $M$ and $T=155$. The direction of the arrow
    indicates increasing $M$.}
  \label{effectofM}
\end{figure}

The role of $M$ in EHD is analogous to that of the Prandtl number in
Rayleigh-B{\'e}nard convection. In figure~\ref{effectofM}(a), it is
shown that the variation of $M$ exerts no influence on the linear
stability criterion, $T_c=159.58$ at $C=100$, $Fe=10^5$, $\alpha=2.57$
and $\beta=0$; the same finding has been reported
in~\cite{Atten1972}. For the transient dynamics, however, the same
conclusion does not hold, as evidenced in
figure~\ref{effectofM}(b). The plot describes a trend of increasing
$G_{max}$ with smaller $M$. The slopes at the final time are slightly
different for each $M$, indicating that the asymptotic growth rates
differ slightly (while the linear stability criterion remains the
same).

\begin{figure}
  \centering
  \subfigure {\includegraphics[width=.5\textwidth]
    {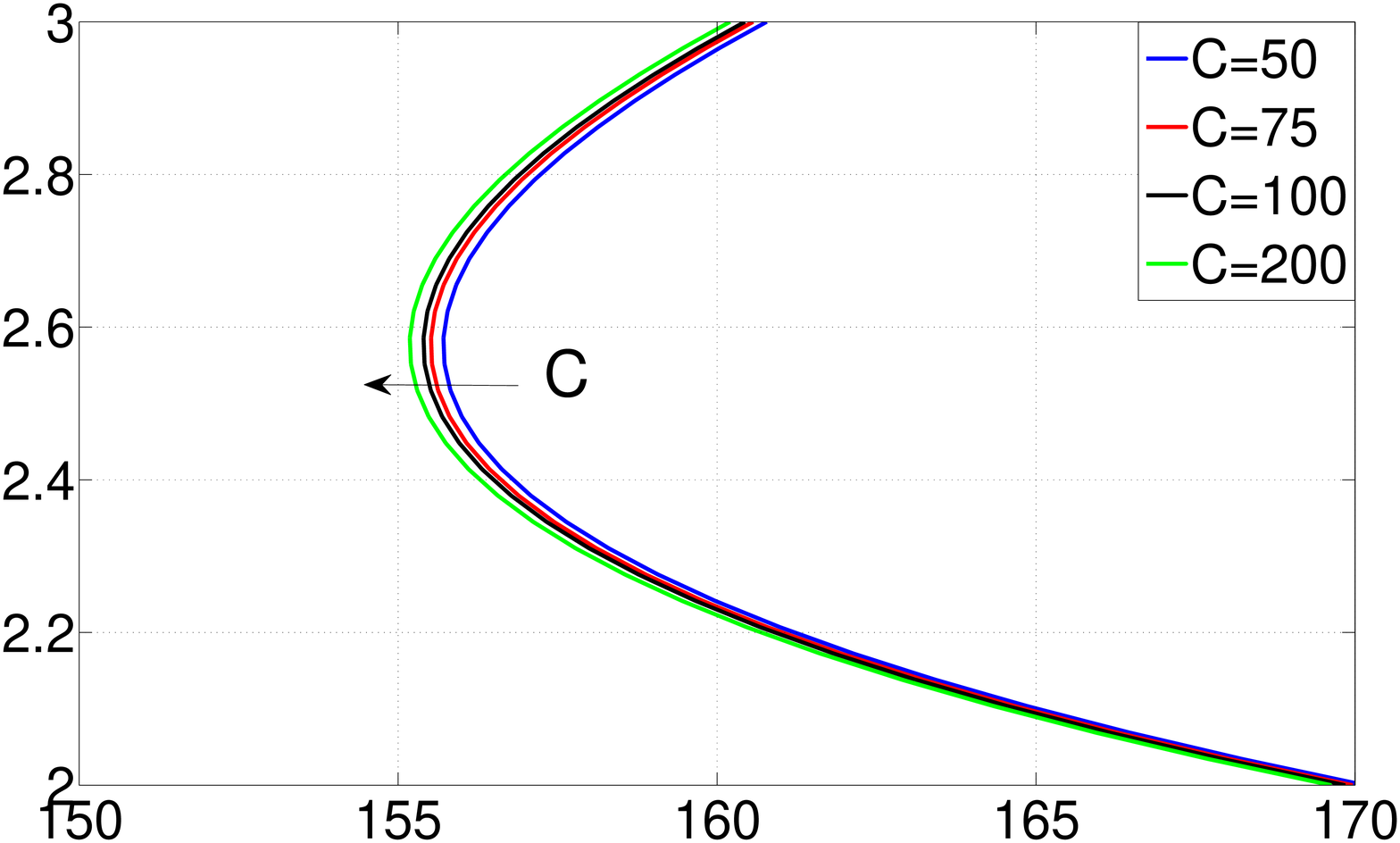}
    \put(-200,100){{\large $(a)$}}
    \put(-190,50){{\large $\alpha$}}
    \put(-95,-5){{\large $T$}}
    \includegraphics[width=.5\textwidth]
                    {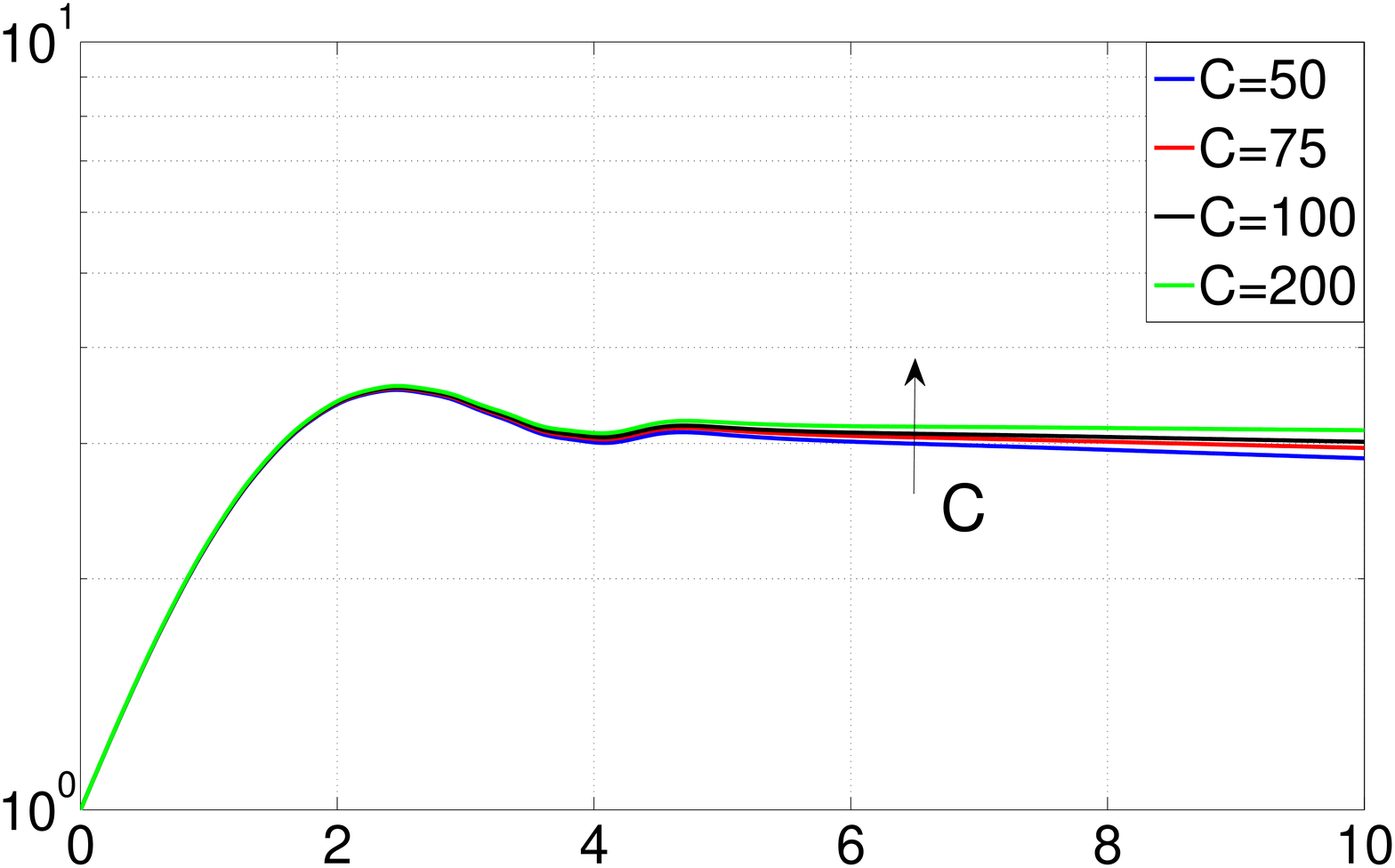}
    \put(-200,100){{\large $(b)$}}
    \put(-185,50){{\large $G$}}
    \put(-95,-5){{\large $t$}} }
  \caption{Effect of $C$. The parameters are $M=100$, $Fe=10^4$,
    $\beta=0$ and $N=250,$ without cross-flow. The direction of the
    arrow indicates increasing $C$. (a) Neutral stability curve for
    four different injection levels $C$ within the SCL regime. (b)
    Transient energy growth for different $C$ at $T=155,
    \alpha=2.57$.}
  \label{effectofC}
\end{figure}

Figure~\ref{effectofC} depicts the influence on $C$, which measures
the intensity of charge injection. \cite{Atten1972,Atten1979} reported a
dependence of the critical value $T_c$ on the parameter $C$. In
figure~\ref{effectofC}(a), we see that, in the SCL regime, increasing
$C$ will yield lower $T_c$. This result can be understood from a
physical argument. Increasing the intensity of charge injection will
lead to a higher concentration of charges between the electrodes. The
linear instability mechanism, as discussed before, relies on the
formation of an electric torque due to convective motions. With higher
charge concentration, the electric torque is stronger. Therefore, a
lower voltage difference is required, which amounts to stating that a
lower $T$ will be sufficient to form an electric torque of comparable
strength. But as we are in the SCL regime (with a value of $C=50$
considered very large), a rise of $C$ to $200$ only yields a minor
decrease in $T_c$. In contrast, the transient dynamics of the
perturbation energy $G$ appears not to be influenced by a change in $C$
for early times; for example, see the time interval $t \in [0, 3]$
in figure~\ref{effectofC}b.

\subsection{EHD with cross-flow} \label{EHDwithcrossflow}

When cross-flow is considered, the property of the linearized system
changes due to the presence of a base shear in the flow. Especially,
this shear will render the linearized operator `more non-normal'. We
first note that, in the modal stability analysis, Squire's theorem
still holds for EHD-Poiseuille flow, that is, a two-dimensional
instability will be encountered first. This can be easily verified by a
perfect analogy with standard viscous
theory~\citep{Schmid2001}. Moreover, there are two sets of scales in
the EHD problem with cross-flow. To study the influence of the
cross-flow on the electric and the charge fields, the values of $M$
and $T$ are kept in the vicinity of the values in the previous section: the scale of
the electric field will be considered primarily, whereas, when we
examine effects of the electric field exerted on the canonical Poiseuille
flow, we take the value of the free parameter $Re$ around $5772$,
i.e., the linear stability criterion for pressure-driven flow; the
latter choice introduces a scale based on the hydrodynamics. In both
cases, we will enforce the relation $Re=T/M^2$, which results in the
Reynolds number $Re$ to be rather low in the former case (denoted as
the low-$Re$ case) and relatively high in the latter case (referred to
as the high-$Re$ case).

\subsubsection{EHD: low $Re$}\label{EHDlowRe}

We have demonstrated that nonmodal effects in hydrostatic EHD are not
significant. In the presence of cross-flow, given that the Reynolds
number in this section is considered small, we expect the nonnormality
to be rather moderate as well. For this reason, we will mainly focus
on the modal stability characteristics for the low-$Re$ case.

\begin{figure}
  \centering
  \subfigure {\includegraphics[width=.5\textwidth]
    {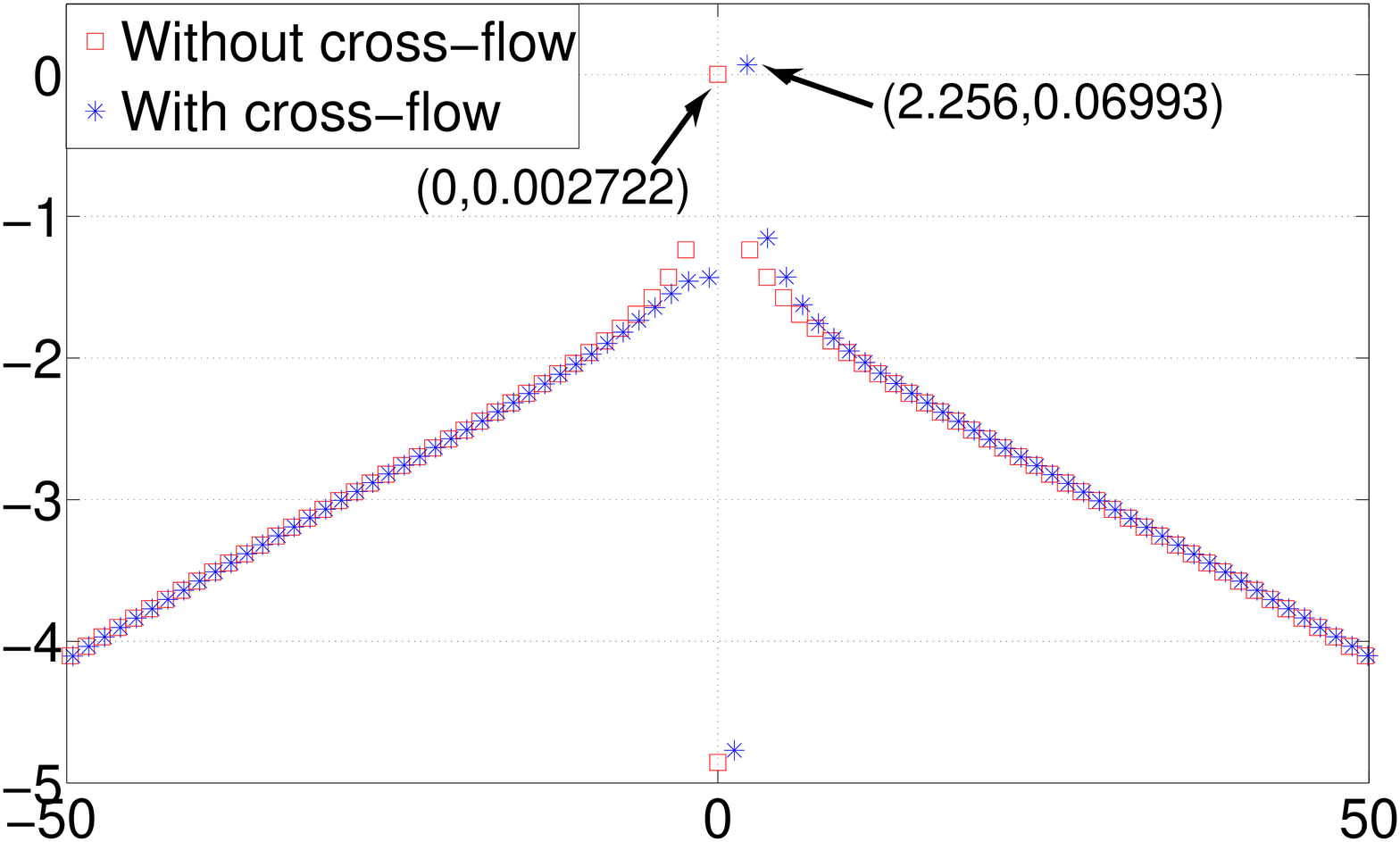}
    \put(-200,100){{\large $(a)$}}
    \put(-190,50){{\large $\omega_i$}}
    \put(-95,-5){{\large $\omega_r$}}
    \includegraphics[width=.5\textwidth]
                    {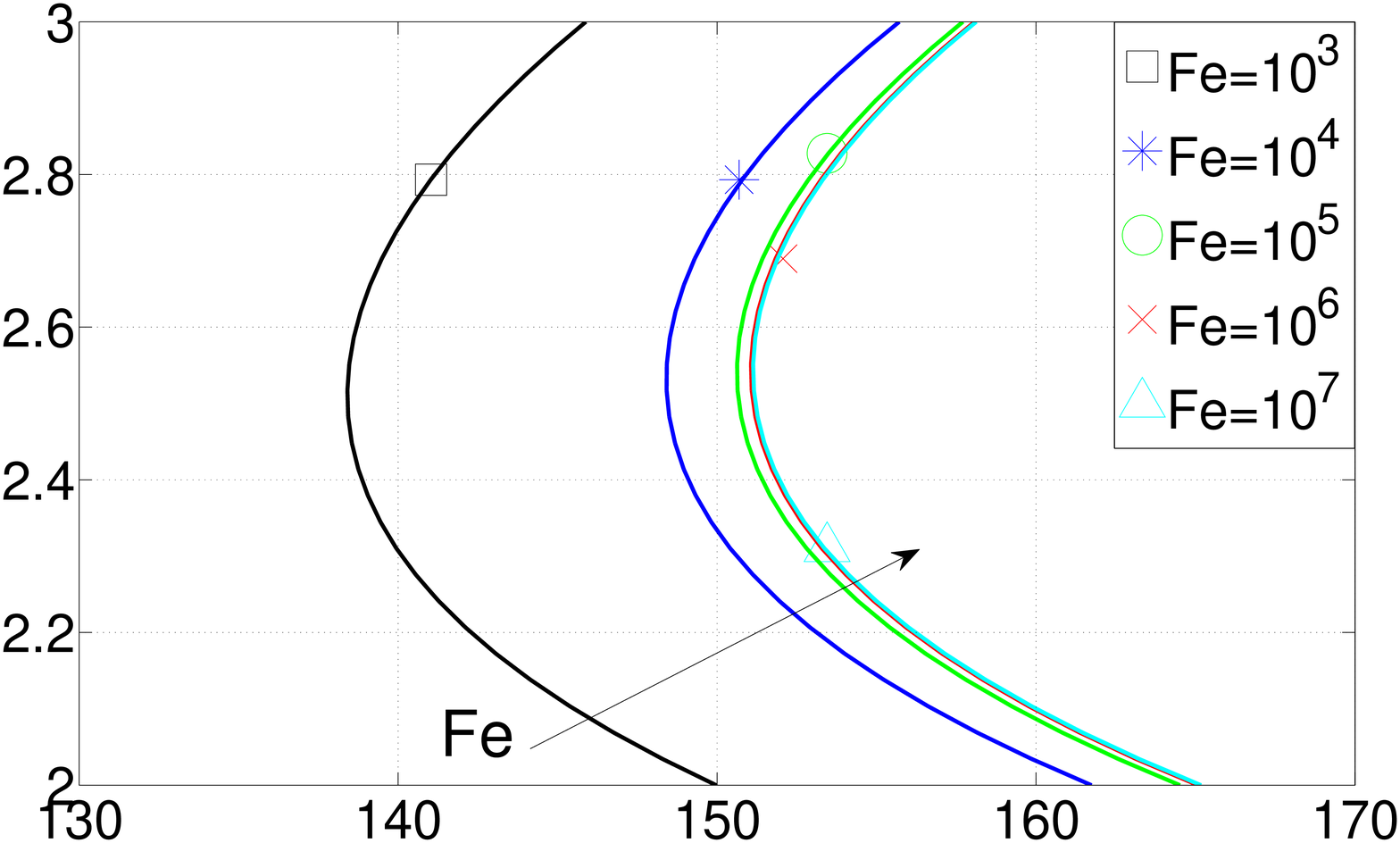}
    \put(-200,100){{\large $(b)$}}
    \put(-190,50){{\large $\alpha$}}
    \put(-95,-5){{\large $T$}} }
  \caption{(a) Spectra for the case with and without cross-flow for
    $C=50$, $M=100$, $Fe=10^5$, $T=160$, $Re=T/M^2=0.016$ (for
    cross-flow), $\alpha=2.57$, $\beta=0$ and $N=250$. (b) Effect of
    $Fe$ on the neutral stability curve. The parameters are
    $C=50$, $M=100$, $\beta=0$ and $N=250,$ with cross-flow. The
    direction of the arrow indicates increasing $Fe$. }
  \label{PoiseuilleeffectofFe}
\end{figure}

In figure~\ref{PoiseuilleeffectofFe}(a), it is observed that the
symmetry of the hydrostatic EHD spectrum is now broken due to the
presence of cross-flow. The most unstable perturbation travels at a
positive phase speed $u_p=\omega_r/\alpha=2.256/2.57=0.8778,$ induced
by cross-flow convection (the centerline velocity of the cross-flow is
$1,$ as we set $Re=T/M^2$ in equation~(\ref{baseflowequation})).  In
figure~\ref{PoiseuilleeffectofFe}b, we show the neutral stability
curve for $C=50$, $M=100$, $\beta=0$ and $N=250$, which can be
directly compared to the results in figure~\ref{effectofFe}. Note that
since $Re=T/M^2$ is enforced, the Reynolds number $Re$ is not
identical for each point, but generally small. We see that, with
cross-flow, the critical $T_c$ decreases compared to the no-cross-flow
case; this indicates that the flow is more unstable in the presence of
a low-$Re$ cross-flow compared to the results in
figure~\ref{effectofFe}(a). To investigate the reason behind this
destabilization, we again resort to an energy analysis in
section~\ref{crossflowlowRe}. Previously, an energy analysis for EHD
with cross-flow has been studied in~\cite{Castellanos1992}.

\begin{figure}
  \centering
  \subfigure {\includegraphics[width=.5\textwidth]
    {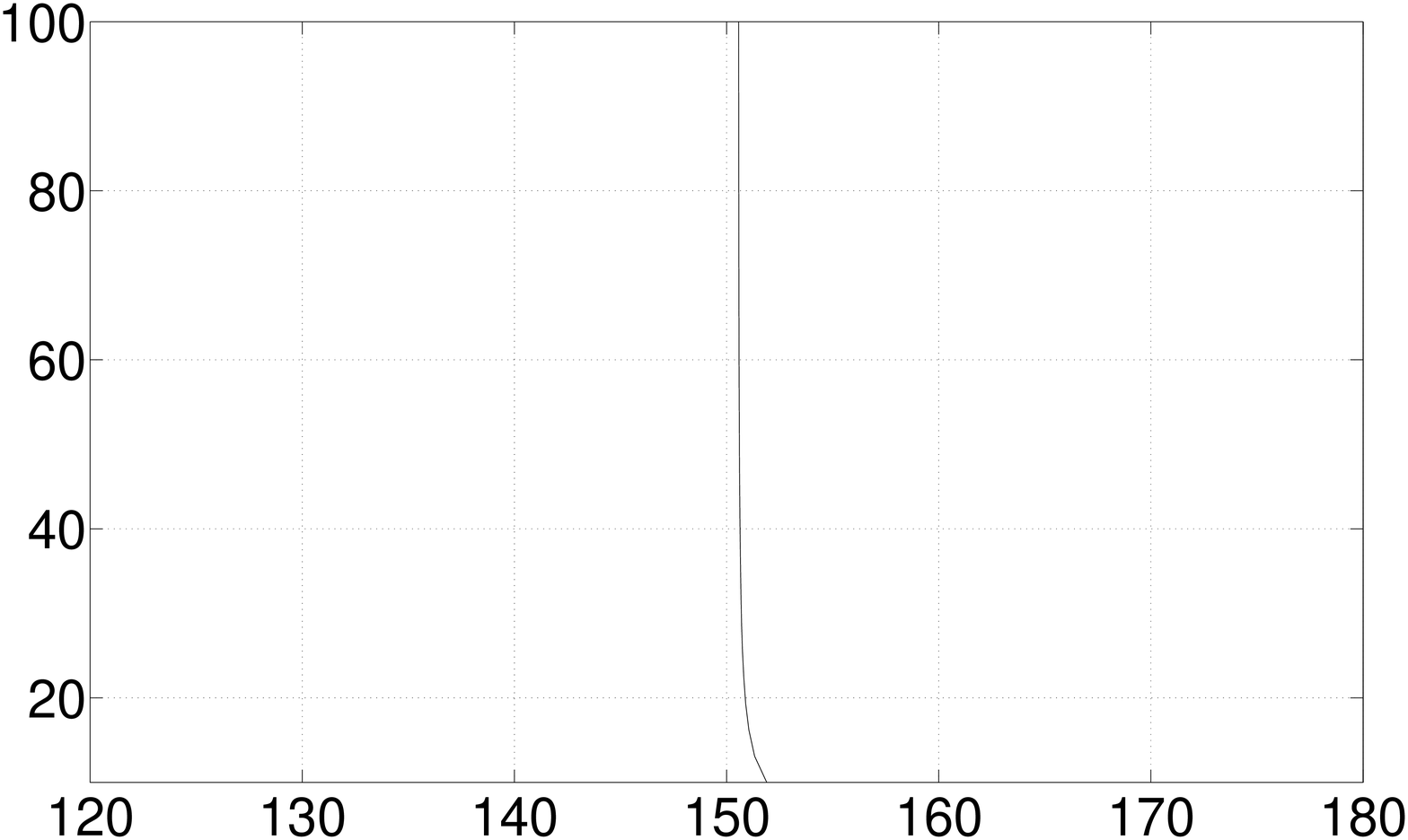}
    \put(-200,100){{\large $(a)$}}
    \put(-190,50){{\large $M$}}
    \put(-95,-5){{\large $T$}}
    \includegraphics[width=.5\textwidth]
                    {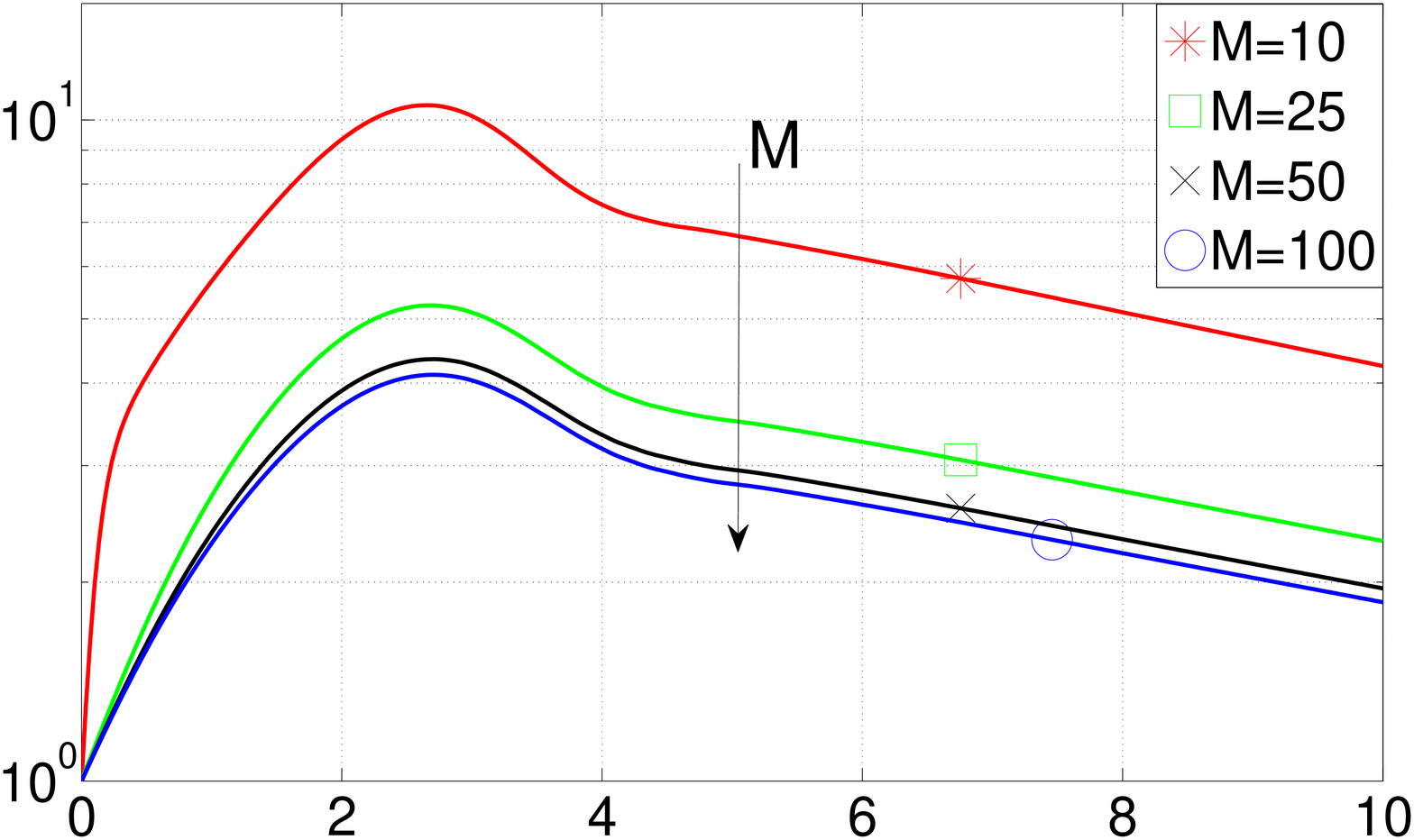}
    \put(-200,100){{\large $(b)$}}
    \put(-190,50){{\large $G$}}
    \put(-95,-5){{\large $t$}} }
  \caption{Effect of $M$. (a) Neutral stability curve with cross-flow
    for $C=100$, $Fe=10^5$, $\alpha=2.57$, $\beta=0$ and $N=250$. (b)
    Transient energy growth for different $M$ and $T=145$. The
    direction of the arrow indicates increasing $M$.}
  \label{PoiseuilleeffectofM}
\end{figure}

Even though varying $M$ has no effect on the linear stability when
$\bar{U}=0,$ as has been discussed briefly in the previous section, in
the presence of cross-flow, changing $M$ does influence the linear
stability. This is displayed in figure~\ref{PoiseuilleeffectofM}(a),
where we see that effects of $M$ are only discernable when $M$ is
small. We will discuss this issue further in the energy analysis
section~\ref{crossflowlowRe}. Considering non-normal linear stability,
transient energy growth $G$ is still small, even though slightly
higher than in the no-cross-flow case.


\subsubsection{EHD: high $Re$}\label{EHDhighRe}

\begin{figure}
  \centering
  \subfigure{
    \includegraphics[width=.5\textwidth]{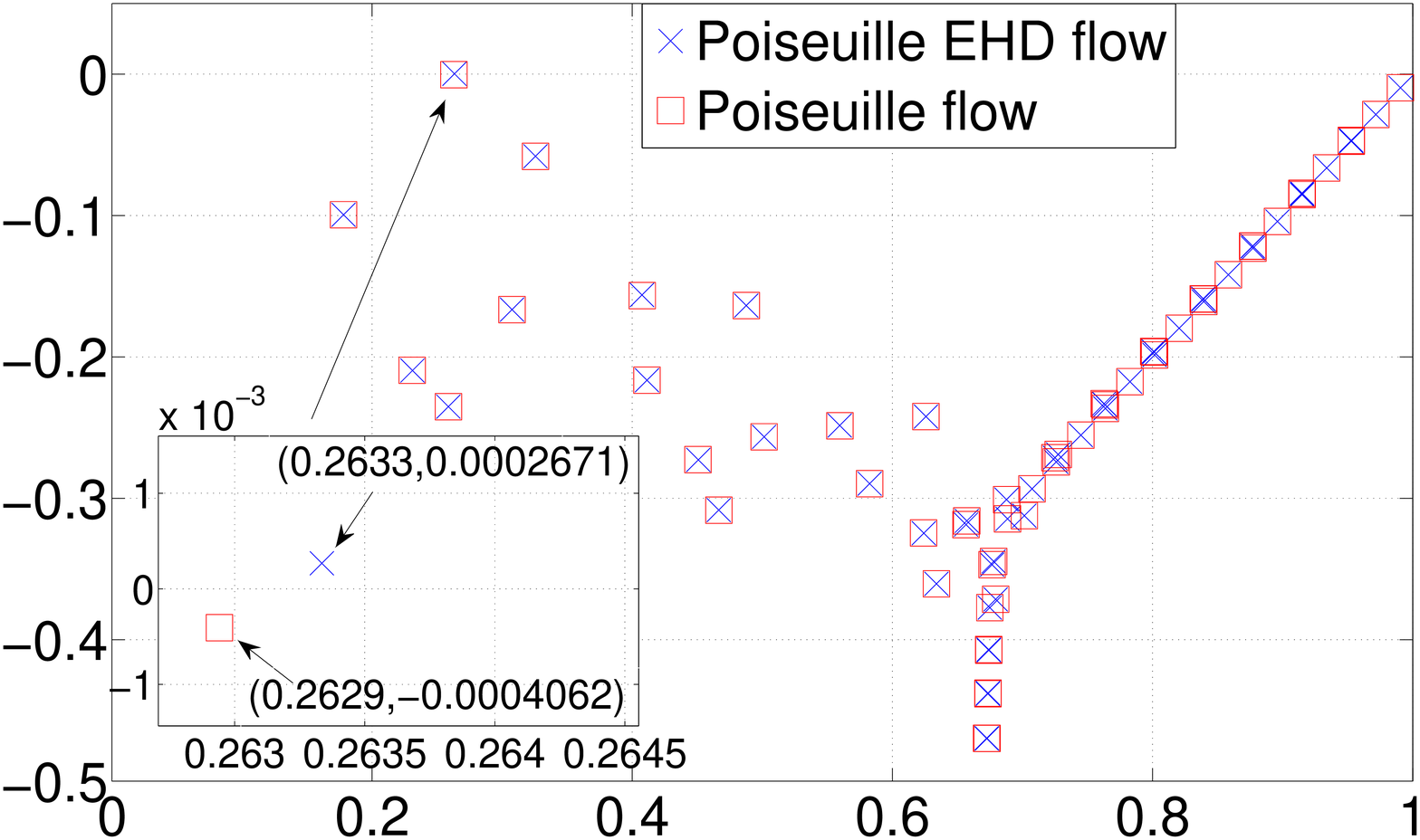}
    \put(-200,100){{\large $(a)$}}
    \put(-190,50){{\large $\omega_i$}}
    \put(-95,-5){{\large $\omega_r$}}
    \includegraphics[width=.5\textwidth]{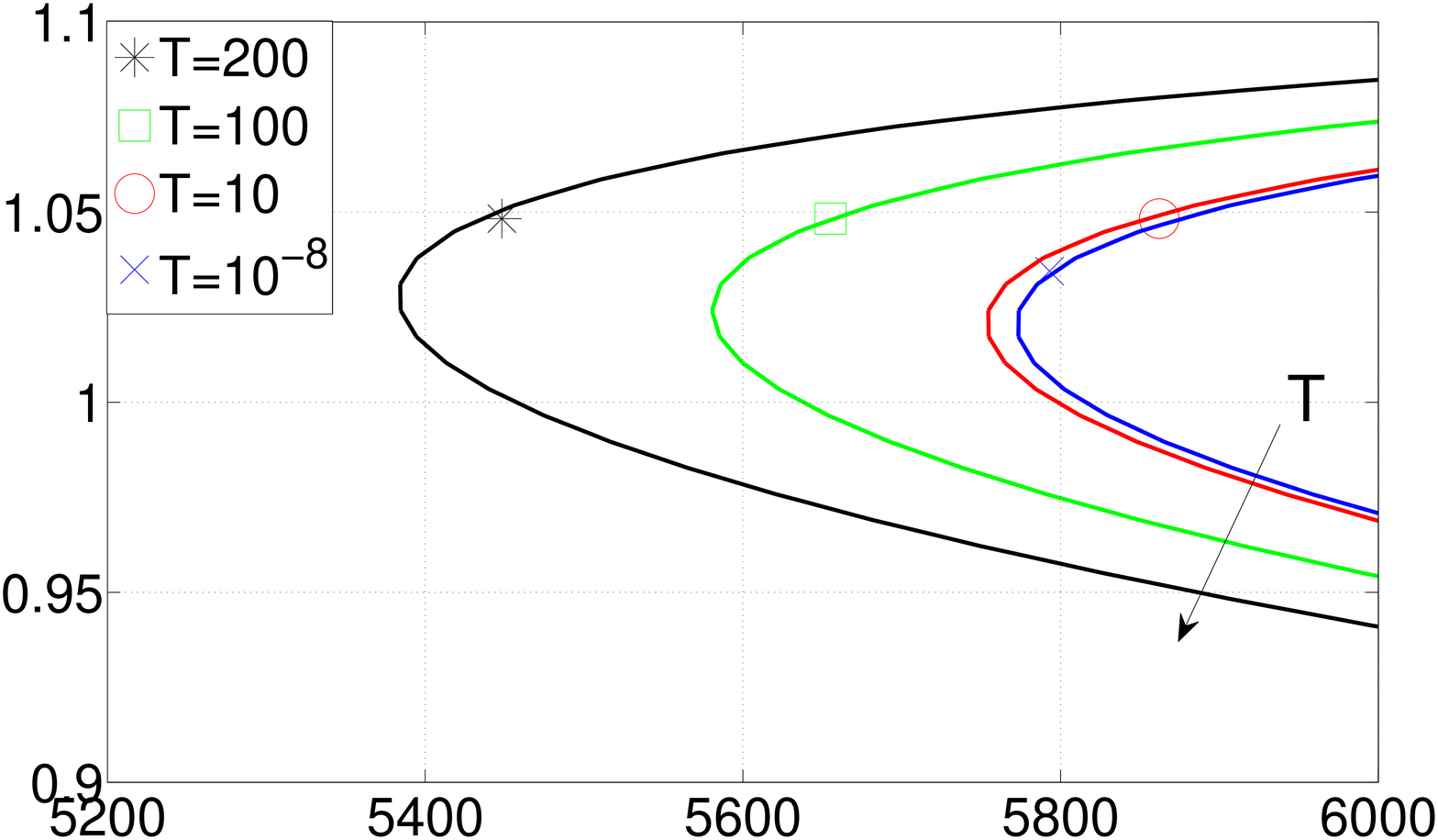}
    \put(-200,100){{\large $(b)$}}
    \put(-190,50){{\large $\alpha$}}
    \put(-95,-5){{\large $Re$}} }
  \subfigure {
    \includegraphics[width=.5\textwidth]{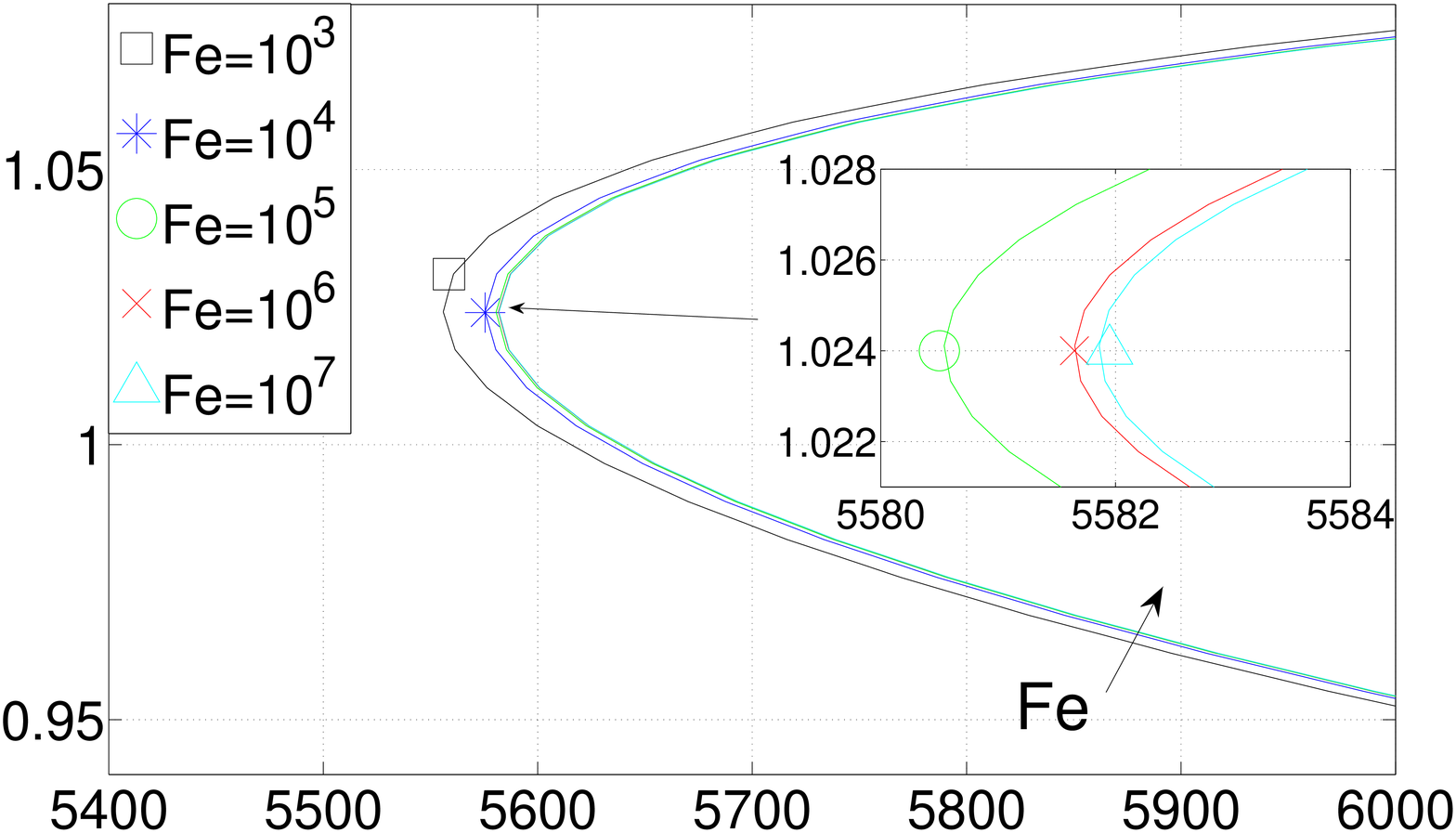}
    \put(-200,100){{\large $(c)$}}
    \put(-190,50){{\large $\alpha$}}
    \put(-95,-5){{\large $Re$}}  }
  \caption{(a) Comparison of two spectra for $T=10^{-8}$ and
    $T=200$ at $C=100, Re=5600, Fe=10^5, \alpha=1, \beta=0$. (b)
    Effect of $T$ on the neutral stability curve with
    cross-flow at $C=100$, $Fe=10^5$, $M=\sqrt{T/M^2}$, $\beta=0$. The
    direction of the arrow indicates increasing $T$. (c) Effect of
    $Fe$ on the neutral stability curve with cross-flow at
    $C=100, T=100, M=\sqrt{T/M^2},\beta=0$. The direction of the arrow
    indicates increasing $Fe$.}
  \label{HighRePoiseuilleeffectofT}
\end{figure}

In this section, we consider the flow governed by the inertial scale,
i.e., in the high-$Re$ regime. To discuss the results more properly, the Reynolds number $Re$
will be the free parameter, and the governing momentum equation is
given by~(\ref{nd_ehd2_highRe}), see section~\ref{linearstabilityproblem}.
The modal stability is examined in
figure~\ref{HighRePoiseuilleeffectofT}. In subfigure (a), changes in
the spectrum due to the additional electric field are visible. It
appears that the core modes, wall modes and center modes do not change
appreciably, except that the growth rate of the most unstable mode
increases. In subfigure (b), we plot the neutral stability curve for
varying $T$. The pure hydrodynamic linear stability limit $Re=5772.2$
is recovered by considering a minute value for $T$ such as $T=10^{-8}$
(we could have taken $T=0$, but to be compatible with
equation~(\ref{nd_ehd2}) and the discussion based on that equation in
other literature, we assign to $T$ a negligibly small value). With
increasing $T$, the system becomes more unstable, as the critical
linear stability criterion becomes smaller. The reason for this is
obviously due to the effect of the electric field transferring energy
into the velocity fluctuations, while at the same time modifying the
canonical channel flow; see table~\ref{tb:energyPoiT} in the energy
analysis section~\ref{crossflowhighRe} at $C=100, Fe=10^5, Re=5500,
\alpha=1$ and $\beta=0$.

We also investigate the effect of charge diffusion $Fe$ on the flow
stability. The results concerning the neutral stability curve are
shown in figure~\ref{HighRePoiseuilleeffectofT}(c) at $C=100, T=100,
M=\sqrt{T/M^2},\beta=0$. For small charge diffusion (large $Fe$), the
critical Reynolds number $Re$ is only slightly affected by changes in
$Fe$. Only when $Fe=10^3$ does the critical Reynolds number $Re$ drop
noticeably, though the destabilization effect is still small. It thus
can be concluded that charge diffusion has only a small influence on
the dynamics of EHD cross-flow at high $Re$. This is due to the
inertial scale we are considering. As we have seen in the hydrostatic
EHD flow, the effect of charge diffusion is significant, considering
the electric scale, i.e., at relatively small (or zero) Reynolds
numbers.

It is well known that in high-$Re$ Poiseuille flow the two-dimensional
Orr mechanism is not the principal mechanism for perturbation energy
growth over a finite time horizon. The flow is expected to become
turbulent within a short time interval, even though the asymptotic
growth rate of the linear system is negative. The non-normal nature of
the linearized Navier-Stokes operator for channel flow --- in physical
terms, due to the base flow modulation by spanwise vorticity tilting
into the wall-normal direction --- suggests that transient disturbance
growth during the early phase should be considered primarily.

\begin{figure}
  \centering
  \subfigure {\includegraphics[width=.5\textwidth]
    {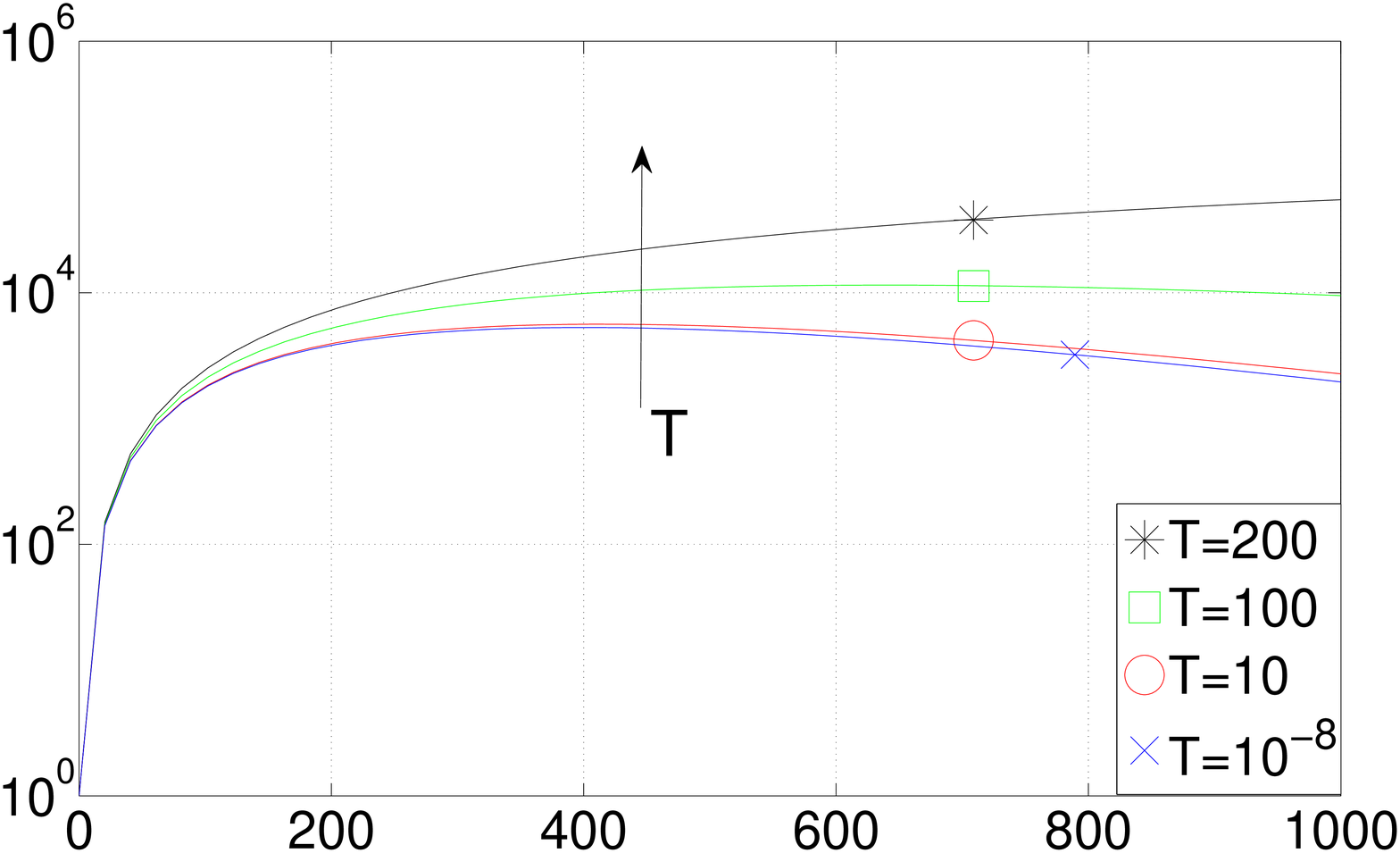}
    \put(-200,100){{\large $(a)$}}
    \put(-190,50){{\large $G$}}
    \put(-95,-5){{\large $t$}}
    \includegraphics[width=.5\textwidth]
                    {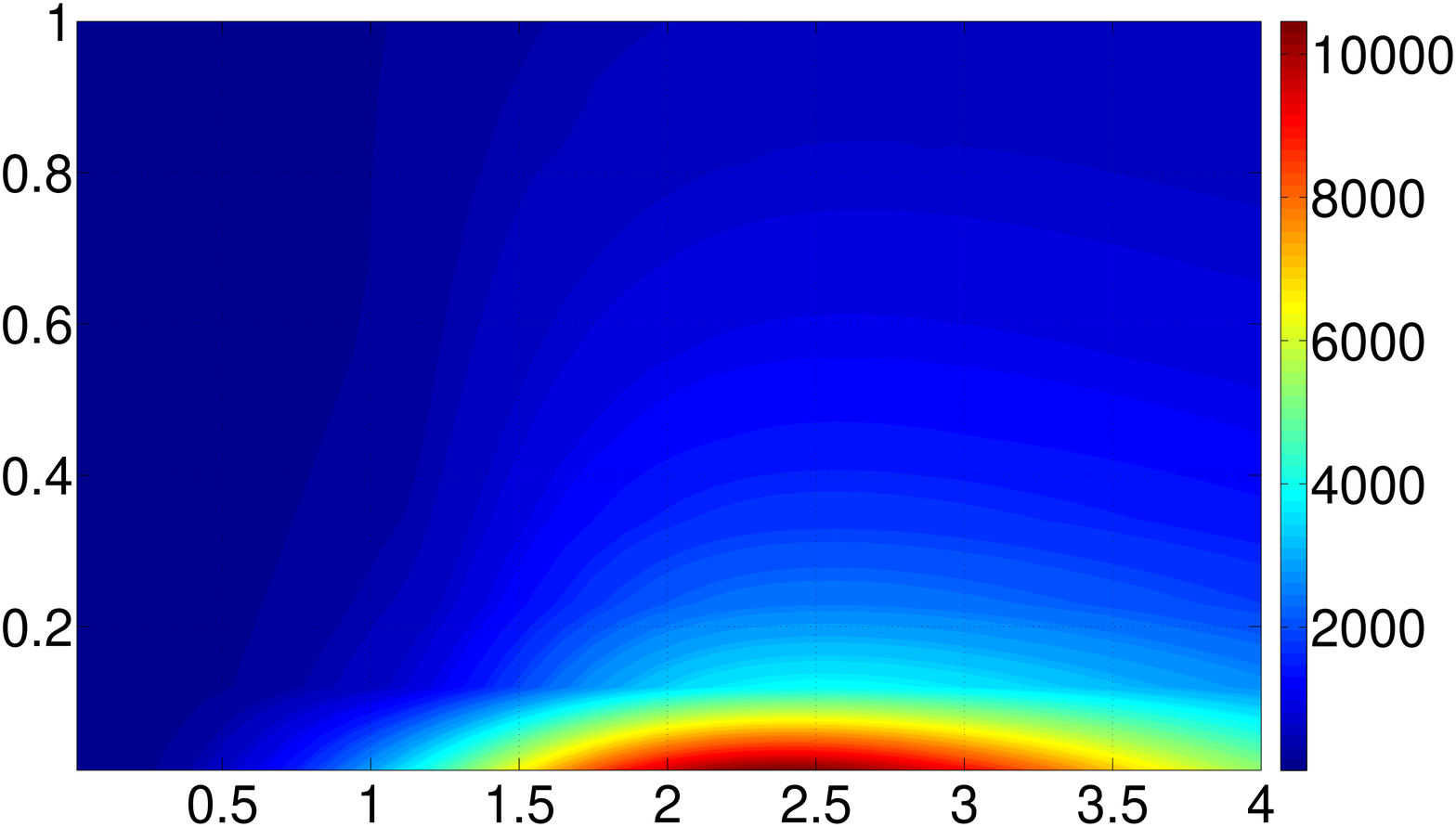}
    \put(-200,100){{\large $(b)$}}
    \put(-190,50){{\large $\alpha$}}
    \put(-95,-5){{\large $\beta$}} }
  \subfigure {\includegraphics[width=.5\textwidth]
    {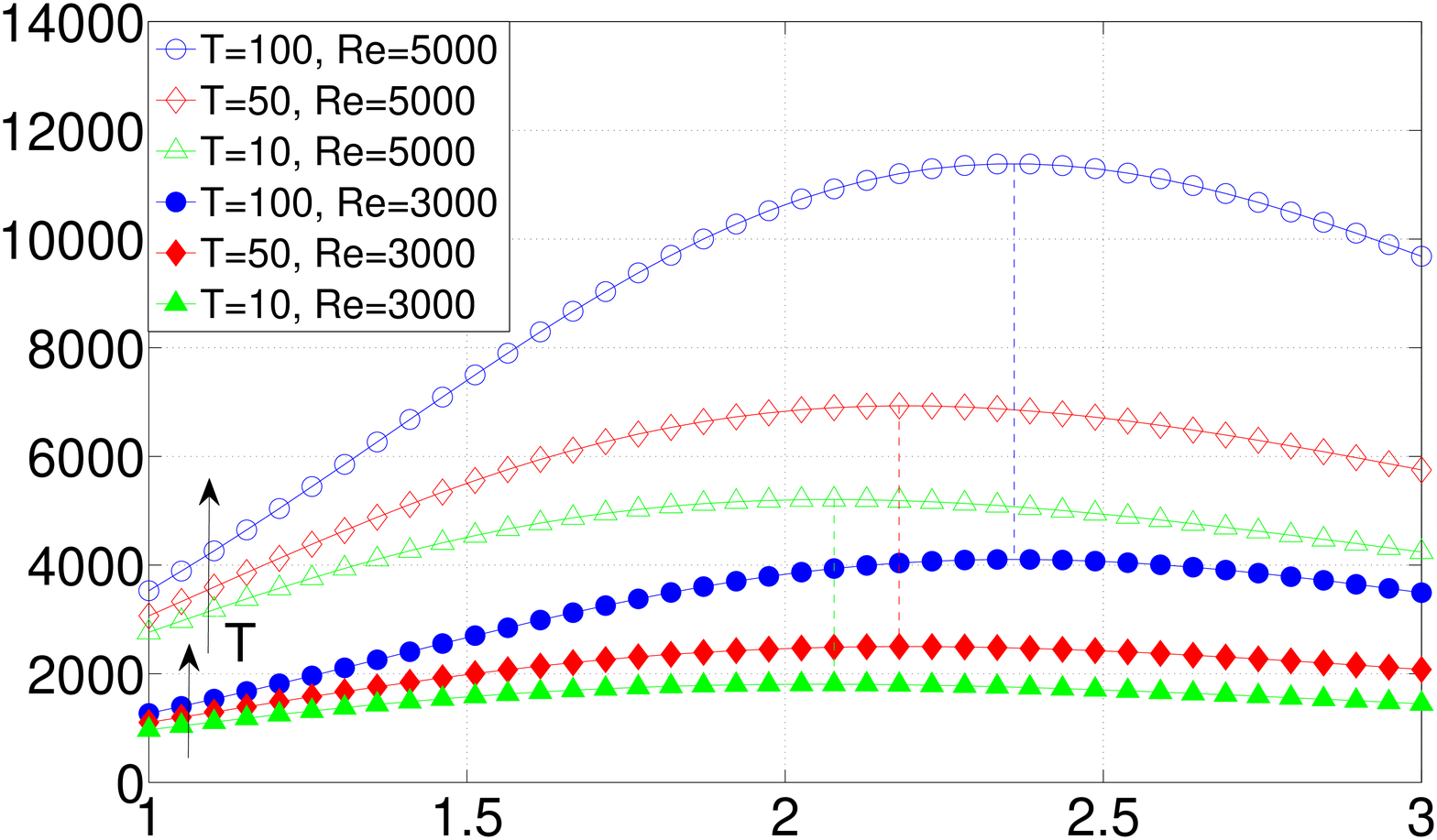}
    \put(-200,100){{\large $(c)$}}
    \put(-200,50){{\large $G$}}
    \put(-95,-5){{\large $\beta$}}   }
  \caption{Effect of $T$. The direction of the arrow indicates
    increasing $T$. (a) Transient energy growth for $C=100$,
    $Re=5200$, $Fe=10^5$, $\alpha=0$, $\beta=2$. (b) Contours of
    transient growth $G$ in the $\alpha$-$\beta$-plane at $C=100,
    Re=5000, Fe=10^5, T=100$. (c) Transient growth $G$ as a function
    of $\beta$ at $C=100$, $Fe=10^5$, $\alpha=0$.}
  \label{HighRePoiseuilleeffectofT_tg}
\end{figure}

In figure~\ref{HighRePoiseuilleeffectofT_tg}(a), we present the
transient growth $G$ for different $T$. Mainly, the effect of
increasing $T$ is to enhance transient growth. The optimal initial
condition which achieves maximum transient growth is shown in
figure~\ref{HighRePoiseuilleeffectofT_tg}(b). The optimal wavenumbers
for the pure hydrodynamic case, independent of $Re$, are found to be
$\alpha=0$ and $\beta=2.05,$ suggesting streamwise-independent
vortices as the most amplified structures~\citep{Schmid2001}. For
high-$Re$ EHD with cross-flow, the maximum transient growth is still
found to favor streaks ($\alpha=0$), but with a different optimal
spanwise wavenumber of $\beta=2.36$ at $C=100, Re=5000, Fe=10^5,
T=100$ (see
figure~\ref{HighRePoiseuilleeffectofT_tg}(b)). Interestingly, for a
different value of $T$, i.e., a different amount of potential drop
across the electrodes, the optimal wavenumber would be different. For
example, at $T=50$ the optimal $\beta=2.18,$ while at $T=10$ the
optimal $\beta=2.08$, as shown in
figure~\ref{HighRePoiseuilleeffectofT_tg}(c). It seems that, for
smaller $T$, approaching the regime of pure hydrodynamics, the optimal
$\beta$ is converging towards $\beta=2.05$. The independence of maximum
transient growth on $Re$ for various $\beta$ still holds in the limit of high-$Re$ EHD
flow, as indicated by the dashed lines connecting the peaks of the two
curves for $Re=5000$ and $Re=3000$ in
figure~\ref{HighRePoiseuilleeffectofT_tg}(c). In the nonlinear regime of EHD Poiseuille flow,
the influence of the electric field on the streaks has been reported
in~\cite{Soldati1998}. These authors reported the spanwise
spacing of the low-speed streaks are about $105\pm15$ in wall units, which 
is different from the average spanwise spacing of the streaks 
in Poiseuille flow, that is $100$ in wall units, see \cite{Butler1993} for example. 
Thus, to some extent, our results that the spanwise spacing of the streaks changes in the linear EHD cross-flow
agree qualitatively with these findings. These authors also found that the cross-flow is
weakened by the electric field. This does not stand in conflict with
the current results, because enhanced transient growth due to the
electric field, as found here, only indicates that, in the linear
phase, transition to turbulence is more rapid when compared to
canonical channel flow; no conclusions can be drawn for the flow
behavior in the nonlinear regime. 
To more fully understand how the
electric field influences streaks and streamwise vortices in the
nonlinear phase of transition, a more comprehensive study of the role
played by EHD in the formation and dynamics of a self-sustaining cycle
\citep{JIMENEZ1999} is called for.

\begin{figure}
  \centering
  \subfigure {\includegraphics[width=.5\textwidth]
    {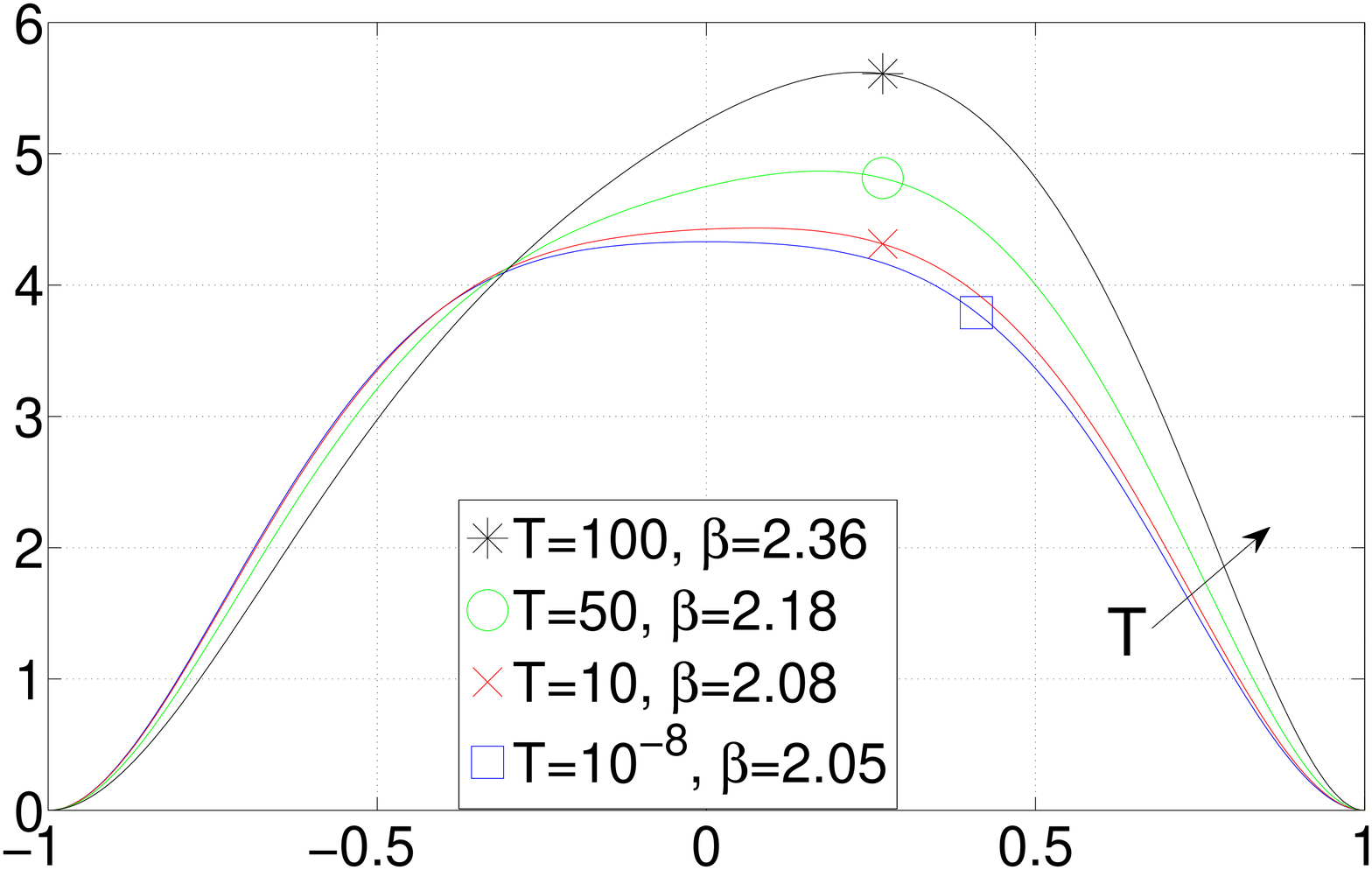}
    \put(-200,100){{\large $(a)$}}
    \put(-190,55){{\large $v$}}
    \put(-95,-5){{\large $y$}}
    \includegraphics[width=.5\textwidth]
                    {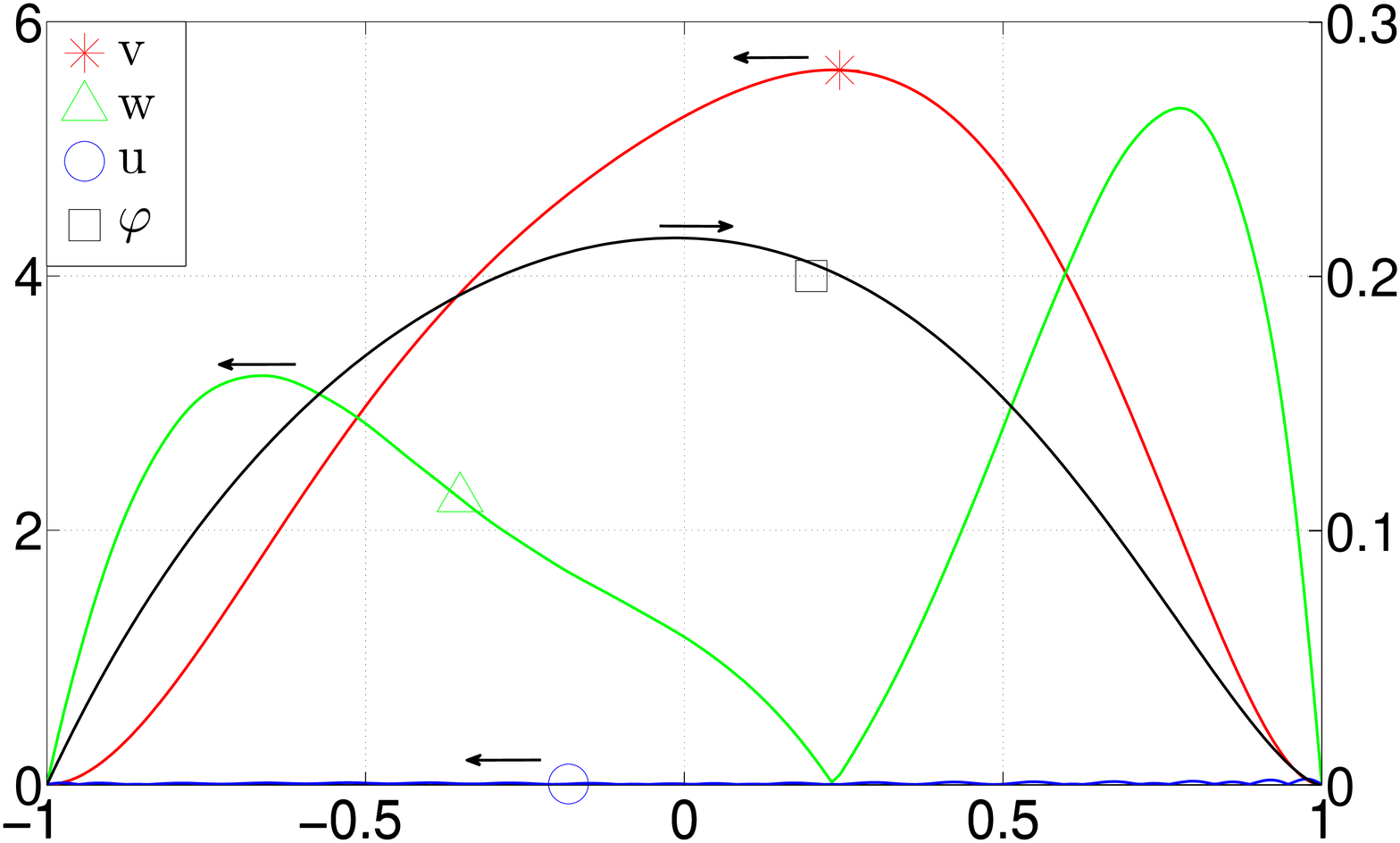}
    \put(-200,100){{\large $(b)$}}
    \put(-185,10){{\large $u$}}
    \put(-185,59){{\large $w$}}
    \put(-185,95){{\large $v$}}
    \put(-5,50){{\large $\varphi$}}
    \put(-95,-5){{\large $y$}} }
  \subfigure {\includegraphics[width=.5\textwidth]
    {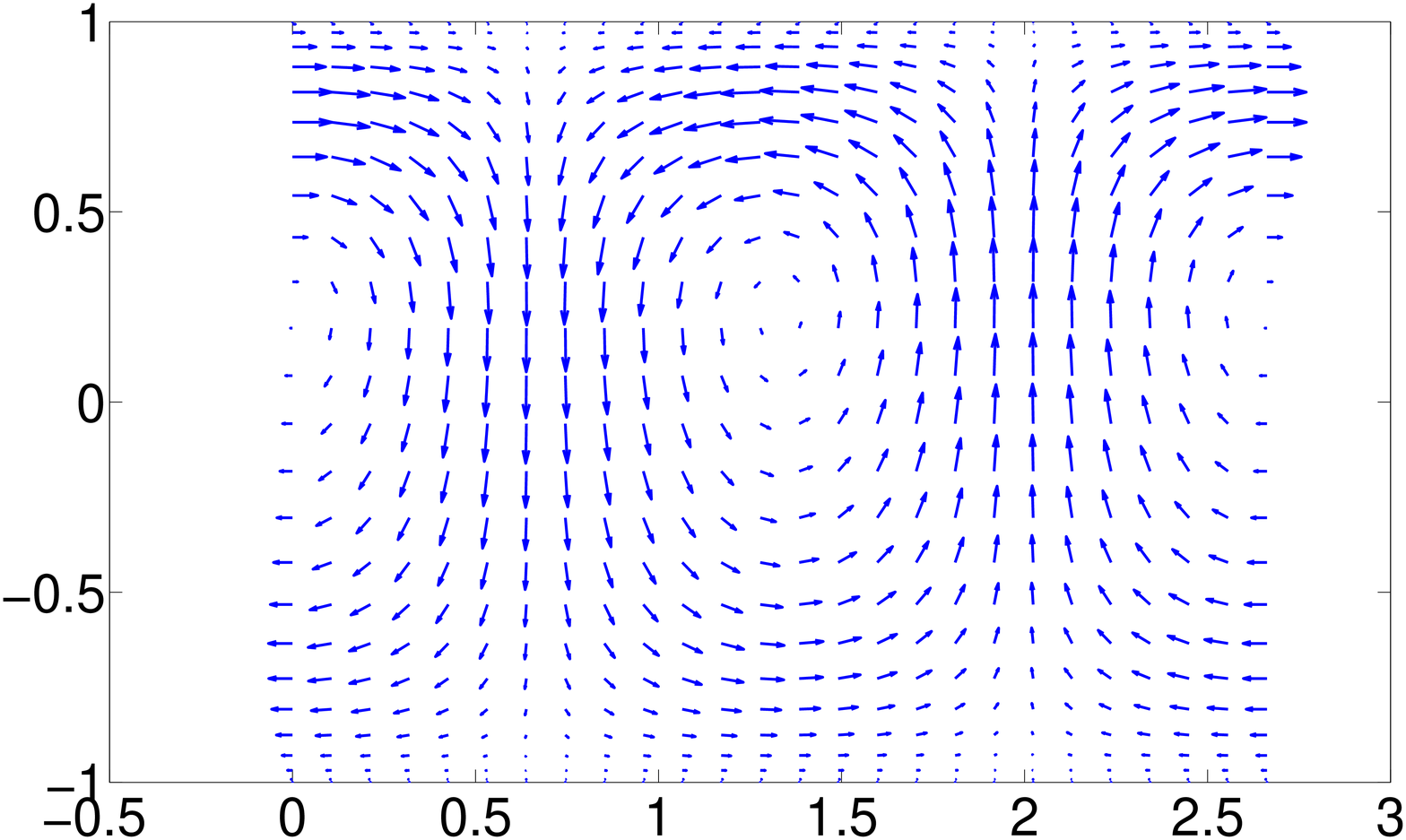}
    \put(-200,100){{\large $(c)$}}
    \put(-190,55){{\large $y$}}
    \put(-95,-5){{\large $z$}}
    \put(-45,-5){{\small $2\pi/\beta$}}
    \includegraphics[width=.5\textwidth]
                    {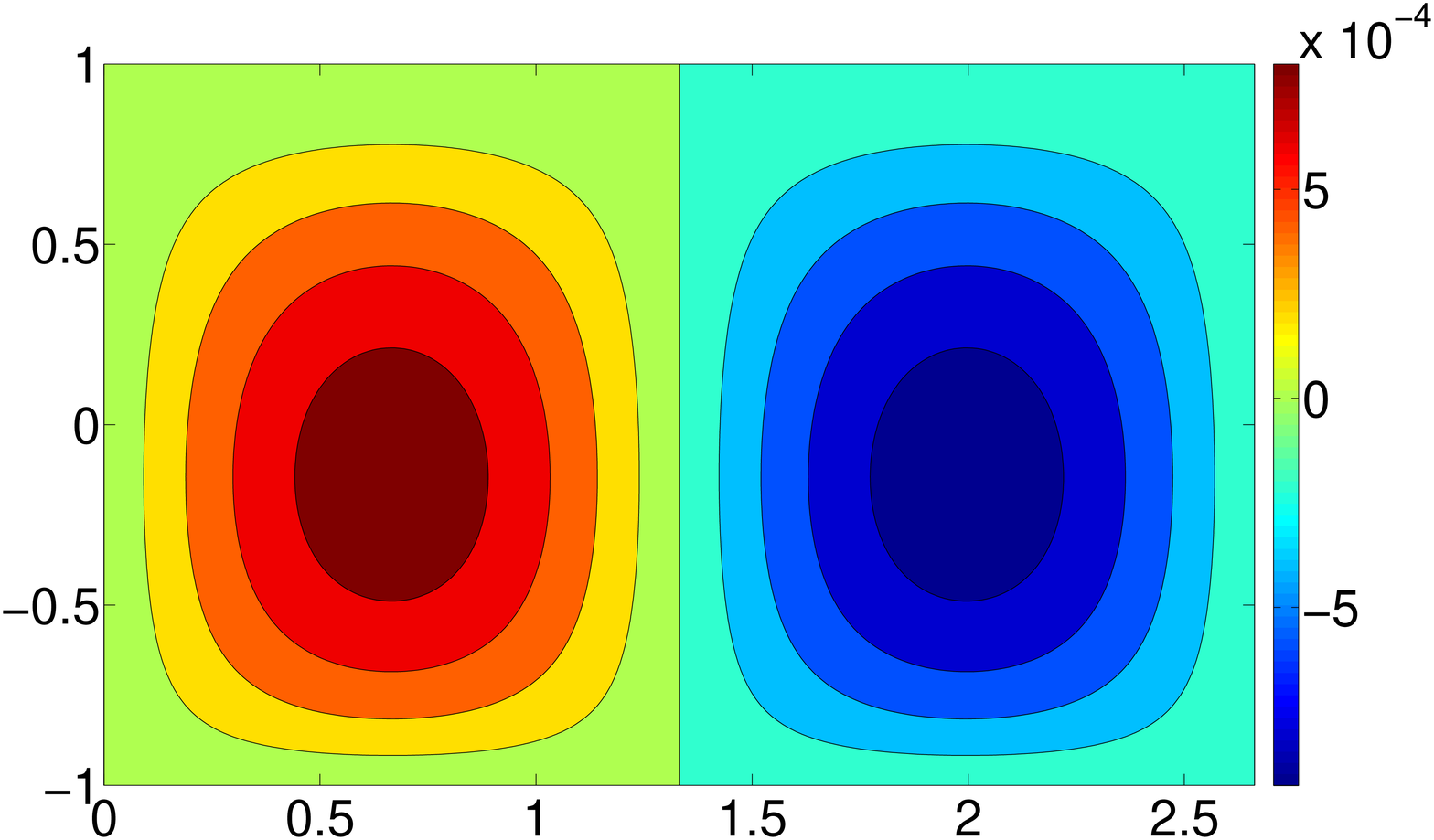}
    \put(-200,100){{\large $(d)$}}
    \put(-190,55){{\large $y$}}
    \put(-95,-5){\large {$z$}}   }
  \caption{Effect of $T$. (a) The optimal initial condition for $v$ as
    a function of $T$ and $\beta$ at $C=100$, $Fe=10^5$, $Re=5000$,
    $\alpha=0$. (b) The optimal initial conditions of velocity and
    potential at $C=100$, $Re=5000$, $Fe=10^5$, $T=100$, $\alpha=0$,
    $\beta=2.36$. (c) Velocity vectors of the optimal initial
    condition in the cross-stream plane; same parameters as in
    (b). (d) Contour of the optimal response for $\varphi$; same
    parameters as in (b).}
  \label{HighRePoiseuilleeffectofT_tg_eigenvector}
\end{figure}

To further investigate the effect of $T$, we plot in
figure~\ref{HighRePoiseuilleeffectofT_tg_eigenvector} the optimal
initial conditions which achieve $G_{max}$ in a given finite time for
parameters $C=100$, $Re=5000$, $Fe=10^5$, $\alpha=0$ and different
values of $T$ with its corresponding optimal $\beta$. In subfigure
(a), the optimal initial conditions for $v$ for various $T$ are
presented. The symmetry of the optimal $v$ with respect to the flow
centerline $y=0,$ when the flow is close to the pure-hydrodynamics
limit, is broken due to the action of the electric field in the
wall-normal direction as $T$ increases. Since the electrode with
higher potential is at $y=1$ in our case, the optimal initial conditions for $w$
and $v$ are tilted towards $y=1$ (see also subfigure (b), which additionally shows
the optimal initial condition for $\varphi$). In subplot (c), the
formation of streamwise vortices in the $y$-$z$-plane is shown; their
centers are shifted upwards by the electric field. In subfigure (d),
the optimal response of $\varphi$, taking the form of waves in the
spanwise direction, is displayed. Recalling that the nonmodal
transient growth is due to base-flow modulations arising from the
tilting of spanwise into wall-normal vorticity, we can state that the
variation of the optimal spanwise wavenumber for different $T$ is the
direct result of the three-dimensional nature of the non-normal
linearized operator under the influence of a constant electric field
pointing in the wall-normal coordinate direction.

\section{Results of energy analysis}\label{energyanalysis}

\subsection{Asymptotic energy analysis} \label{energyNoC}

The dynamics of the disturbance energy (of the velocity fluctuations)
in the limit of an infinite time horizon is examined in this
section. The governing equation for the energy evolution is obtained
by multiplying the linearized equation~(\ref{nd_ehd2}b) by the complex
conjugate velocity $v_i^\dag$, i.e.,
\begin{equation}
  v_i^\dag \frac{\partial v_i}{\partial t} + v_i^\dag  v_j
  \frac{\partial \bar{U}_i}{\partial x_j}  + v_i^\dag  \bar{U}_j
  \frac{\partial v_i}{\partial x_j} =-v_i^\dag  \frac{\partial p}{\partial v_i} +
  v_i^\dag  \frac{M^2}{T} \frac{\partial^2 v_i}{\partial x_j^2}+v_i^\dag M^2
  (\frac{\partial \bar{\phi}}{\partial x_i}\frac{\partial^2 \varphi}{\partial x_j x_j} +
  \frac{\partial^2 \bar{\phi}}{\partial x_j x_j}\frac{\partial \varphi}{\partial x_i}),
\end{equation}
taking the complex conjugate of the obtained equation, and averaging
the two equations, which leaves us with
\begin{eqnarray}
  \frac{\partial \mathcal{E}}{\partial t}  &=& - \frac{1}{2} (v_i^\dag  v_j  + v_j^\dag  v_i )
  \frac{\partial \bar{U}_i}{\partial x_j}    - \frac{M^2}{T}
  \frac{\partial v_i^\dag}{\partial x_j}\frac{\partial v_i}{\partial x_j} -
  \frac{M^2}{2} \frac{\partial \bar{\phi}}{\partial x_i}
  (\frac{\partial \varphi^\dag}{\partial x_j}\frac{\partial v_i}{\partial x_j} +
  \frac{\partial \varphi}{\partial x_j}\frac{\partial v_i^\dag}{\partial x_j}) \nonumber \\
  &&-\frac{M^2}{2}\frac{\partial^2 \bar{\phi}}{\partial x_i \partial x_j}
  (v_i^\dag \frac{\partial \varphi}{\partial x_j} +
  v_i \frac{\partial \varphi^\dag}{\partial x_j}  ) - \frac{M^2}{2}
  \frac{\partial^3 \bar{\phi}}{\partial x_i \partial x_i \partial x_j}
  (\varphi v_j^\dag + \varphi^\dag v_j) \nonumber \\
  &&+ \frac{\partial}{\partial x_j} \left[ -\frac{1}{2} v_i v_i^\dag  \bar{U}_j -
    \frac{1}{2}(v_i^\dag p + v_ip^\dag )\delta_{ij} +  \frac{M^2}{2T}
    ( v_i^\dag\frac{\partial v_i}{\partial x_j} +
    v_i\frac{\partial v_i^\dag}{\partial x_j}) \right. \nonumber \\
    &&+ \left. \frac{M^2}{2}
   \frac{\partial \bar{\phi}}{\partial x_i}
    ( \frac{\partial \varphi}{\partial x_j}v_i^\dag +
    \frac{\partial \varphi^\dag}{\partial x_j}v_i ) +
    \frac{M^2}{2}\frac{\partial^2 \bar{\phi}}{\partial x_i x_i}
    ( \varphi v_j^\dag + \varphi^\dag v_j) \right] ,
\end{eqnarray}
where $\mathcal{E}=v_i^\dag v_i/2$ is the perturbation energy density
of the hydrodynamic part in the spectral space. The terms in the square
brackets are the transport terms which, in case of periodic as well as 
no-slip and no-penetration boundary conditions
, exert no influence on the energy balance. Therefore, after
integrating the above equation over the control volume $\Omega$, we
obtain
\begin{eqnarray}
  \int_{\Omega}\frac{\partial \mathcal{E}}{\partial t} \ dV =
  &-&\underbrace{ \int_{\Omega}\frac{1}{2} (v_i^\dag  v_j  + v_j^\dag  v_i )
    \frac{\partial \bar{U}_i}{\partial x_j} \ dV }_{{\mathsf{Pr}}}
  -\underbrace{\int_{\Omega} \frac{M^2}{T} \frac{\partial v_i^\dag}{\partial x_j}
    \frac{\partial v_i}{\partial x_j} \ dV}_{{\mathsf{VD}}} \nonumber \\
  &-&\underbrace{\int_{\Omega}\frac{M^2}{2} \frac{\partial \bar{\phi}}{\partial x_i}
    \left(\frac{\partial \varphi^\dag}{\partial x_j}\frac{\partial v_i}{\partial x_j} +
    \frac{\partial \varphi}{\partial x_j}\frac{\partial v_i^\dag}
         {\partial x_j}\right)\ dV}_{{\mathsf{VE1}}} \nonumber \\
  &-&\underbrace{\int_{\Omega}\frac{M^2}{2}
    \frac{\partial^2 \bar{\phi}}{\partial x_i \partial x_j}
    \left(v_i^\dag \frac{\partial \varphi}{\partial x_j} + v_i
    \frac{\partial \varphi^\dag}{\partial x_j} \right) \ dV}_{{\mathsf{VE2}}}
  \nonumber \\
  &-&\underbrace{ \int_{\Omega}\frac{M^2}{2}
    \frac{\partial^3 \bar{\phi}}{\partial x_i \partial x_i \partial x_j}
    (\varphi v_j^\dag + \varphi^\dag v_j) \ dV}_{{\mathsf{VE3}}}.
  \label{energybudget}
\end{eqnarray}
Since the boundary conditions are periodic in the wall-parallel
coordinate directions, it is legitimate to consider the control
``volume'' $\Omega$ only in the $y$-direction, that is,
$\Omega=[-1,1]$ and $dV = dy.$ The first term on the right-hand side
of equation~(\ref{energybudget}) represents energy production from the
mean shear (${\mathsf{Pr}}$), which is zero in the hydrostatic case;
the second term describes viscous dissipation (${\mathsf{VD}}$); the
third to fifth terms are the energy transfer terms between the
velocity fluctuation field and the electric field (${\mathsf{VE1}}$,
${\mathsf{VE2}}$, ${\mathsf{VE3}},$ respectively). The Einstein 
summation convention does not apply for the subscripts of ${\mathsf{VE1}}_{ij}$, for example;
the term ${\mathsf{VE1}}_{21}$,
for instance, represents $\displaystyle{\int_{\Omega}}
\displaystyle{\frac{M^2}{2}} \displaystyle{\frac{\partial
    \bar{\phi}}{\partial y}} \left(\frac{\partial
  \varphi^\dag}{\partial x}\frac{\partial v}{\partial x} +
\frac{\partial \varphi}{\partial x}\frac{\partial v^\dag}{\partial
  x}\right)\ dV$.

As has been discussed and verified for polymeric flows
in~\cite{Zhang2013}, the time variation of the normalized perturbation
energy density should be equal to the twice the asymptotic growth rate
of linear disturbances, i.e.,
\begin{equation}
  R_e = \frac{\displaystyle{\int_{\Omega}}
    \displaystyle{\frac{\partial \mathcal{E}}{\partial t}}\ dV }
  {\displaystyle{\int_{\Omega}} \mathcal{E} dV} = 2\omega_i,
  \label{2omega}
\end{equation}
where $\omega_i$ denotes the growth rate of the least stable mode. We
will validate this relation in the following sections and use it as an
{\it{a posteriori}} check for our results.

\subsubsection{EHD without cross-flow}

We apply the energy analysis for hydrostatic EHD flow with different
values of charge diffusion coefficients $Fe$ to probe how the electric
field interacts with the velocity fluctuations. Quantitative results
are listed in table~\ref{tb:energyHS}, with the notation
${\mathsf{VD}} =
{\mathsf{VD}}_{11}+{\mathsf{VD}}_{12}+{\mathsf{VD}}_{21}+{\mathsf{VD}}_{22}$
and, likewise, ${\mathsf{VE}} =
{\mathsf{VE1}}_{21}+{\mathsf{VE1}}_{22}+{\mathsf{VE2}}+{\mathsf{VE3}}$. 
No spanwise dependence as $\beta=0$. Immediately,
one can make several direct observations. First, viscous dissipation
is always negative for the hydrodynamics. Second, since the EHD flow
is hydrostatic, there is no production from the mean shear,
${\mathsf{Pr}}=0$. The only terms that can lead to growths in the
hydrodynamic disturbance energy density $\mathcal{E}$ are linked to
the energy transfer from the electric field, ${\mathsf{VE}}$. The most
efficient mechanism seems to be related to the term
${\mathsf{VE1}}_{21}$, which represents the interaction between the
streamwise perturbed electric field and the wall-normal velocity shear
under the constant effect of the wall-normal base electric field. The
term ${\mathsf{VE3}}$ is even negative, indicating that the electric
field can absorb energy from the perturbed hydrodynamic field by an
out-of-phase configuration between $\varphi$ and $v$ (in the
energy-budget equation for the perturbed electric field, one would
find the exact same term with opposite sign). Regarding the effect of
charge diffusion, with increasing $1/Fe$ (increasing charge diffusion)
from the right column to the left in the table, the total energy
transfer ${\mathsf{VE}}$ diminishes, but, at the same time, the
hydrodynamic diffusion is also dissipating less energy into
heat. Furthermore, even though ${\mathsf{VE}}$ (and thus
${\mathsf{VE1}}_{22}$, ${\mathsf{VE2}}$ and ${\mathsf{VE3}}$)
decreases with rising charge diffusion, the primary mechanism of
energy transfer ${\mathsf{VE1}}_{21}$ transfers more energy from the
electric field to the hydrodynamic fluctuations, together with a less dissipation,
leading to an
unstable flow for the chosen parameters. Therefore, it seems that the
effect of charge diffusion is to catalytically enhance the efficiency
of the most productive energy transfer mechanism between the perturbed
electric field and the hydrodynamics, expressed by the term
${\mathsf{VE1}}_{21}$, and result in a lower dissipation. 
As a consequence, increasing charge diffusion
leads to a more unstable flow.

\begin{table}
  \centering
  \begin{tabular}{l r r r r}
    terms & $Fe=10^4$ & $Fe=10^5$ & $Fe=10^6$ & $Fe=10^7$ \\ \hline
    ${\mathsf{VD}}_{11}$ & -254.9836 & -255.3751 & -255.4628 & -255.4781 \\
    ${\mathsf{VD}}_{12}$ & -481.9379 & -486.1119 & -487.0231 & -487.1773 \\
    ${\mathsf{VD}}_{21}$ & -570.6289 & -570.2374 & -570.1497 & -570.1344 \\
    ${\mathsf{VD}}_{22}$ & -254.9836 & -255.3751 & -255.4628 & -255.4781 \\
    ${\mathsf{VE1}}_{21}$ & 1154.1206 & 1152.8793 & 1152.6403 & 1152.5936 \\
    ${\mathsf{VE1}}_{22}$ & 428.695 & 431.8134 & 432.4984 & 432.6197 \\
    ${\mathsf{VE2}}$ & 30.0141 & 30.7628 & 30.9299 & 30.9614 \\
    ${\mathsf{VE3}}$ & -50.2196 & -48.3506 & -47.9797 & -47.9186 \\
    ${\mathsf{VD}}$ & -1562.5341 & -1567.0994 & -1568.0984 & -1568.2679 \\
    ${\mathsf{VE}}$ & 1562.6102 & 1567.1049 & 1568.089 & 1568.2561 \\
    ${\mathsf{Pr}}$ & 0 & 0 & 0 & 0 \\
    $R_e$ & 0.0761 & 0.0054 & -0.0094 & -0.0119 \\
    $2 \omega_{i}$ & 0.0761 & 0.0054 & -0.0094 & -0.0119
  \end{tabular}
  \caption{Energy budget for modal instability of hydrostatic EHD flow
    for different values of the charge diffusion. The results are
    normalized as in equation~(\ref{2omega}). The parameters are
    $C=50$, $M=100$, $T=160$, $\alpha=2.57$, $\beta=0$ for hydrostatic
    flow.}
  \label{tb:energyHS}
\end{table}

It is instructive to assess the effect of $M$ on the linear stability
(see section~\ref{EHDwo}) with the help of the energy-budget
equation~(\ref{energybudget}).  For a vanishing time derivative of the
energy density, the factor $M^2$ on the right-hand side can be
eliminated for the hydrostatic case $\bar{U}=0$. In the case of
cross-flow, however,  with the same reasoning $M$ will have an influence on the linear
stability criterion.

\subsubsection{EHD with low-$Re$ cross-flow} \label{crossflowlowRe}

Table~\ref{tb:energyPoi} shows the results for $C = 50$, $M = 100$, $T
= 160$, $Re = T/M^2 = 0.016$, $\alpha= 2.57$, $\beta = 0$ with
cross-flow for different $Fe$. Compared to the case without
cross-flow, the results are quite similar. However, it is interesting
to note that the fluctuation energy production from the mean shear
${\mathsf{Pr}}$, even though rather small, is negative, indicating
that the perturbed flow field transfers energy to the base
flow. Recalling the results in figure~\ref{PoiseuilleeffectofM} of
section~\ref{EHDlowRe}, a change of $M$ does not have a strong effect
on the rate of change of the disturbance energy density ${\mathsf{E}}$
since ${\mathsf{Pr}}$ is very small.

In the case of EHD flow with a weak cross-flow ($Re=0.016$), the main
mechanism for transferring energy into the hydrodynamic subsystem is
still based on the potential difference across the two electrodes ---
the same as for the no cross-flow case. This can be confirmed by
inspecting table~\ref{tb:energyPoi}: ${\mathsf{VE1}}_{21}$ is the
dominant energy transfer term.

\begin{table}
  \centering
  \begin{tabular}{l r r r r}
    terms & $Fe=10^4$ & $Fe=10^5$ & $Fe=10^6$ & $Fe=10^7$ \\ \hline
    ${\mathsf{VD}}_{11}$ & -254.1325 & -254.1337 & -254.1318 & -254.1309 \\
    ${\mathsf{VD}}_{12}$ & -475.0779 & -476.292 & -476.5068 & -476.5348 \\
    ${\mathsf{VD}}_{21}$ & -571.48 & -571.4788 & -571.4807 & -571.4816 \\
    ${\mathsf{VD}}_{22}$ & -254.1325 & -254.1337 & -254.1318 & -254.1309 \\
    ${\mathsf{VE1}}_{21}$ & 1158.7772 & 1157.62 & 1157.3579 & 1157.3025 \\
    ${\mathsf{VE1}}_{22}$ & 422.786 & 423.6012 & 423.7687 & 423.7989 \\
    ${\mathsf{VE2}}$ & 26.3245 & 26.363 & 26.3747 & 26.3794 \\
    ${\mathsf{VE3}}$ & -52.8799 & -51.3981 & -51.1094 & -51.0629 \\
    ${\mathsf{VD}}$ & -1554.8229 & -1556.0382 & -1556.2511 & -1556.2782 \\
    ${\mathsf{VE}}$ & 1555.0078 & 1556.1861 & 1556.3919 & 1556.4179 \\
    ${\mathsf{Pr}}$ & -0.0079038 & -0.0080187 & -0.0080507 & -0.0080566 \\
    $R_e$ & 0.1769 & 0.1399 & 0.1327 & 0.1316 \\
    $2 \omega_{i}$ & 0.1769 & 0.1399 & 0.1327 & 0.1316
  \end{tabular}
  \caption{Energy budget for modal instability of EHD flow with low-$Re$
    cross-flow for different values of the charge diffusion $Fe$. The
    results are normalized as in equation~(\ref{2omega}). The
    parameters are $C=50$, $M=100$, $T=160$, $Re=T/M^2=0.016$,
    $\alpha=2.57$, $\beta=0$ with cross-flow.}
  \label{tb:energyPoi}
\end{table}

\subsubsection{EHD with high-$Re$ cross-flow} \label{crossflowhighRe}

\begin{table}
  \centering
  \begin{tabular}{l r r r r}
    terms \ $(\times 10^{-4})$ & $T=10^{-8}$ & $T=10$ & $T=100$ & $T=200$ \\ \hline
    ${\mathsf{VD}}_{11}$ & -2.643 & -2.643 & -2.6429 & -2.6429 \\
    ${\mathsf{VD}}_{12}$ & -138.22 & -138.24 & -138.42 & -138.61 \\
    ${\mathsf{VD}}_{21}$ & -0.99338 & -0.99339 & -0.99343 & -0.99347 \\
    ${\mathsf{VD}}_{22}$ & -2.643 & -2.643 & -2.6429 & -2.6429 \\
    ${\mathsf{VE1}}_{21}$ & 3.1782 $\!\cdot 10^{-10}$ & 0.31785 & 3.1814 & 6.3689 \\
    ${\mathsf{VE1}}_{22}$ & 7.3458 $\!\cdot 10^{-10}$ & 0.73467 & 7.355 & 14.729 \\
    ${\mathsf{VE2}}$ & 7.0501 $\!\cdot 10^{-11}$ & 0.070507 & 0.70555 & 1.4122 \\
    ${\mathsf{VE3}}$ & -1.2043 $\!\cdot 10^{-10}$ & -0.12044 & -1.2055 & -2.4133 \\
    ${\mathsf{VD}}$ & -144.5 & -144.52 & -144.7 & -144.89 \\
    ${\mathsf{VE}}$ & 10.025 $\!\cdot 10^{-10}$ & 1.0026 & 10.036 & 20.096 \\
    ${\mathsf{Pr}}$ & 132.35 & 132.06 & 129.41 & 126.44 \\
    $R_e$ & -12.153 & -11.463 & -5.2526 & 1.6542 \\
    $2 \omega_{i}$ & -12.153 & -11.463 & -5.2526 & 1.6542
  \end{tabular}
  \caption{Energy budget for modal instability of EHD flow with a
    high-$Re$ cross-flow for different values of the stability
    parameter $T$. The results are normalized as in
    equation~(\ref{2omega}). The parameters are $C=100$, $Fe=10^5$,
    $Re=5500$, $\alpha=1$ and $\beta=0$ with cross-flow.}
  \label{tb:energyPoiT}
\end{table}

The energy analysis for the EHD Poiseille flow at $C=100$, $Fe=10^5$,
$Re=5500$, $\alpha=1$ and $\beta=0$ is summarized in
table~\ref{tb:energyPoiT}. Note that the production ${\mathsf{Pr}}$ is
dimishing with increasing $T$. On the other hand, ${\mathsf{VE}}$
increases with larger values of $T$, compensating and exceeding the
decrease of ${\mathsf{Pr}}$ at higher $T$. This is consistent with the
results in figure~\ref{HighRePoiseuilleeffectofT}: higher values of
$T$ yield a more unstable flow. However, the principal mechanism
underlying the flow instability is still linked to the production
${\mathsf{Pr}}$. The electric field only assumes a secondary role in
destabilizing the flow, at least for the parameters considered in this
case. Unlike the hydrostatic case where ${\mathsf{VE1}}_{21}$ is
responsible for the dominant energy transfer, in the presence of
cross-flow ${\mathsf{VE1}}_{22}$ becomes the most efficient agent
transferring fluctuation energy ${\mathcal{E}}$ between the electric
field and the perturbed velocity field.

\subsection{Transient energy analysis} \label{transientenergyanalysis}

To investigate the cause for the increase of nonmodal growth with $T$, recall figure~\ref{HighRePoiseuilleeffectofT_tg}, we formulate and perform an energy analysis for the initial-value problem of equation~(\ref{lineartimeevolution}). We consider the energy density evolution over a finite time horizon following~\cite{Butler1992}
\begin{eqnarray}
  \frac{1}{\vert \Omega \vert}\int_{\Omega}\frac{\partial \mathcal{E}}{\partial t}
  \ dV &=& \frac{1}{\vert\Omega\vert}\int_{-1}^{1}\int_{0}^{a}\int_{0}^{b}
  \frac{\partial}{\partial t} \left(\frac{u^2+v^2+w^2}{2}\right)\ dV \nonumber \\
  &=& \frac{1}{\vert\Omega\vert}\int_{-1}^{1}\int_{0}^{a}\int_{0}^{b}
   \underbrace{\left[ -uv\frac{d U}{d y}\right.}_{{\mathsf{Pr}}}
    - \underbrace{\frac{1}{Re}\frac{\partial u_i}{\partial x_j}
      \frac{\partial u_i}{\partial x_j}}_{{\mathsf{VD}}} \\
    && \phantom{123456} - \underbrace{\frac{T}{Re}
      \frac{\partial \bar{\phi}}{\partial y} \frac{\partial \varphi}{\partial x_j}
      \frac{\partial v}{\partial x_j}}_{{\mathsf{VE1}}}
    - \underbrace{\frac{T}{Re}\frac{\partial^2 \bar{\phi}}{\partial y^2} v
      \frac{\partial \varphi}{\partial y} }_{{\mathsf{VE2}}}
    - \underbrace{\left.\frac{T}{Re}\frac{\partial ^3 \bar{\phi}}{\partial y^3}
      \varphi v \right]}_{{\mathsf{VE3}}} \ dz\ dx\ dy \nonumber
\end{eqnarray}
where $\vert\Omega\vert=2ab$, $a=2\pi/\alpha$ and $b=2\pi/\beta$. In
the above equation, we label, as before, the first term on the
right-hand side as ${\mathsf{Pr}}$ (production from the mean shear),
the second term as ${\mathsf{VD}}$ (viscous dissipation), the third to
fifth terms collectively as ${\mathsf{VE}}$ (energy density transfer
between the perturbed velocity field and the perturbed electric field)
and the sum of all five terms as ${\mathsf{Total}}$. In this temporal
evolution problem, the initial condition is the optimal one, following
the procedure in section~\ref{EHDhighRe}. 

\begin{figure}
  \centering
  \subfigure {\includegraphics[width=.5\textwidth]
    {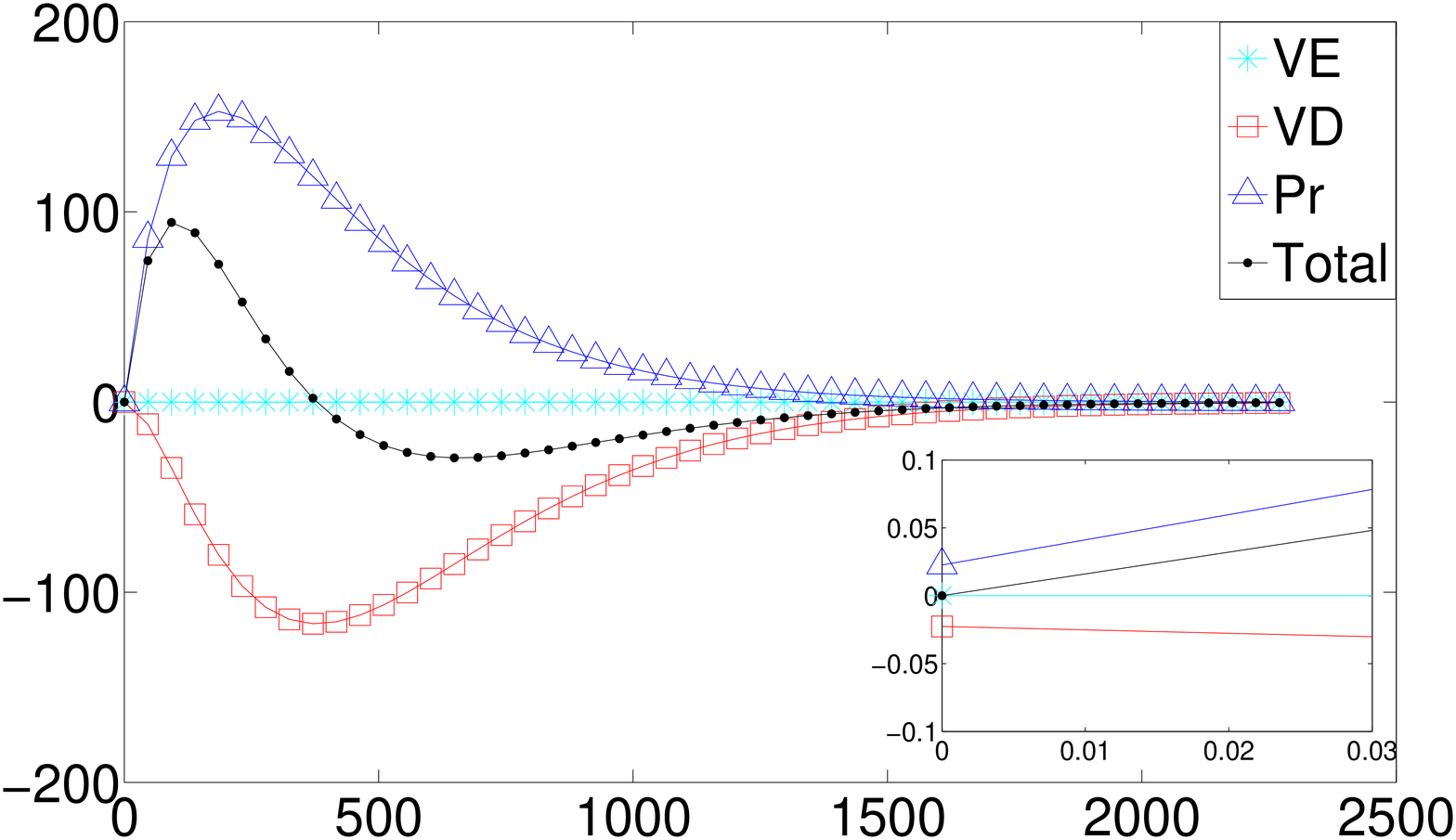}
    \put(-200,100){{\large $(a)$}}
    \put(-190,50){{\large $E$}}
    \put(-95,-5){{\large $t$}}
    \includegraphics[width=.5\textwidth]
                    {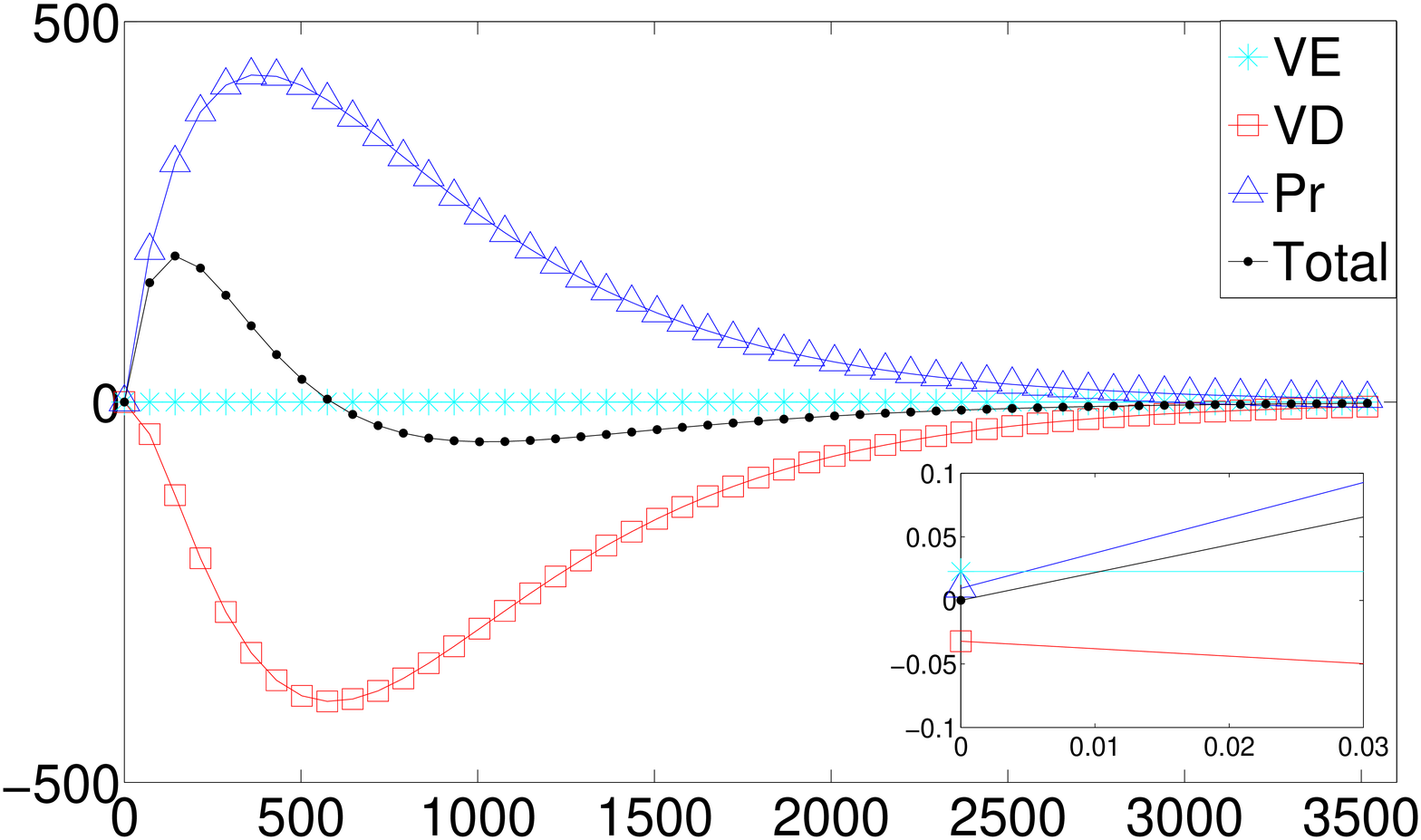}
    \put(-200,100){{\large $(b)$}}
    \put(-190,50){{\large $E$}}
    \put(-95,-5){{\large $t$}} }
  \caption{Energy analysis over a finite time horizon at $C=100$,
    $Re=5000$, $Fe=10^5$, $\alpha=0.$ (a) $T=10^{-8}$, $\beta=2.05;$
    (b) $T=100$, $\beta=2.36$.}
  \label{EnergyanalysisTG}
\end{figure}

Results of our energy analysis are presented in
figure~\ref{EnergyanalysisTG}. In subfigure (a) and its inset, the
pure hydrodynamic result is shown, where the production
${\mathsf{Pr}}$ counteracts the viscous dissipation
${\mathsf{VD}}$. In subplot (b) for EHD cross-flow, we observe that the term
${\mathsf{VE}}$ is insignificant, even though at $T=100$; this is in
contrast to both the linear modal stability criterion (see
figure~\ref{HighRePoiseuilleeffectofT}) and the overall nonmodal
transient growth (see figures~\ref{HighRePoiseuilleeffectofT_tg}
and~\ref{HighRePoiseuilleeffectofT_tg_eigenvector}) where its effect is
not negligible. Furthermore, production ${\mathsf{Pr}}$ increases by a
factor of $2 \sim 3$ compared to the pure hydrodynamic flow. These
results seem to indicate that, concerning the nonmodal analysis, the
effect of the additional electric field on the canonical channel flow
is incidental, i.e., the perturbation velocity energy is only
indirectly influenced by the electric field; in fact, the electric
field does not induce a substantial energy transfer directly into
velocity fluctuations at all and its effect is to enhance the lift-up mechanism, therefore the production.

Examining more closely the inset in figure~\ref{EnergyanalysisTG}(b),
we see that the term ${\mathsf{VE}}$ surpasses ${\mathsf{Pr}}$ only in
the very beginning of the time horizon. This is due to the high-$Re$
regime we are investigating. As discussed earlier, $Re={T}/{M^2}$
represents the ratio of the momentum relaxation time $L^{*2}/\nu^*$ to
the charge relaxation time $L^{*2}/(K^*\Delta \phi_0^*)$. With the
maximum ${\mathsf{Total}}$ energy achieved at $t_{max} \approx 144$ in
figure~\ref{EnergyanalysisTG}(b), we can estimate the time horizon in
the inset by observing that $144/5000=0.029$, a value close to the
time scale depicted in the inset of (b). Moreover, the minimal energy
growth due to the electric force validates our previous observation
that the purely EHD-induced non-normality is rather small (see
appendix~\ref{appinputoutput} for a direct proof of this statement via
an input-output formulation). In the case of other complex flows at
high Reynolds numbers, a similar conclusion can be draw, for
instance, in viscoelastic flows~\citep{Zhang2013,Brandt2014}, polymer
stretching cannot induce disturbance growth when the fluid inertia is
prevalent.

\section{Discussion and conclusions} \label{discussion}

In this article, we have presented a comprehensive linear stability analysis
of charge-injection-induced electrohydrodynamic flows between two plate electrodes, 
covering the hydrostatic as well as the cross-flow case,
employing modal as well as nonmodal tools. We intend to 
examine whether the linear framework is sufficient for describing 
the transition to turbulence of EHD flow in the early phase of perturbation 
evolution. It is hoped that the results presented above and summarized 
below would help to understand better the EHD flow 
instability and its transition mechanism and shed light on its flow physics 
as well as flow control design.

\subsection{EHD without cross-flow}
In the hydrostatic case, the often-omitted charge diffusion is taken
into account and found to have a non-negligible effect, particularly
on the critical linear stability parameter $T_c$ with SCL injection
--- a finding running contrary to a common assumption in previous
studies. In those studies, a linear stability analysis predicts a
critical value of $T_c \approx 161$ in the strong injection
limit. This result is reproduced in our computations for a negligible
value of $1/Fe,$ but even for a moderate amount of charge diffusion
the flow quickly becomes more unstable. Hence, we suggest that charge
diffusion be accounted for in linear stability analyses and numerical
simulations whenever the real physics indicates charge diffusion that can not be neglected, 
as it improves not only the model of the flow physics but also
the robustness of the numerics as well. In fact, the common use of 
total variation diminishing (TVD) 
schemes \citep{Harten1983} in direct numerical simulations of EHD flow, which introduces
{\it{artificial}} numerical diffusion, seems unnecessary when
{\it{true}} physical charge diffusion could be included.

The longstanding discrepancy of the critical stability parameter $T_c$
between the experimental and theoretical value, however, could not be
resolved by our analysis: even though $T_c$ in the SCL limit drops to
$140$ at $Fe=10^3$ (a physical value according to~\cite{Perez1989}), a
substantial gap remains to the experimentally measured parameter of
$T_c \approx 110.$ Motivated by the researches in subcritical channel flow,
we examine other mechanisms for early transition to
turbulence, specifically, transient growth due to the non-normality
of the linearized EHD operator. Nonmodal stability theory has been
successfully applied to the variety of wall-bounded shear flows in an
attempt to explaining aspects of the transition process. In the case
of hydrostatic EHD flow, our calculations seem to indicate that
transient energy growth, as defined in
equation~(\ref{energydefinition}), is not significant, reaching gains
of $\sim 10$ at most: the flow instability is rather dictated by the
asymptotic growth rate of the least stable mode.
These results seem to indicate that below the critical $T_c$ the significant energy 
growth observed in the real EHD flow is not of a linear nature, 
otherwise the linear framework would succeed to detect it.
It might be hypothesized that the major energy growth mechanism 
in subcritical hydrostatic EHD follows a nonlinear route; nevertheless, 
it is only after performing a full nonlinear simulation of subcritical EHD flow that 
can one conclude whether its energy growth mechanism is truly nonlinear or not.
Besides, these results also seem to shed some light on the flow control of hydrostatic EHD.
It is now well-established that in canonical channel flow, where the perturbation energy 
is found to grow linearly in the early phase, a linear flow control strategy is sufficient to 
abate the perturbation development
\citep{Kim2003,Kim2007}. Due to the limited early perturbation energy growth in hydrostatic EHD, 
we thus suggest that different flow control methods be examined and applied in addressing
the flow control of such flow. 
There might exist another possibility for the limited transient growth. 
As discussed by~\cite{Atten1974}, the correct prediction of the linear
stability criterion might require a closer comparison between the
experimental conditions and the mathematical model. In this light, one
may suggest a re-examination of the charge creation and transport
processes, as the current charge creation model does not seem to
accommodate any efficient energy transfer from the electric to the
flow field, during the linear phase. 

\subsection{EHD with cross-flow}
The flow instability and the transition to turbulence in canonical or 
complex channel (for example, EHD, MHD or polymeric) flows are currently
not well understood. The study on the complex channel flow, serving as 
a supplement to the investigation of the canonical flows, focuses on the 
flow modification under the influence of external fields, for example, 
electric field, magnetic field or polymer stress field. The study of such
 flows will not only improve our understanding of these particular flow 
 configurations, but also, more importantly, help us to better understand, 
 during the linear, transition and turbulent phases, the dynamics of the important flow 
 structures, for instance, the streak formation and attenuation, by probing 
 the interaction between the fluids (or flow structures) and the external fields. 
 For example, in the point of view of flow control, the research on polymer 
 turbulence drag reduction reveals the mechanism how the auto-generation 
 cycles of turbulence is modified in the presence of polymer molecules \citep{DUBIEF2004}. 
 This has led to an even broader picture of the dynamics of turbulence. 
 Similarly, in the case of EHD, our goal is to understand how the flow changes 
 in response to the electric effects and provide a physical interpretation. 
 Below we present the results of EHD cross-flow. We differentiated low-$Re$ and
high-$Re$ cases. For low-$Re$ flow, the effects of $M$ and $Fe$ are
similar to those of the hydrostatic flow, with the linear stability
criterion being smaller for low-$Re$ cross-flow when compared to
hydrostatic flow.

The high-$Re$ case is more interesting. In both modal and nonmodal
stability analyses, the canonical channel flow becomes more unstable,
once an electric field is applied between the two electrodes. From an
input-output and an energy analysis we found, however, that the energy
growth directly related to the electric field is not significant and
that the effect of the electric field on the flow instability is
indirect. In general, in high-$Re$ channel flow, the maximum transient
growth is achieved by vortices aligning along the streamwise
coordinate direction and generating streamwise streaks via an
efficient energy growth mechanism known as lift-up. These optimal
streamwise vortices are symmetric with respect to the channel
centerline for standard Poiseille flow. In contrast to other complex
flows, in EHD flows the electric field, which always points in the
wall-normal direction, actively participates in the formation of the
streamwise rolls by accelerating the downward-moving fluid (note that,
in our setting, the injector is at $y=1$). Consequently, this yields
stronger transient growth via the lift-up mechanism, when compared to
the common channel flow. In other words, the electric field provides
wall-normal momentum.  As has been discussed in~\cite{Landahl1980} and
recently reviewed by~\cite{Brandt2014}, the presence of wall-normal
momentum will cause any three-dimensional, asymptotically stable or
unstable shear flow to exhibit energy growth during a transient
phase. In the present study, the role of the electric field is to
provide the shear flow with such a source of wall-normal momentum and
to strengthen the lift-up mechanism for EHD flow with high-$Re$
cross-flow. Besides, we also find that the optimal wavenumbers for
maximum transient growth increase under a stronger electric
effect. Since the electric field will help to establish streamwise
vortices, it may constitute a good actuator for drag reduction
techniques, using the two-dimensional rolls together with a flow
control strategy as described in~\cite{Schoppa1998,Soldati1998}.

\begin{acknowledgments}
F.M. was supported by the Italian Ministry for University and Research
under grant PRIN 2010. The authors would like to thank Emanuele
Bezzecchi for his initial input. M.Z. would like to thank Prof. Luca
Brandt of the Royal Institute of Technology (KTH), Sweden and Dr. Peter Jordan of Universit\'e de Poitiers, France.
\end{acknowledgments}

\begin{appendix}
\section{Code validation}\label{validation}

\begin{figure}
  \centering
  \subfigure{\includegraphics[width=.5\textwidth]
    {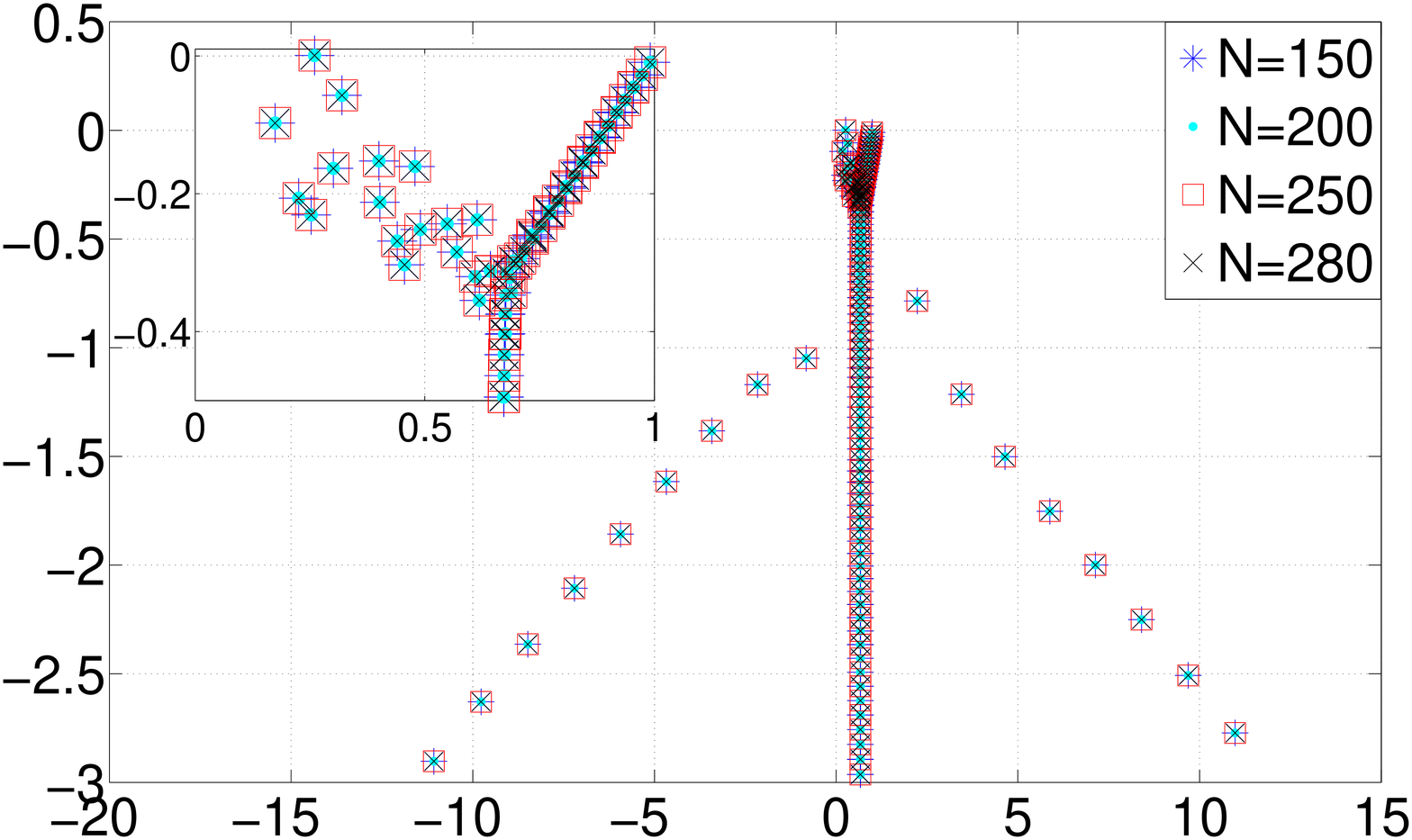}
    \put(-200,100){{\large $(a)$}}
    \put(-190,50){{\large $\omega_i$}}
    \put(-95,-5){{\large $\omega_r$}}
    \includegraphics[width=.5\textwidth]
                    {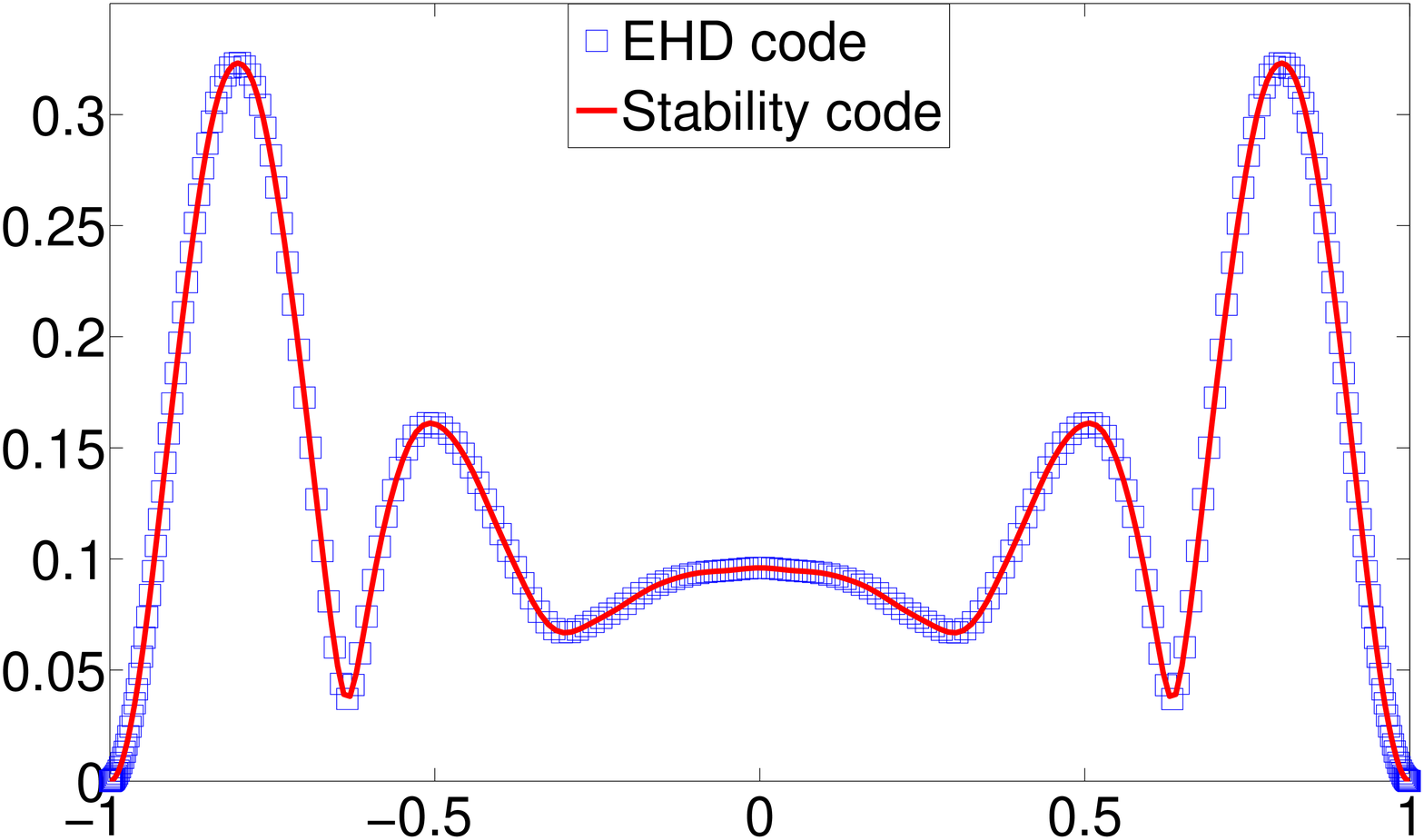}
                    \put(-200,100){{\large $(b)$}}
                    \put(-190,50){{\large $v$}}
                    \put(-95,-5){{\large $y$}}}
  \caption{Code validation. (a) Resolution check for EHD flow with
    cross-flow at $C=50$, $Fe=2000$, $Re=6000$, $T=100$,
    $M=\sqrt{T/Re}=0.1291$, $\alpha=1$ and $\beta=0$. (b) Eigenvector
    component $v$ for the most unstable mode (normalized to have the
    same peak value for the two codes). The parameters are $C=50$,
    $M=10^{-11}$, $Fe=2000$, $Re=6000$, $T=M^2\cdot Re$, $\alpha=1$,
    $\beta=0$ and $N=250$ for the EHD code, and $Re=6000$, $\alpha=1$,
    $\beta=0$ and $N=250$ for the hydrodynamic stability code.}
  \label{resolutioncheck}
\end{figure}

We first perform a resolution check to examine the convergence of the
results. The parameters in this case are $C=50$, $Fe=2000$, $Re=6000$,
$T=100$, $M=\sqrt{T/Re}=0.129$, $\alpha=1$ and $\beta=0$. The
eigenspectra for four different grid resolutions $N$ are shown in
figure~\ref{resolutioncheck}(a). The most unstable mode in these cases
are listed in table~\ref{unstablemodecheck}. Satisfactory convergence,
with increasing $N$, is observed.

Secondly, the EHD eigenvector, from using~(\ref{iterativemethod}), is
examined against a verified, pure hydrodynamic stability code
employing the same spectral collocation method and solving the
Orr-Sommerfeld-Squire system, see~\cite{Schmid2001}, as shown in
figure \ref{resolutioncheck} (b). The parameters for the EHD code are
$C=50$, $M=10^{-11}$, $Fe=2000$, $Re=6000$, $T=M^2\cdot Re$,
$\alpha=1$, $\beta=0$ and $N=250$ . The parameters for the
hydrodynamic stability code are $Re=6000$, $\alpha=1$, $\beta=0$ and
$N=250$. We see that the iteratively solved EHD eigenvector is the
same as the pure hydrodynamic one, which is solved directly by the
Matlab routine {\tt{eig}}. For the computation of the transient
amplification $G$ in equation~(\ref{TGequation}), it is legitimate to
include only the first several, most unstable
modes~\citep{Schmid2001}, i.e., eigenmodes corresponding to
eigenvalues with imaginary part smaller than a certain $\omega_i^c$
are discarded, see table~\ref{numberofmodes} for a validation of this
approach. The reason for a minor increase of $G,$ as more modes are
included, is due to the newly incorporated eigenvectors, not because
of an insufficiently refined grid.

\begin{table}
  \centering
  \begin{tabular}{ll}
    $N$ & most unstable mode \\ \hline 
    $150$ \qquad \qquad & $0.260023637950300 + 0.000652797815269i$ \\  
    $200$ & $0.260023637089882 + 0.000652796819289i$ \\ 
    $250$ & $0.260023637069851 + 0.000652796791624i$ \\ 
    $280$&$0.260023637052960 + 0.000652796810920i$ \\ 
  \end{tabular}
  \caption{Code validation. Resolution check for the most unstable
    eigenvalue of EHD flow with cross-flow; with same parameters as
    in~figure~\ref{resolutioncheck}.}
  \label{unstablemodecheck}
\end{table}

\begin{table}
  \centering
  \begin{tabular}{rcccrc}
    $\qquad \qquad \omega_i^c$ & $G$ & \phantom{12} & \phantom{12} & $\qquad \qquad \omega_i^c$ & $G$ \\ \hline
    $-3.5$ & $3.312404$ & & & $-0.5$  & $1.138543e+04$ \\
    $-5.5$ & $3.434333$ & & & $-3.5$  & $1.174459e+04$ \\
    $-7.5$ & $3.434351$ & & & $-10.5$ & $1.175565e+04$\\
  \end{tabular}
  \caption{$G$ versus different cut-off growth rates $\omega_i^c$. The
    first two columns represent hydrostatic EHD flow at $C=50$,
    $Fe=10^5$, $M=100$, $T=155$, $\alpha=2.5$, $\beta=0$, $N=250$. The
    last two columns represent EHD flow with cross-flow at $C=100$,
    $Fe=10^5$, $Re=5000$, $T=100$, $\alpha=0$, $\beta=2.36$,
    $N=250$. Refer to the text for the definition of $\omega_i^c$.}
\label{numberofmodes}
\end{table}

With the eigenvalue problem reliably solved as shown above, we present
validation for the specific flows considered here. In the case of
hydrostatic flow, our results for $Fe=10^7$, approximating the case
of zero charge diffusion, $T_c=160.67$ and $\alpha=2.57$ at $C=50$ (see figure~\ref{effectofFe}(a) in
section~\ref{EHDwo}), are very
close to the linear stability criterion reported in~\cite{Atten1972},
$T_c=160.75$ and $\alpha=2.569$ in the case of $C \rightarrow \infty$, 
where a coupled flow and electric system with neglected charge 
diffusion has been considered. This additionally implies that a value of $C$ 
higher than $50$ can well approximate the space-charge-limit.

\begin{figure}
  \centering
  \subfigure {\includegraphics[width=.5\textwidth]{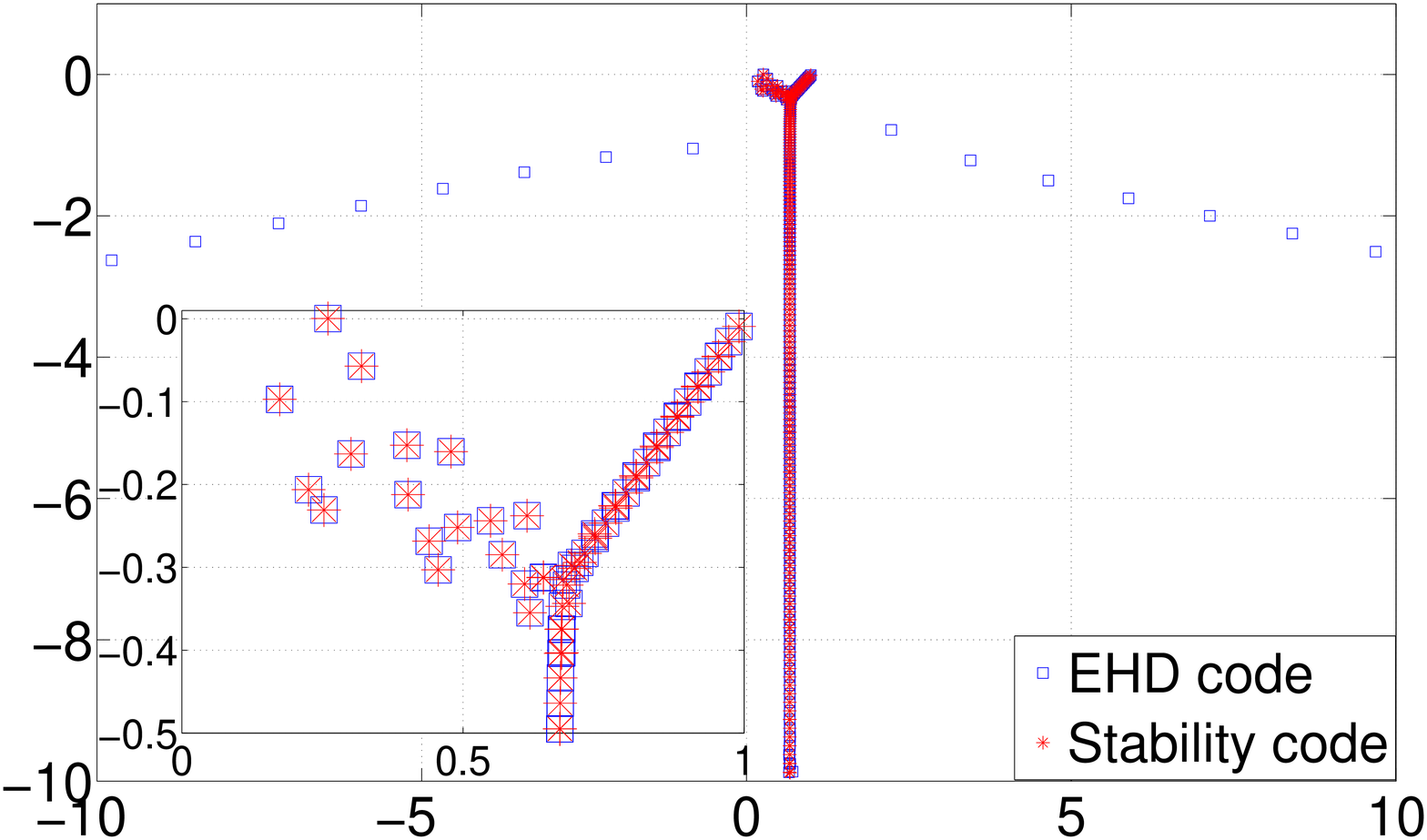}
    \put(-200,100){{\large $(a)$}}
    \put(-190,50){{\large $\omega_r$}}
    \put(-95,-5){{\large $\omega_i$}}
    \includegraphics[width=.5\textwidth]{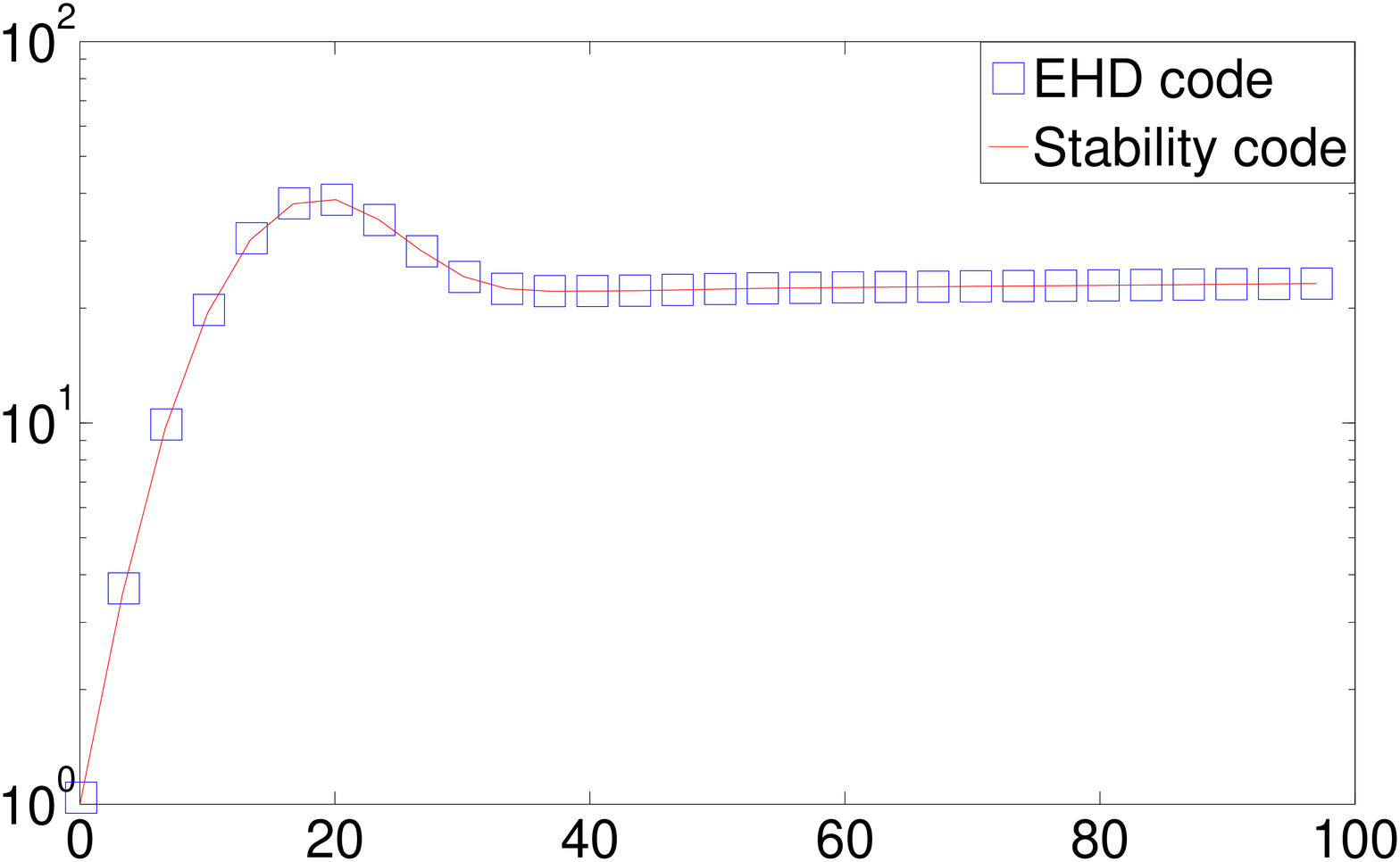}
    \put(-200,100){{\large $(b)$}}
    \put(-190,50){{\large $G$}}
    \put(-95,-5){{\large $t$}}   }
  \caption{Comparison between EHD-Poiseuille flow and canonical
    Poiseuille flow. The parameters are $C=50$, $M=10^{-11}$,
    $Fe=2000$, $Re=6000$, $T=M^2\cdot Re$, $\alpha=1$, $\beta=0$ and
    $N=250$ for the EHD code, and $Re=6000$, $\alpha=1$, $\beta=0$ and
    $N=250$ for the purely hydrodynamic stability code. (a) The
    eigenvalue spectrum. (b) Transient amplification of initial energy.}
  \label{spectrumcomparison}
\end{figure}

In the presence of cross-flow, since there exist no quantitative
results for eigenvalues and eigenvectors of the EHD problem in the
literature, we partially verify our results by examining the pure
hydrodynamic limit of the EHD-linearized problem, i.e., with electric
effects being very small. This comparison is made with the stability
code. The parameters are identical to the ones chosen above for the
comparison of the eigenvectors. The Poiseuille base flow is
$\bar{U}=1-y^2$ in both codes. It is obvious that, with these selected
parameters, the governing equation~(\ref{ehdeq3}c) for $\varphi$ is
void of the coupling with $v,$ since $\bm{L}_{v\varphi}$ in
equation~(\ref{ehdeq1}a) is negligible. Therefore, the hydrodynamics
equations for $v$ and $\eta$ in~(\ref{ehdeq1}a) and~(\ref{ehdeq2}b)
must reproduce the results of the stability code. This match is shown
in figure~\ref{spectrumcomparison}. In subfigure (a), the spectra of
two codes are seen to collapse, even in the intersection region of the
three eigenbranches, which is known to be sensitive due to the high
non-normality of the linearized
system~\citep{Schmid2001}. Additionally, the blue eigenmodes in (a)
not matched by the red hydrodynamic modes are the supplementary
eigenvalues linked to the presence of an electric field. The most
unstable eigenvalue is shown in table~\ref{codecomparison}. In
subfigure (b), transient growth using an eigenvector expansion with
$n=71$ eigenmodes is shown. A quantitative comparison of the maximum
transient growth $G_{max}$ and its corresponding time $t_{max}$ is
presented in in table~\ref{codecomparison}. Agreement up to the fourth
digit is achieved.
The computations of $t_{max}$ and $G_{max}$ involve $n=71$
eigenfunctions, each one solved with the iterative method. Even though
each individual mode may be prone to small inaccuracies,
figure~\ref{spectrumcomparison}(b) illustrates that transient growth
(a multi-modal phenomenon) can be reliably and robustly computed using
the eigenvector expansion outlined above.

\begin{table}
  \centering
  \begin{tabular}{ccc}
    \phantom{1234567890}& EHD code & hydrodynamic stability code\\ \hline
    $\omega_{max,r}$ & $0.259815871017297$ & $0.259815871062631$\\
    $\omega_{max,i}$ & $0.000323088678313$ & $0.000323088655527$\\
    $t_{max}$ & $18.86745$ & $18.87514$ \\
    $G_{max}$ & $38.92401$ & $38.93307$  \\
  \end{tabular}
  \caption{Code validation. The parameters are the same as in
    figure~\ref{spectrumcomparison}.}
  \label{codecomparison}
\end{table}

\section{Input-output formulation}\label{appinputoutput}

An input-output formulation can reveal additional information on
prevalent instability mechanisms by considering different types of
forcings (input) and responses (output)~\citep{JOVANOVIC2005}. To
demonstrate that the transient growth due to perturbative $\varphi$ is
small, we compare the full responses to perturbations consisting of
(i) all variables $v, \eta$ and $\varphi$, (ii) both $v$ and $\eta$,
and (iii) only $\varphi$. We thus define for these three cases
different input filters $\bm{B}$, where $\bm{B}_{in1} = \bm{I}_{3N \times 3N}$ for
the first case, while for the second and third cases we have
\begin{equation}
  B_{in2} = \begin{pmatrix} \bm{I}_{N\times N}& \bm{0}\\
    \bm{0}& \bm{I}_{N\times N}\\ \bm{0}& \bm{0} \end{pmatrix}, \qquad \qquad
  B_{in3} = \begin{pmatrix} \bm{0}\\ \bm{0}\\ \bm{I}_{N\times N} \end{pmatrix}.
\end{equation}
The output filter $\bm{C}_{out}$ is $\bm{I}_{3N\times 3N}$ for all three cases:
we examine the flow response in all velocities and the electric
field. Consequently, the energy weight matrices should be redefined
with $\bm{\mathcal{M}}_{out} = \bm{C}_{out}\bm{\mathcal{M}} \bm{C}_{out}^T$ and
$\bm{\mathcal{M}}_{in} = \bm{B}_{in}^T\bm{\mathcal{M}} \bm{B}_{in}$. After
applying a Cholesky decomposition to these energy weight matrices, we
obtain $\bm{F}_{out}$ and $\bm{F}_{in}$ for a formulation based on the
$L_2$-norm. Finally, the maximum transient growth $G$ over a finite
time interval is given by
\begin{eqnarray}
  G(t) &=& \underset{\bm{\gamma}_0}{\max}\frac{||\bm{\gamma}_{out}(t)||_{E_{out}}}
  {||\bm{\gamma}_{in}(0)||_{E_{in}}}=\underset{\bm{\gamma}_0}{\max}
  \frac{||\bm{\mathcal{T}}\bm{\gamma}_{in}(0)||_{E_{out}}}
       {||\bm{\gamma}_{in}(0)||_{E_{in}}}=\underset{\bm{\gamma}_0}{\max}
       \frac{||\bm{F}_{out}\bm{\mathcal{T}} \bm{\gamma}_{in}(0)||_2}
            {||\bm{F}_{in}\bm{\gamma}_{in}(0)||_2} \notag \\
            &=&\underset{\bm{\gamma}_0}{max}
            \frac{||\bm{F}_{out}\bm{\mathcal{T}}\bm{F}_{in}^{-1}\bm{F}_{in}\bm{\gamma}_{in}(0)||_2}
                 {||\bm{F}_{in}\bm{\gamma}_{in}(0)||_2}=
                 ||\bm{F}_{out}\bm{\mathcal{T}} \bm{F}_{in}^{-1}||_{2} \notag \\
                 &=& ||\bm{F}_{out}\bm{C}_{out}e^{t\bm{\mathcal{L}}}\bm{B}_{in}\bm{F}_{in}^{-1}||_{2}.
\end{eqnarray}

\begin{table}
  \centering
  \begin{tabular}{ccccccc}
    input &\phantom{1234} & $B_{in1}\{v,\eta,\varphi\}$ &\phantom{12} &
    $B_{in2}\{v,\eta\}$ & \phantom{12} & $B_{in3}\{\varphi\}$ \\ \hline
    $t_{max}$ & & 583.4087 & & 587.7147 & & 552.0443 \\
    $G_{max}$ & & 11765.40 & & 11350.38 & & 428.8926
  \end{tabular}
  \caption{Result from an input-output analysis at $C=100$, $Fe=10^5$,
    $Re=5000$, $T=100$, $\alpha=0$ and $\beta=2.36$.}
  \label{inputoutput}
\end{table}

We report the transient growth results for the above three cases in
table~\ref{inputoutput} at $C=100$, $Fe=10^5$, $Re=5000$, $T=100$,
$\alpha=0$ and $\beta=2.36$. We observe that perturbations solely in
$\varphi$ (case (iii)) exhibit transient growth two orders smaller
than in the other two cases. For cases (i) and (ii) the transient
growth characteristics are nearly identical which suggests that the
nonnormality of the linear operator is mainly related to the
hydrodynamics.

\end{appendix}

\bibliographystyle{jfm}

\begin{thebibliography}{48}
\expandafter\ifx\csname natexlab\endcsname\relax\def\natexlab#1{#1}\fi

\bibitem[Alj {\em et~al.\/}(1985)Alj, Denat, Gosse, Gosse \& Nakamura]{Alj1985}
{\sc Alj, A., Denat, A., Gosse, J.-P., Gosse, B. \& Nakamura, I.} 1985 Creation
  of charge carriers in nonpolar liquids. {\em IEEE Trans. on Electrical
  Insulation\/} {\bf EI-20}~(2), 221--231.

\bibitem[Allen \& Karayiannis(1995)]{Allen1995}
{\sc Allen, P. \& Karayiannis, T.} 1995 Electrohydrodynamic enhancement of heat
  transfer and fluid flows. {\em Heat Recovery Systems and CHP\/} {\bf 15}~(5),
  389 -- 423.

\bibitem[Atten(1974)]{Atten1974}
{\sc Atten, P.} 1974 Electrohydrodynamic stability of dielectric liquids during
  transient regime of space-charge-limited injection. {\em Physics of Fluids\/}
  {\bf 17}~(10), 1822--1827.

\bibitem[Atten(1976)]{Atten1976}
{\sc Atten, P.} 1976 R{\^o}le de la diffusion dans le probl{\`e}me de la
  stabilit{\'e} hydrodynamique d'un liquide di{\`e}lectrique soumis {\`a} une
  injection unipolaire forte. {\em Compt. Rend. Acad. Sci. Paris\/} {\bf 283},
  29--32.

\bibitem[Atten \& Honda(1982)]{Atten1982}
{\sc Atten, P. \& Honda, T.} 1982 The electroviscous effect and its explanation
  {I}-the electrohydrodynamic origin; study under unipolar {D.C.} injection.
  {\em J. Electrostatics\/} {\bf 11}~(3), 225 -- 245.

\bibitem[Atten \& Lacroix(1979)]{Atten1979}
{\sc Atten, P. \& Lacroix, J.~C.} 1979 Non-linear hydrodynamic stability of
  liquids subjected to unipolar injection. {\em J. M{\'e}canique\/} {\bf 18},
  469--510.

\bibitem[Atten \& Moreau(1972)]{Atten1972}
{\sc Atten, P. \& Moreau, R.} 1972 Stabilit{\'e} electrohydrodynamique des
  liquides isolants soumis {\`a} une injection unipolaire. {\em J.
  M{\'e}canique\/} {\bf 11}, 471--520.

\bibitem[Bart {\em et~al.\/}(1990)Bart, Tavrow, Mehregany \& Lang]{Bart1990}
{\sc Bart, S.~F., Tavrow, L.~S., Mehregany, M. \& Lang, J.~H.} 1990
  Microfabricated electrohydrodynamic pumps. {\em Sensors and Actuators A:
  Physical\/} {\bf 21}~(1-3), 193 -- 197.

\bibitem[Boyd(2001)]{Boyd2001}
{\sc Boyd, J.} 2001 {\em Chebyshev and Fourier Spectral Methods\/}. 2nd revised
  edition, Dover Publications.

\bibitem[Bradshaw(1969)]{Bradshaw1969}
{\sc Bradshaw, P.} 1969 The analogy between streamline curvature and buoyancy
  in turbulent shear flow. {\em Journal of Fluid Mechanics\/} {\bf 36},
  177--191.

\bibitem[Brandt(2014)]{Brandt2014}
{\sc Brandt, L.} 2014 The lift-up effect: The linear mechanism behind
  transition and turbulence in shear flows. {\em Eur. J. Mech. B/Fluids\/} {\bf
  47}, 80 -- 96.

\bibitem[Bushnell \& McGinley(1989)]{Bushnell1989}
{\sc Bushnell, D.~M. \& McGinley, C.~B.} 1989 Turbulence control in wall flows.
  {\em Annual Review of Fluid Mechenics\/} {\bf 21}, 1--20.

\bibitem[Butler \& Farrell(1992)]{Butler1992}
{\sc Butler, K.~M. \& Farrell, B.~F.} 1992 Three-dimensional optimal
  perturbations in viscous shear flow. {\em Phys. Fluids\/} {\bf 4}~(8),
  1637--1650.

\bibitem[Butler \& Farrell(1993)]{Butler1993}
{\sc Butler, K.~M. \& Farrell, B.~F.} 1993 Optimal perturbations and streak
  spacing in wall-bounded turbulent shear flows. {\em Physics of Fluids A:
  Fluid Dynamics (1989-1993)\/} {\bf 5}~(3), 774--777.

\bibitem[Castellanos(1998)]{Castellanos1998}
{\sc Castellanos, A.} 1998 {\em Electrohydrodynamics\/}. Springer-Verlag.

\bibitem[Castellanos \& Agrait(1992)]{Castellanos1992}
{\sc Castellanos, A. \& Agrait, N.} 1992 Unipolar injection induced
  instabilities in plane parallel flows. {\em IEEE Trans. on Industry
  Applications,\/} {\bf 28}~(3), 513--519.

\bibitem[Chakraborty {\em et~al.\/}(2009)Chakraborty, Liao, Adler \&
  Leong]{Chakraborty2009}
{\sc Chakraborty, S., Liao, I.-C., Adler, A. \& Leong, K.~W.} 2009
  Electrohydrodynamics: A facile technique to fabricate drug delivery systems.
  {\em Adv. Drug Delivery Rev.\/} {\bf 61}~(12), 1043--1054.

\bibitem[Darabi {\em et~al.\/}(2002)Darabi, Rada, Ohadi \& Lawler]{Darabi2002}
{\sc Darabi, J., Rada, M., Ohadi, M. \& Lawler, J.} 2002 Design, fabrication,
  and testing of an electrohydrodynamic ion-drag micropump. {\em J.
  Microelectromechanical Systems\/} {\bf 11}~(6), 684--690.

\bibitem[Dubief {\em et~al.\/}(2004)Dubief, White, Terrapon, Shaqfeh, Moin \&
  Lele]{DUBIEF2004}
{\sc Dubief, Y., White, C.~M., Terrapon, V.~E., Shaqfeh, E. S.~G., Moin, P. \&
  Lele, S.~K.} 2004 On the coherent drag-reducing and turbulence-enhancing
  behaviour of polymers in wall flows. {\em Journal of Fluid Mechanics\/} {\bf
  514}, 271--280.

\bibitem[Farrell \& Ioannou(1996)]{Farrell1996}
{\sc Farrell, B.~F. \& Ioannou, P.~J.} 1996 Generalized stability theory.
  {P}art {I}: Autonomous operators. {\em Journal of the Atmospheric Sciences\/}
  {\bf 53}~(14), 2025--2040.

\bibitem[F{\'e}lici(1971)]{Felici1971}
{\sc F{\'e}lici, N.} 1971 {DC} conduction in liquid dielectrics ({P}art {II}):
  {E}lectrohydrodynamic phenomena. {\em Direct Current and Power Electronics\/}
  {\bf 2}, 147--165.

\bibitem[Grossmann \& Lohse(2000)]{GROSSMANN2000}
{\sc Grossmann, S. \& Lohse, D.} 2000 Scaling in thermal convection: a unifying
  theory. {\em Journal of Fluid Mechanics\/} {\bf 407}, 27--56.

\bibitem[Harten(1983)]{Harten1983}
{\sc Harten, A.} 1983 High resolution schemes for hyperbolic conservation laws.
  {\em Journal of Computational Physics\/} {\bf 49}~(3), 357 -- 393.

\bibitem[Jim{\'e}nez \& Pinelli(1999)]{JIMENEZ1999}
{\sc Jim{\'e}nez, J. \& Pinelli, A.} 1999 The autonomous cycle of near-wall
  turbulence. {\em Journal of Fluid Mechanics\/} {\bf 389}, 335--359.

\bibitem[Jones(1978)]{Jones1978}
{\sc Jones, T.} 1978 Electrohydrodynamically enhanced heat transfer in liquids-
  a review. {\em Advances in heat transfer\/} {\bf 14}, 107--148.

\bibitem[Jovanovi{\'c} \& Bamieh(2005)]{JOVANOVIC2005}
{\sc Jovanovi{\'c}, M.~R. \& Bamieh, B.} 2005 Componentwise energy
  amplification in channel flows. {\em Journal of Fluid Mechanics\/} {\bf 534},
  145--183.

\bibitem[Kim(2003)]{Kim2003}
{\sc Kim, J.} 2003 Control of turbulent boundary layers. {\em Physics of Fluids
  (1994-present)\/} {\bf 15}~(5), 1093--1105.

\bibitem[Kim \& Bewley(2007)]{Kim2007}
{\sc Kim, J. \& Bewley, T.~R.} 2007 A linear systems approach to flow control.
  {\em Annual Review of Fluid Mechanics\/} {\bf 39}~(1), 383--417.

\bibitem[Kourmatzis \& Shrimpton(2012)]{Kourmatzis2012}
{\sc Kourmatzis, A. \& Shrimpton, J.~S.} 2012 Turbulent three-dimensional
  dielectric electrohydrodynamic convection between two plates. {\em Journal of
  Fluid Mechanics\/} {\bf 696}, 228--262.

\bibitem[Lacroix {\em et~al.\/}(1975)Lacroix, Atten \& Hopfinger]{Lacroix1975}
{\sc Lacroix, J.~C., Atten, P. \& Hopfinger, E.~J.} 1975 Electro-convection in
  a dielectric liquid layer subjected to unipolar injection. {\em Journal of
  Fluid Mechanics\/} {\bf 69}, 539--563.

\bibitem[Landahl(1980)]{Landahl1980}
{\sc Landahl, M.~T.} 1980 A note on an algebraic instability of inviscid
  parallel shear flows. {\em Journal of Fluid Mechanics\/} {\bf 98}, 243--251.

\bibitem[Lee {\em et~al.\/}(2006)Lee, Cho, Huh, Ko, Lee, Jang, Lee, Kang \&
  Choi]{Lee2006}
{\sc Lee, J.-G., Cho, H.-J., Huh, N., Ko, C., Lee, W.-C., Jang, Y.-H., Lee,
  B.~S., Kang, I.~S. \& Choi, J.-W.} 2006 Electrohydrodynamic {(EHD)}
  dispensing of nanoliter {DNA} droplets for microarrays. {\em Biosensors and
  Bioelectronics\/} {\bf 21}~(12), 2240--2247.

\bibitem[Martinelli {\em et~al.\/}(2011)Martinelli, Quadrio \&
  Schmid]{Martinelli2011}
{\sc Martinelli, F., Quadrio, M. \& Schmid, P.~J.} 2011 Stability of planar
  shear flow in presence of electroconvection. In {\em VII Int. Symp. on
  Turbulence and Shear Flow Phenomena\/}.

\bibitem[Melcher(1981)]{Melcher1981}
{\sc Melcher, J.~R.} 1981 {\em Continuum Electromechanics\/}. MIT Press.

\bibitem[Ogilvie \& Proctor(2003)]{Ogilvie2003}
{\sc Ogilvie, G.~I. \& Proctor, M. R.~E.} 2003 On the relation between
  viscoelastic and magnetohydrodynamic flows and their instabilities. {\em
  Journal of Fluid Mechanics\/} {\bf 476}, 389--409.

\bibitem[P{\'e}rez \& Castellanos(1989)]{Perez1989}
{\sc P{\'e}rez, A.~T. \& Castellanos, A.} 1989 Role of charge diffusion in
  finite-amplitude electroconvection. {\em Phys. Rev. A\/} {\bf 40},
  5844--5855.

\bibitem[Saad(2011)]{Saad2011}
{\sc Saad, Y.} 2011 {\em Numerical Methods for Large Eigenvalue Problems\/}.
  SIAM Press.

\bibitem[Schmid(2007)]{Schmid2007}
{\sc Schmid, P.~J.} 2007 Nonmodal stability theory. {\em Annual Review of Fluid
  Mechenics\/} {\bf 39}, 129--162.

\bibitem[Schmid \& Brandt(2014)]{Schmid2014}
{\sc Schmid, P.~J. \& Brandt, L.} 2014 Analysis of fluid systems: Stability,
  receptivity, sensitivity. {\em Appl. Mech. Rev.\/} {\bf 66}~(2), 024803.

\bibitem[Schmid \& Henningson(2001)]{Schmid2001}
{\sc Schmid, P.~J. \& Henningson, D.~S.} 2001 {\em Stability and Transition in
  Shear Flows\/}. Springer Verlag, New York.

\bibitem[Schneider \& Watson(1970)]{Schneider1970}
{\sc Schneider, J.~M. \& Watson, P.~K.} 1970 Electrohydrodynamic stability of
  space-charge-limited currents in dielectric liquids. {I}. {T}heoretical
  study. {\em Physics of Fluids\/} {\bf 13}~(8), 1948--1954.

\bibitem[Schoppa \& Hussain(1998)]{Schoppa1998}
{\sc Schoppa, W. \& Hussain, F.} 1998 A large-scale control strategy for drag
  reduction in turbulent boundary layers. {\em Physics of Fluids\/} {\bf
  10}~(5), 1049--1051.

\bibitem[Soldati \& Banerjee(1998)]{Soldati1998}
{\sc Soldati, A. \& Banerjee, S.} 1998 Turbulence modification by large-scale
  organized electrohydrodynamic flows. {\em Physics of Fluids\/} {\bf 10}~(7),
  1742--1756.

\bibitem[Traor{\'e} \& P{\'e}rez(2012)]{Traore2012}
{\sc Traor{\'e}, P.~H. \& P{\'e}rez, A.~T.} 2012 Two-dimensional numerical
  analysis of electroconvection in a dielectric liquid subjected to strong
  unipolar injection. {\em Physics of Fluids\/} {\bf 24}~(3), 037102.

\bibitem[Trefethen {\em et~al.\/}(1993)Trefethen, Trefethen, Reddy \&
  Driscoll]{Trefethen1993}
{\sc Trefethen, L.~N., Trefethen, A.~E., Reddy, S.~C. \& Driscoll, T.~A.} 1993
  Hydrodynamic stability without eigenvalues. {\em Science\/} {\bf 261}~(5121),
  578--584.

\bibitem[Weideman \& Reddy(2000)]{Weideman2000}
{\sc Weideman, J.~A. \& Reddy, S.~C.} 2000 A {MATLAB} differentiation matrix
  suite. {\em ACM Trans. on Mathematical Software\/} {\bf 26}~(4), 465--519.

\bibitem[Wu {\em et~al.\/}(2013)Wu, Traor\'e, V\'azquez \& P\'erez]{Wu2013}
{\sc Wu, J., Traor\'e, P., V\'azquez, P.~A. \& P\'erez, A.~T.} 2013 Onset of
  convection in a finite two-dimensional container due to unipolar injection of
  ions. {\em Physical Review E\/} {\bf 88}, 053018.

\bibitem[Zhang {\em et~al.\/}(2013)Zhang, Lashgari, Zaki \& Brandt]{Zhang2013}
{\sc Zhang, M., Lashgari, I., Zaki, T.~A. \& Brandt, L.} 2013 Linear stability
  analysis of channel flow of viscoelastic {O}ldroyd-{B} and {FENE-P} fluids.
  {\em Journal of Fluid Mechanics\/} {\bf 737}, 249--279.

\end{thebibliography}

\end{document}